\pdfoutput=1
\documentclass[epj,final]{svjour}
\usepackage{latexsym}
\usepackage{amssymb}
\usepackage{amsmath}
\usepackage{textcomp}
\usepackage{graphics}
\usepackage{epsfig}
\usepackage{float}
\usepackage{placeins}
\usepackage{longtable}
\usepackage{graphicx}
\usepackage{multirow}
\usepackage{mciteplus}
\usepackage{subfigure}
\usepackage{bm}

 \usepackage[pdftitle={Optical absorption in boron clusters B$_{6}$ and B$_{6}^{+}$: A first principles 
configuration interaction approach by Ravindra Shinde},pdfauthor={Ravindra Shinde \& Alok Shukla}]{hyperref}
 \usepackage[all]{hypcap}
  \hypersetup{
  unicode=false,          
  pdftoolbar=true,        
  pdfmenubar=true,        
  pdffitwindow=true,     
  pdfstartview={FitH},    
  pdfsubject={Computational Condensed Matter},   
  pdfnewwindow=true,      
  pdfcreator={Latex},
  colorlinks=true,       
  linkcolor=red,          
  citecolor=blue,        
  urlcolor=blue,           
  }

\begin{document}
\title{Optical absorption in boron clusters B$_{6}$ and B$_{6}^{+}$: A first principles configuration interaction approach}
\author{Ravindra Shinde \and Alok Shukla} 

\institute{Department of Physics, Indian Institute of Technology Bombay, Mumbai, Maharashtra, India 400076. 
\email{ravindra.shinde@iitb.ac.in, shukla@phy.iitb.ac.in}}
%
\date{Received: date / Revised version: date}
%

\abstract{
The linear optical absorption spectra in neutral boron cluster B$_{6}$ and cationic 
 B$_{6}^{+}$ are calculated using a first principles correlated electron approach.
 The geometries of several low-lying isomers of these clusters were optimized at the
coupled-cluster singles doubles (CCSD) level of theory. With
these optimized ground-state geometries, excited states of
different isomers were computed using the singles
configuration-interaction (SCI) approach.
The many body wavefunctions of various excited states have been analysed
and the nature of optical excitation involved are found to be of collective, plasmonic type.
} 

\authorrunning{Ravindra Shinde and Alok Shukla}
\titlerunning{Optical absorption in boron clusters B$_{6}$ and B$_{6}^{+}$: A first principles CI approach}
\maketitle

\section{Introduction}
\label{intro}
The cluster science is now a fast emerging field with
novel properties and tremendous potential for applications. 
A cluster of atoms, or molecules can have few atoms to
thousands of atoms in it, with various structural forms,
such as spheres, tubes, planar structures \emph{etc}.
Most of the time, the properties exhibited by clusters are completely different than their
bulk counterpart \cite{alonso_book, jena_book}.
Boron, in particular, is as interesting as carbon, because
of its short covalent radius and ability to form any structure due to catenation.
Boron also exhibits structures like nanotubes, fullerenes,
planar sheets etc. Some of the structures of boron have
been studied in the context of hydrogen storage \cite{cabria_alonso}.
Both $\sigma$ and $\pi$ aromaticity is observed in many planar boron
clusters. Some planar boron structures are also
found to be analogous to hydrocarbons \cite{kiran_wang}.
A circular B$_{19}^{-}$ cluster,
with a unit of B$_{6}$ wheel in the center behaves as a Wankel motor,
\emph{i.e.} the inner B$_{6}$ wheel rotating opposite to the outer
B$_{13}$ ring \cite{boron-19-rotor,boron19-natchem}. The all-boron
clusters are promising candidates as inorganic ligands \cite{coord-chem-review,boron-8-ligand}. 

There have been various experimental and theoretical investigations of
bare boron clusters. Electronic structures of boron quasi-planar structures, 
tubes, wheels and rings were studied
experimentally using photo-emission spectroscopy by
Wang and co-workers \cite{kiran_wang,tube_wang, wheel_wang, ring_wang}. 
They also gave a density functional theory (DFT) based explanation of their electronic
structure. Alexandrova \emph{et. al.} explored the structural and electronic properties
and chemical bonding in B$_{6}^{-}$ and B$_{6}$ using anion photoelectron spectroscopy and \emph{ab initio} calculations
\cite{structure-bonding-b6}.
La Placa, Roland and Wynne studied abundance
spectrum of boron clusters generated using laser ablation
technique \cite{la_placa}. Ionic boron clusters were experimentally studied by 
Hanley, Whitten and Anderson \cite{hanley_whitten}. They studied the fragmentation
of bigger clusters into smaller ones by studying the collision-induced dissociation of boron clusters. 
Boustani, on the other hand, gave a
theoretical description of electronic structures of small
bare boron clusters using configuration interaction
method, although with a relatively small Gaussian basis set \cite{boustani_prb97}.
Recently a DFT based study of bare boron clusters was
done by Ati\c{s}, \"{O}zdogan and G\"{u}ven\c{c} \cite{turkish_boron}. 
As far as optical properties of boron clusters are
concerned, there are very few reports available. Marques
and Botti studied the optical absorption spectra of B$_{20}$, B$_{38}$,
B$_{44}$, B$_{80}$ and B$_{92}$ using time-dependent DFT \cite{marques_fullerene}. 
To the best of our knowledge, no other experimental results are available
on the optical absorption of boron clusters.

Our group has recently studied the optical absorption
in boron clusters B$_{n}$ (n=2 -- 5) using a large-scale multi-reference 
configuration interaction method \cite{nano_life}. Since this
method is quite extensive and scales as N$^6$, where N is the
number of orbitals used in the calculations, it cannot be
applied for large clusters, or clusters with no symmetry.
However good insights can be achieved with relatively
less extensive method known as singles configuration
interaction (SCI), containing only one electron excitations
from the Hartree-Fock ground state. This method has been
extensively used for the study of the excited states and optical absorption in
various other systems \cite{sci-oligofluorenes,sci-phenylene,sci-spectra-jcp,cpl-indo-sci,sci-si29-apl,sci-c60-prb}. 
Since optical absorption
spectra is very sensitive to the structural geometry, the
optical absorption spectroscopy along with the extensive
calculations of optical absorption spectra, can be used to
distinguish between distinct isomers of a cluster. In this
report, we present an extensive calculations of the linear
optical absorption spectrum of low-lying isomers of
B$_{6}$ and B$_{6}^{+}$ clusters with different structures. This study along with
experimental absorption spectra can lead to identification
of these distinct isomers. Also, in the interpretation of the measured spectra,
the theoretical insight of the excited states of clusters play an important role \cite{vlasta-cis-excited}.

The remainder of this paper is organised as follows. Next section describes the theoretical and computational
details of the work, followed by section \ref{sec:results}, in which results are presented and discussed.
In the last section we present our conclusions, and explore the scope for future work. 
Detailed information regarding the excited states contributing to the optical absorption is presented in the appendix.

\section{Theoretical and Computational Details}
\label{sec:theory}

Different possible arrangements and orientations of atoms of the B$_{6}$ cluster 
(both neutral and cationic) were randomly selected for the initial configurational 
search of geometries of isomers. For a given spin multiplicity, the geometry optimization was done at a 
correlated level, i.e. at singles doubles coupled cluster (CCSD) level with 6-311G(d,p) basis set 
as implemented in {\tt Gaussian 09} \cite{gaussian09}. 
Since neutral cluster can have singlet or higher spin multiplicity, the optimization was repeated
for different spin configurations to get the lowest energy isomer. Similarly for cationic clusters with
odd number of electrons, spin multiplicities of 2 and 4 were considered in the optimization.
In total, we have obtained 11 neutral B$_{6}$  and 8 cationic B$_{6}^{+}$ low-lying isomers.
These optimized geometries of neutral B$_{6}$ cluster, as shown in Fig. \ref{fig:geometries-neutral}, 
are found to be in good agreement with other available reports. Figure \ref{fig:geometries-cationic}
shows the corresponding geometries of cationic B$_{6}^{+}$ cluster. 
The unique bond lengths, point group symmetry and the electronic ground states are given 
in respective sub-figures.

For finite systems, such as clusters or quantum dots, the ratio $\frac{a}{\lambda}$ $\ll$ 1, where $a$ is system size 
and $\lambda$ is incident wavelength of light. In this case, the optical absorption cross section of the system 
corresponding to linear absorption can be computed within the electric dipole approximation,
\begin{equation} 
\sigma_{n}(\omega) = 4\pi\alpha\sum_{i} \frac{\omega_{in} \rvert \langle i|\hat{e}. \vec{\bar r} | n\rangle \rvert^2 \gamma}{(\omega_{in}-\omega)^2+\gamma^2}
\label{eqn:opt}
\end{equation}
where, $\alpha$ is fine structure constant, $\omega_{in}$ is frequency corresponding to energy difference of final state 
$|i\rangle$ and initial state $|n\rangle$. $\vec{\bar r}$ denotes position vector and $\gamma$ is linewidth of absorption.

The excited state energies of isomers are obtained using the \emph{ab initio} SCI approach. 
In this method, different configurations are constructed by replacing one of the occupied molecular orbitals
 in the Hartree-Fock ground state by a virtual orbital. 
Excited states of the system will have a linear combination of all such substituted configurations,
with corresponding variational coefficients. The energies of the excited states will then be obtained 
by diagonalizing the Hamiltonian in this configurational space \cite{meld}.
 The dipole matrix elements are calculated using the ground state and the excited state wavefunctions.
This is subsequently used for calculating the optical absorption 
cross section assuming Lorentzian lineshape, with some
artificial finite linewidth (\emph{cf.} Eqn. \ref{eqn:opt}). The contribution of wavefunction
of the excited states to the absorption peaks gives an
insight into the nature of optical excitation.

In one of the earlier reports published elsewhere\cite{nano_life}, 
we have extensively studied the dependence of basis sets,
freezing of 1s$^{2}$ chemical core and energy cutoff on virtual orbitals on the 
computed photoabsorption spectra of neutral boron clusters \cite{nano_life}.
We have shown that the optical absorption spectra
of small boron clusters do not change even if we freeze the chemical core of 
boron atoms. Also, a careful investigation was done for obtaining 
the energy cutoff of virtual orbitals to be included in the active space. 
Hence, in the present calculations, we have taken
augmented correlation consistent polarized valence double
zeta Gaussian basis set \cite{emsl_bas1,emsl_bas2}. We also froze the chemical core,
and kept only those virtual orbitals with energies of 1 Hartree or less.


\section{Results and Discussion}
\label{sec:results}
In this section, we discuss the structure and energetics of 
various isomers of neutral and cationic B$_{6}$ cluster, followed by
discussion of results of computed absorption spectra and nature of 
photo-excitations.

In the many-particle wavefunction analysis of excited states contributing to the various peaks
, we have used following convention. For doublet systems $H_{1}$ denotes
the singly occupied molecular orbital. For triplet systems, two singly occupied molecular orbitals
are denoted by $H_1$ and $H_2$, while $H$ and $L$ stands for highest occupied molecular 
orbital and lowest unoccupied molecular orbital respectively. For quartets, the third singly
occupied molecular orbital is denoted by $H_3$.

\subsection{B$_{6}$}

\begin{figure*}
\begin{center}
\subfigure[C$_{2h}$, $^{3}A_{u}$ \newline Planar ring (Tp)]
{\psfig{figure=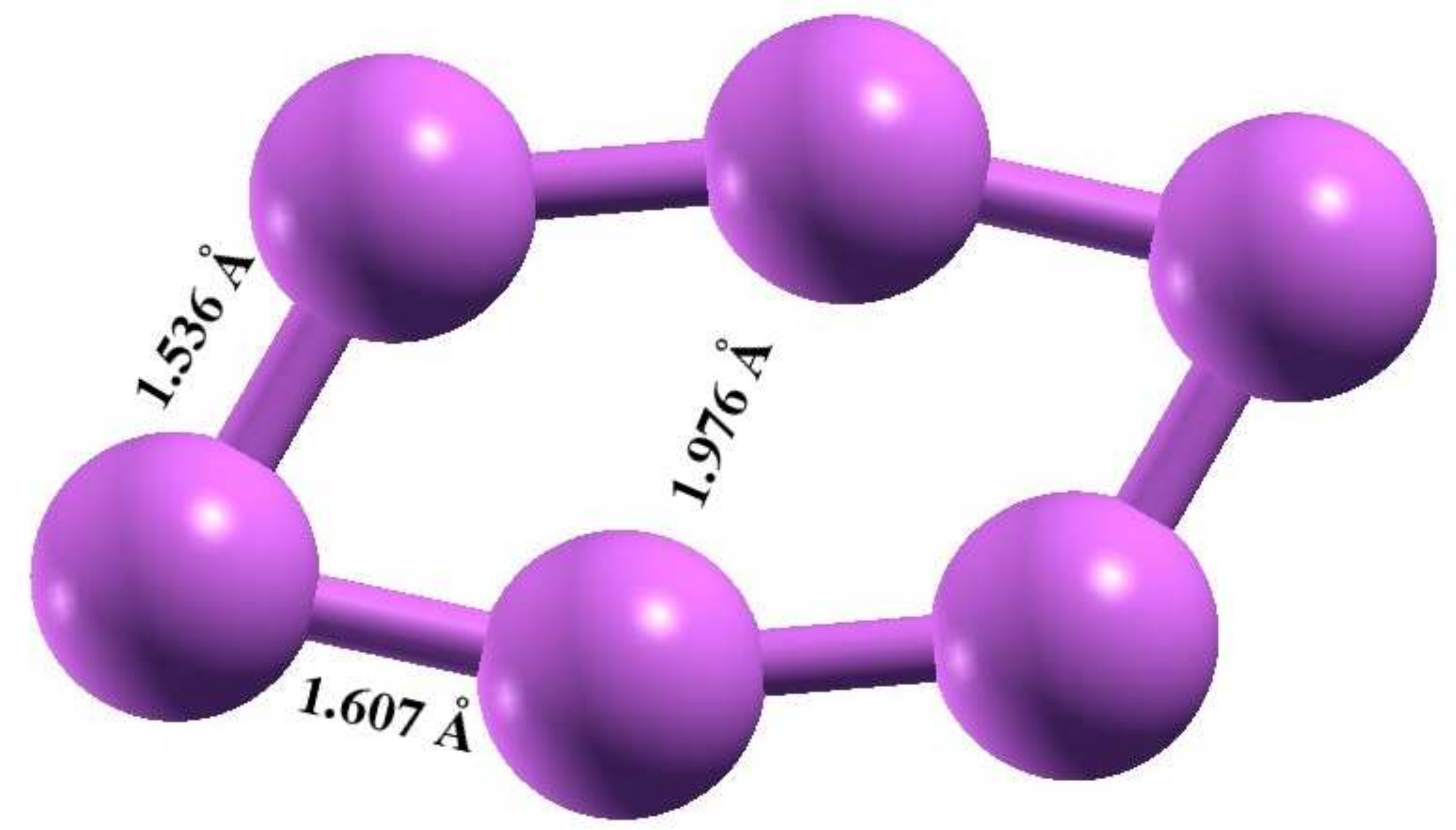,scale=0.12}} \hspace{0.5cm}
\subfigure[C$_{2v}$, $^{3}B_{1}$ \newline Incomplete Wheel]
{\psfig{figure=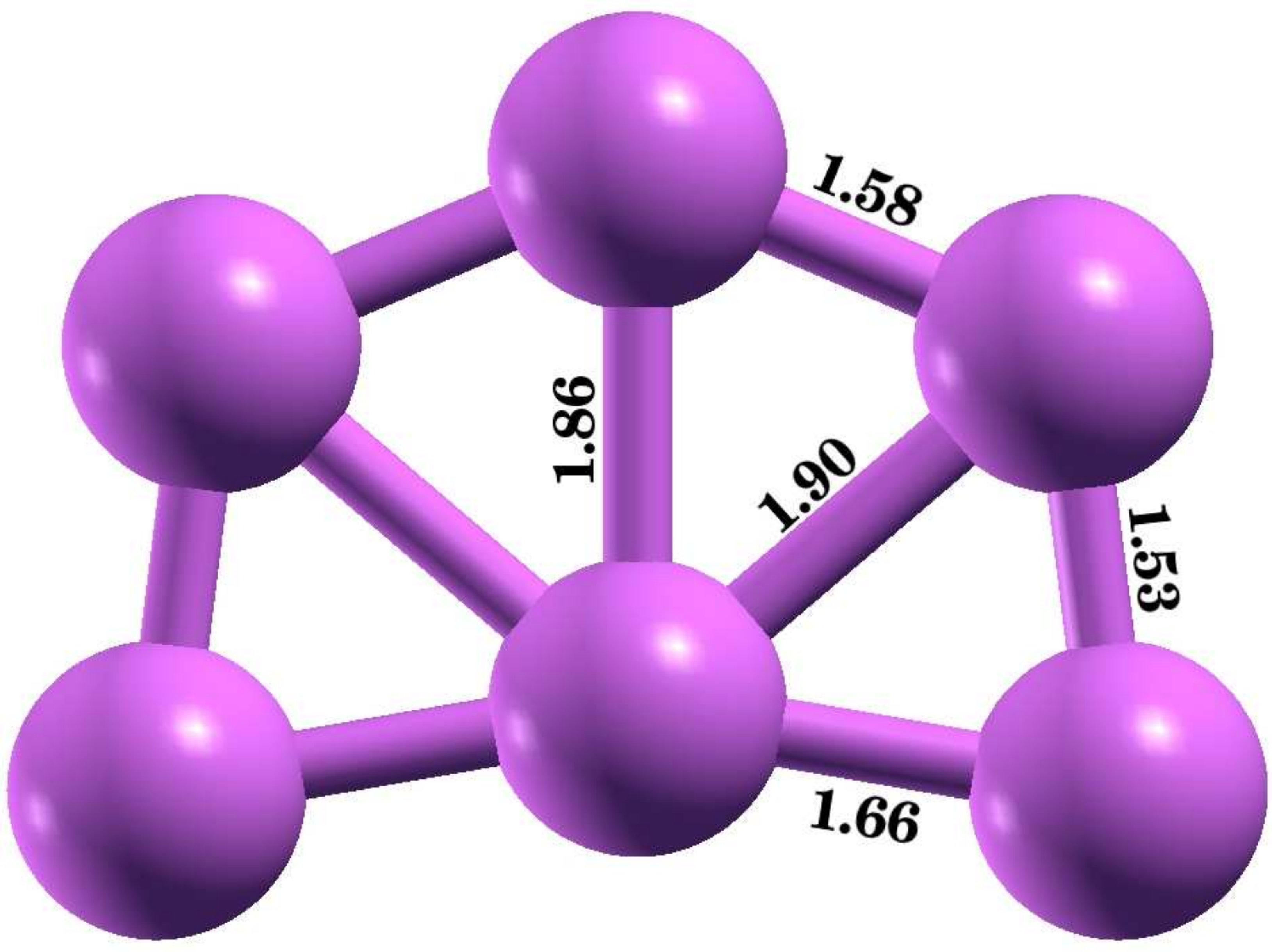,scale=0.085}} \hspace{0.5cm}
\subfigure[C$_{5v}$, $^{1}A_{1}$ Bulged wheel]
{\psfig{figure=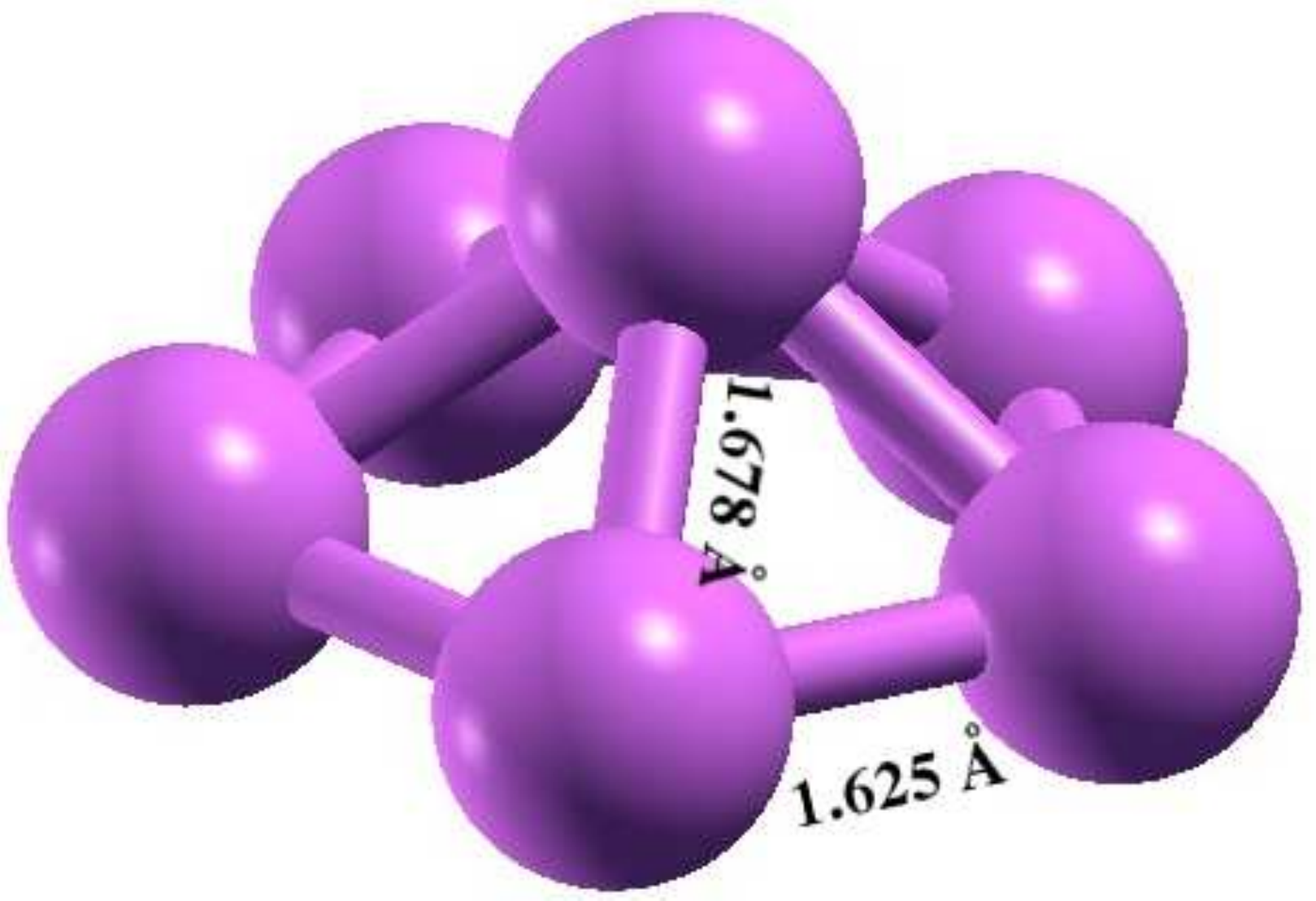,width=2.5cm}} \hspace{0.5cm}
\subfigure[C$_{s}$, $^{1}A^{'}$ \newline Planar ring (Sg)]
{\psfig{figure=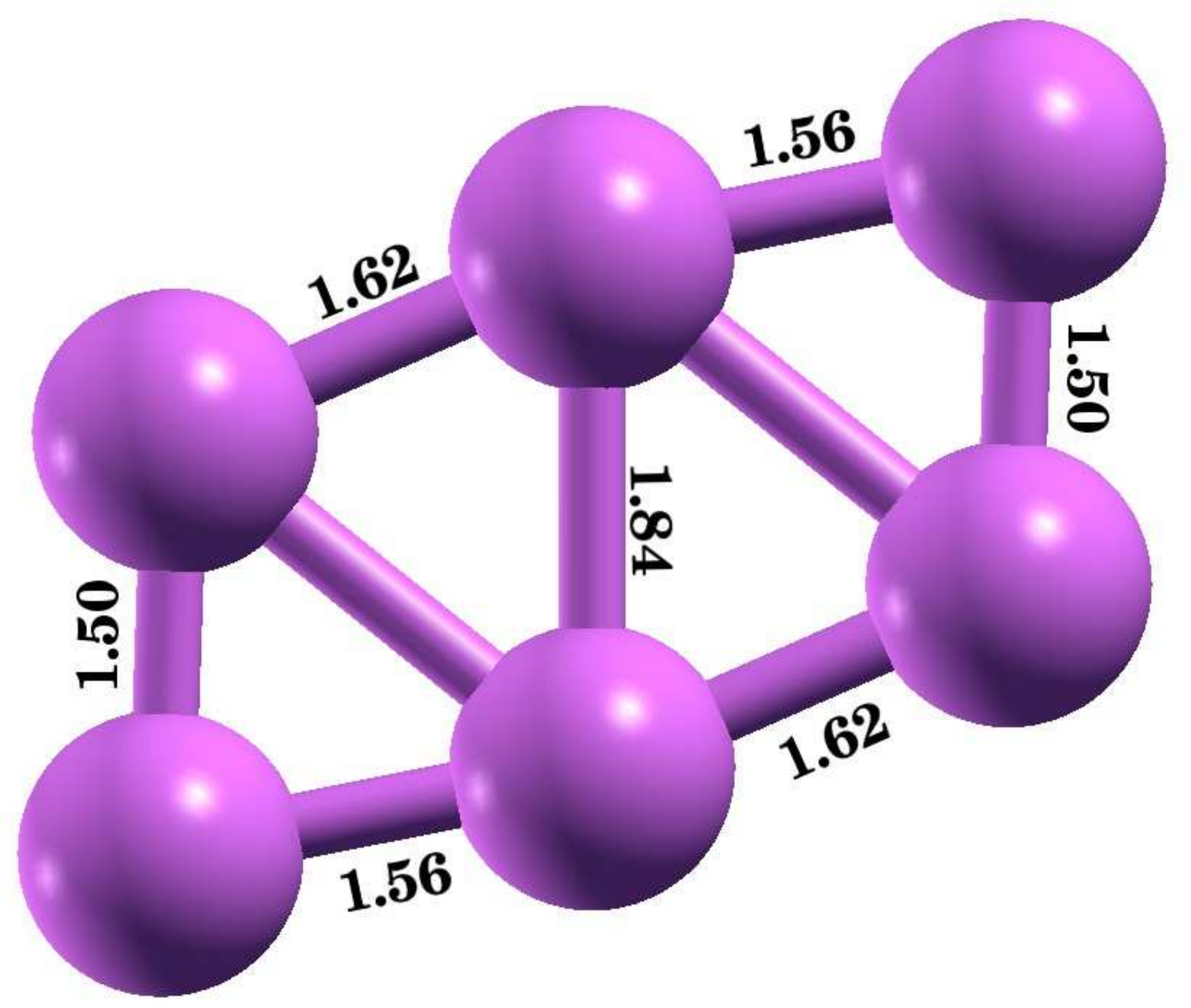,scale=0.085}} \hspace{0.5cm}
\subfigure[O$_{h}$, $^{3}A_{1g}$ Octahedron]
{\psfig{figure=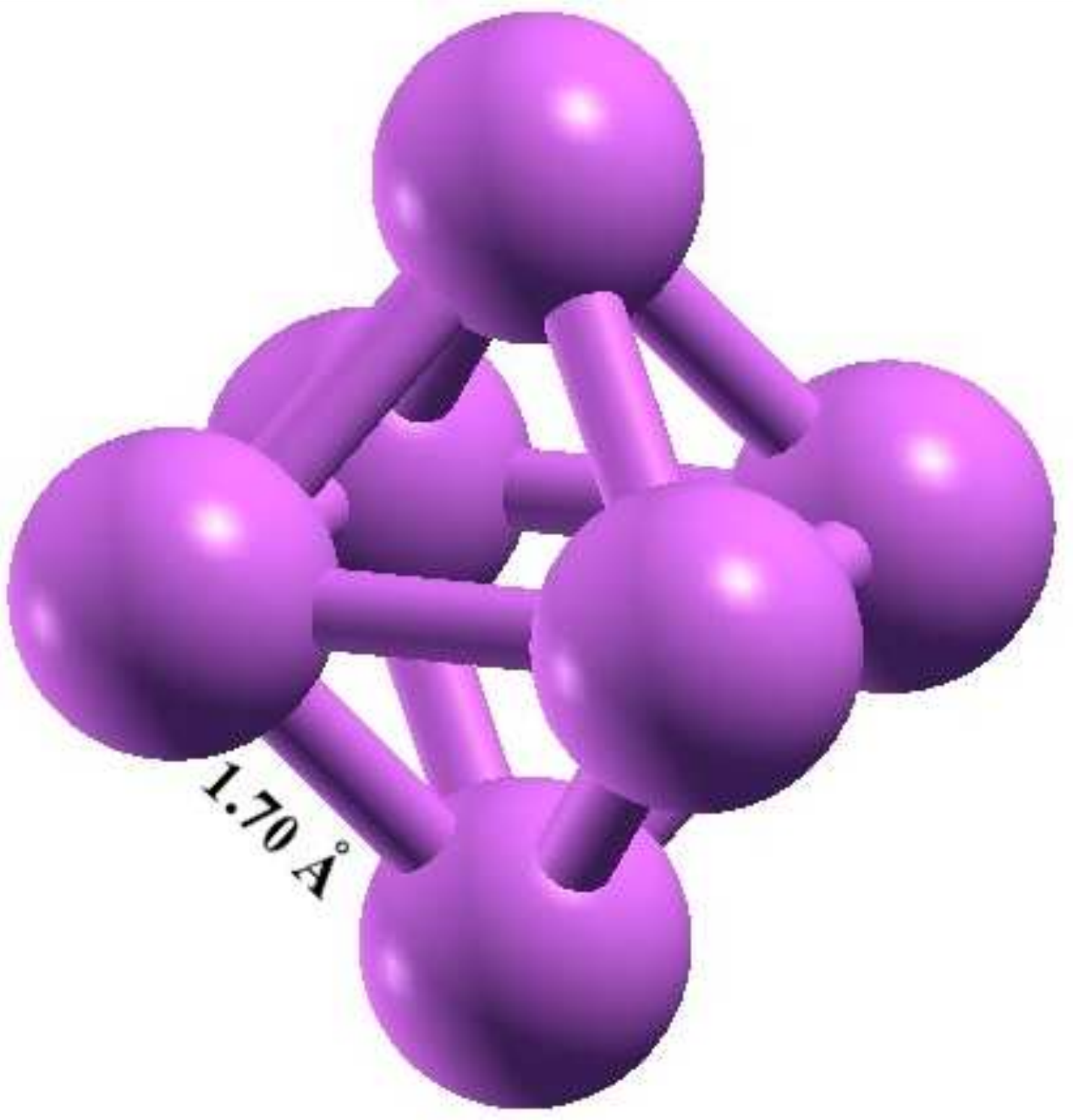,width=2.2cm}}  \\
\subfigure[C$_{1}$, $^{3}A$ Threaded tetramer]
{\psfig{figure=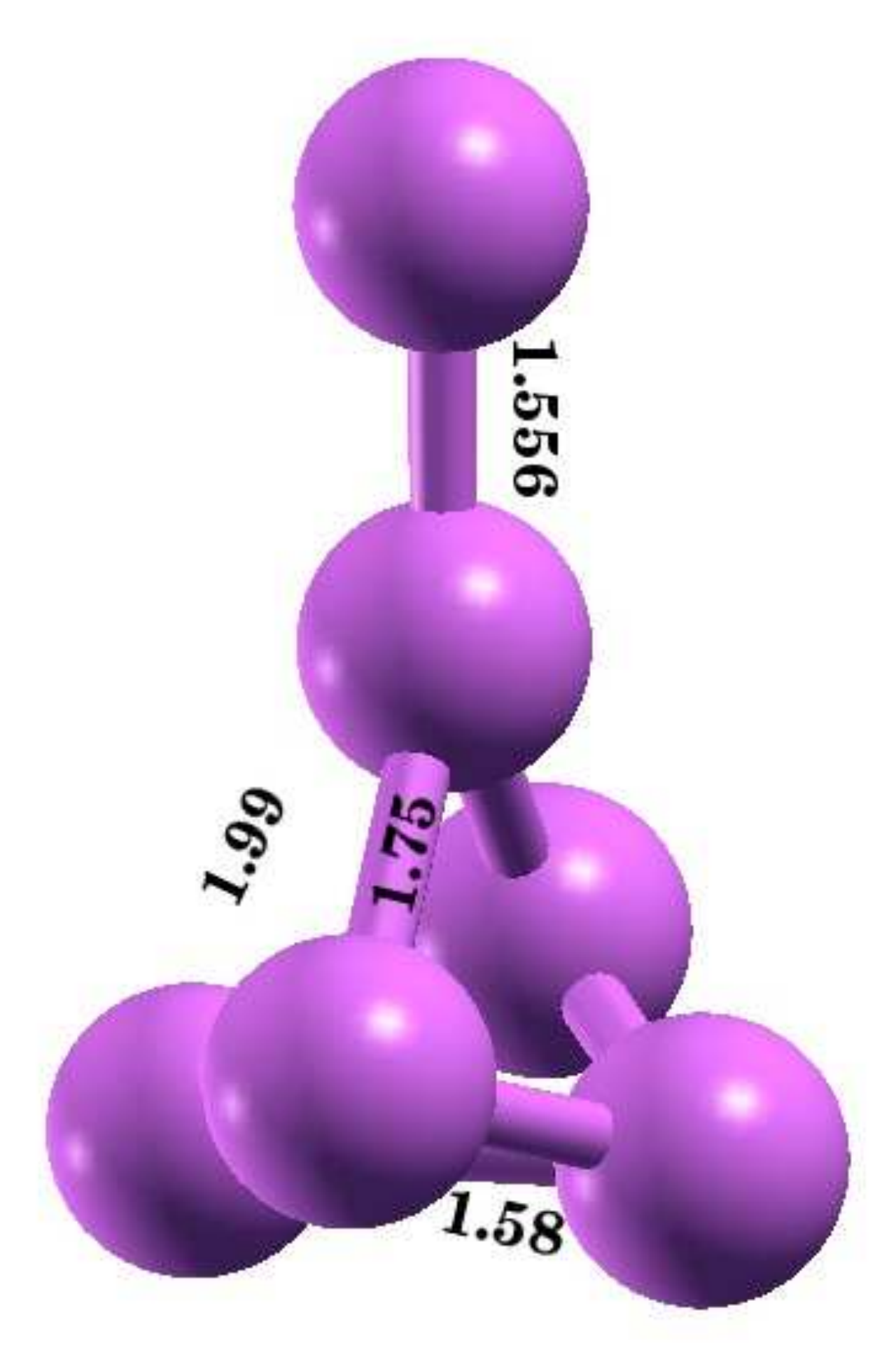,width=2cm}} \hspace{0.5cm}
\subfigure[C$_{2v}$, $^{3}B_{1}$ Threaded trimer]
{\psfig{figure=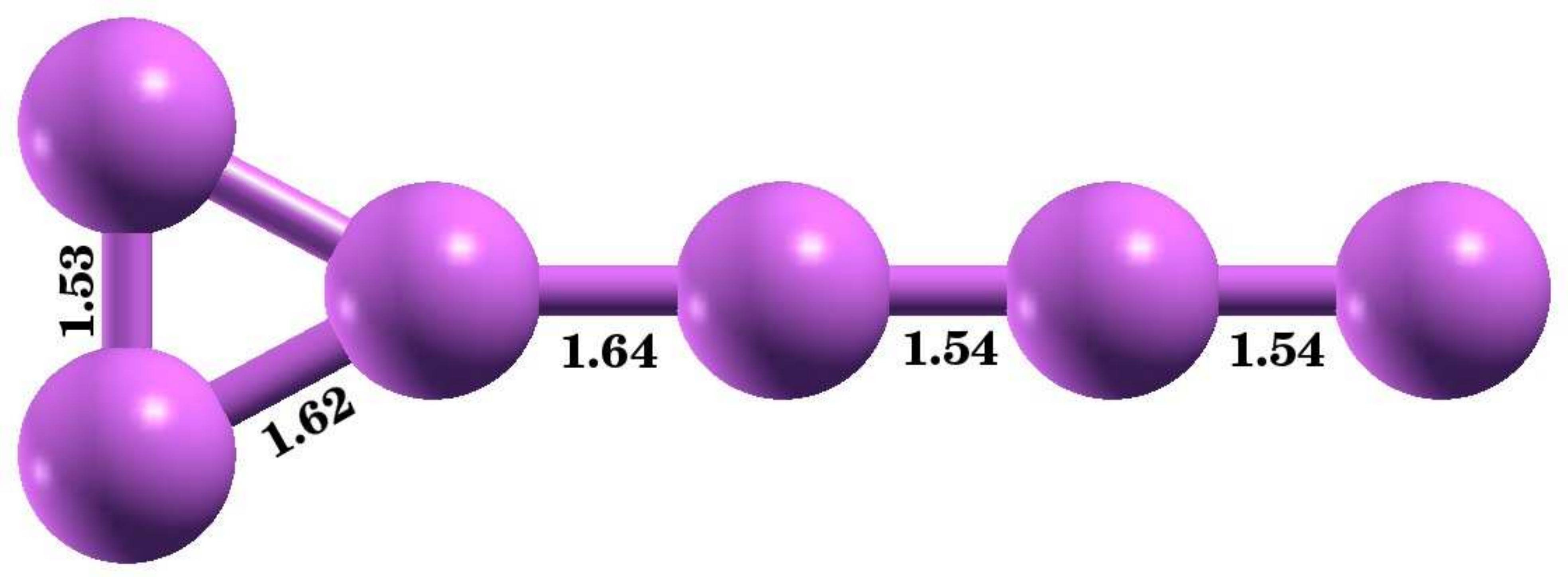,scale=0.13}}  \hspace{0.5cm}
\subfigure[C$_{1}$, $^{1}A$ \newline Twisted trimers]
{\psfig{figure=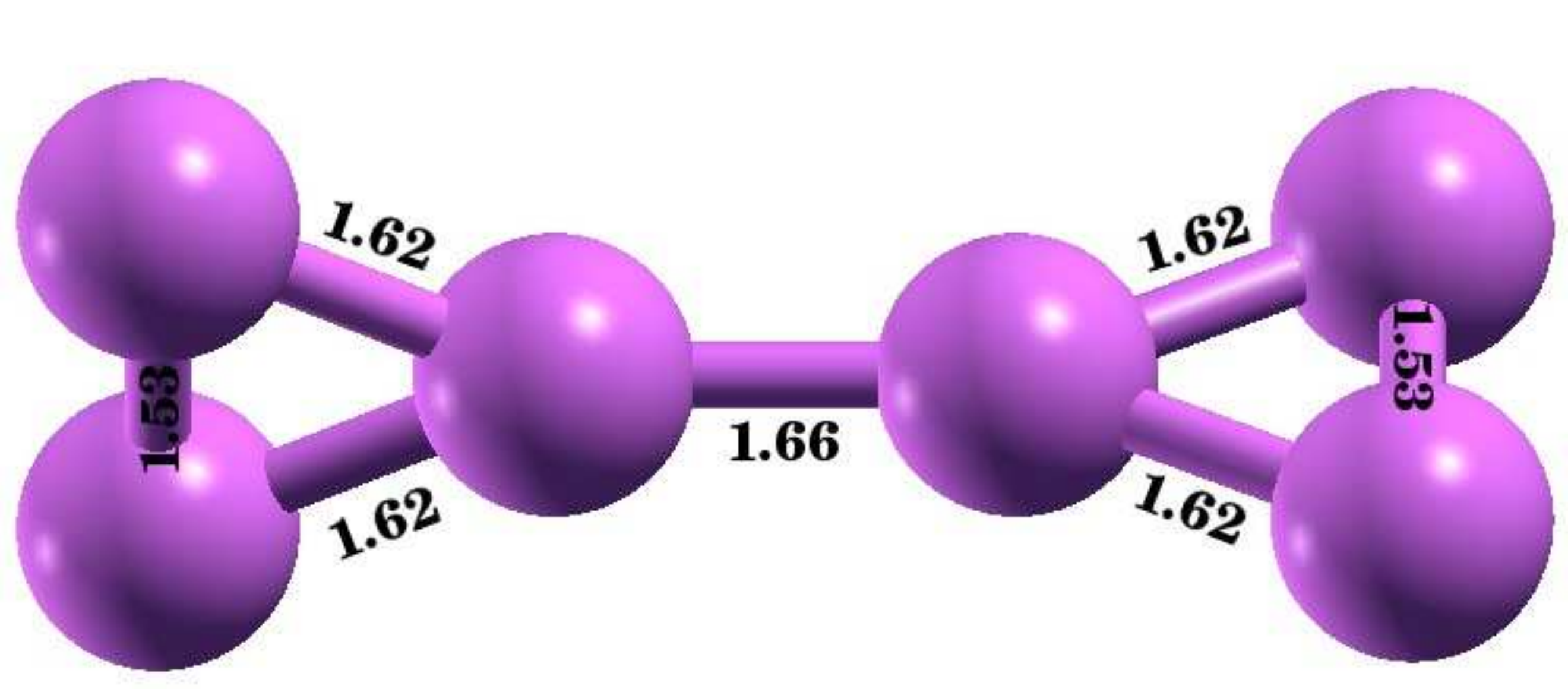,scale=0.15}}  \hspace{0.5cm}
\subfigure[D$_{2h}$, $^{1}A_{g}$ Planar trimers]
{\psfig{figure=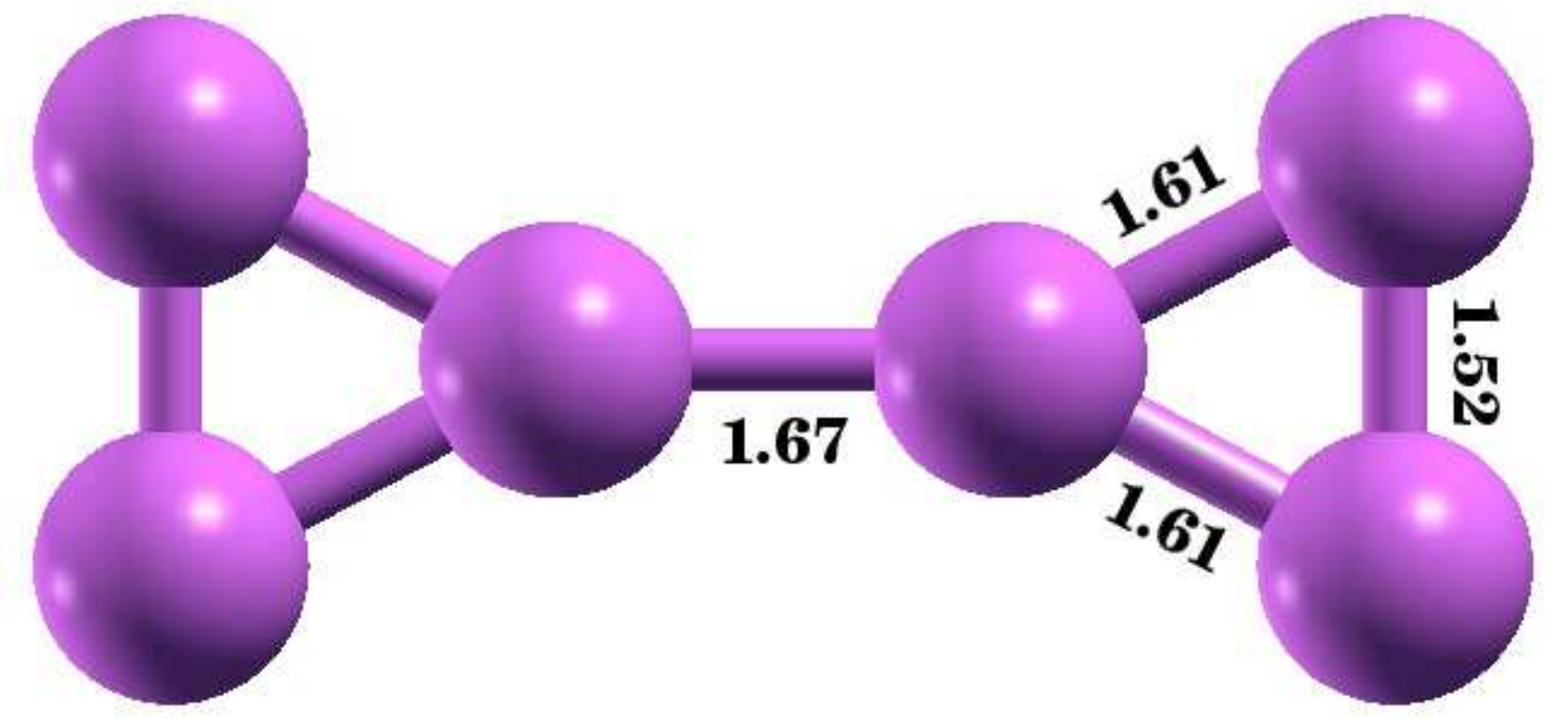,scale=0.16}} \hspace{0.5cm}
\subfigure[C$_{1}$, $^{1}A$ Convex bowl]
{\psfig{figure=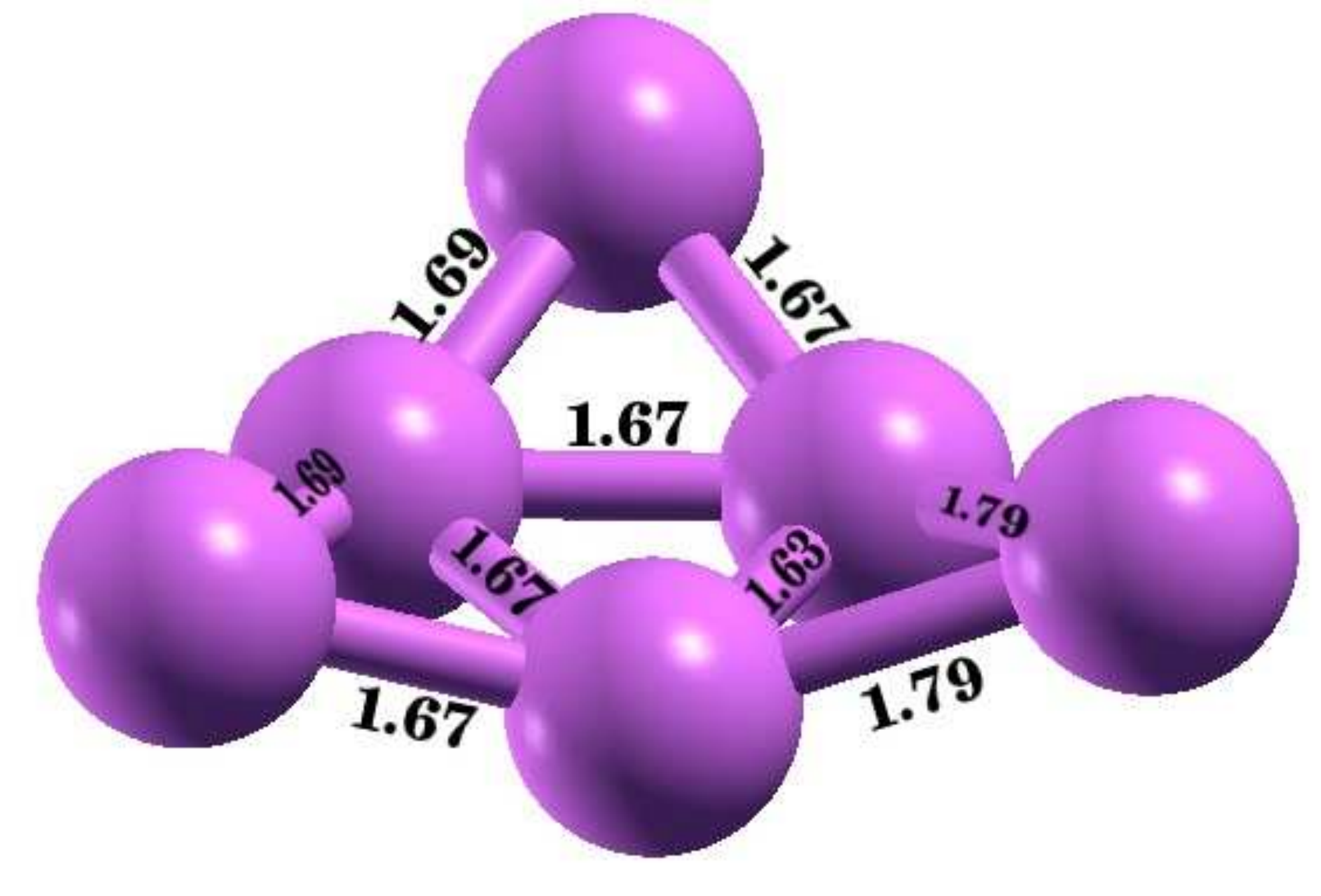,scale=0.15}} \hspace{0.5cm}
\subfigure[D$_{\infty h}$, $^{1}\Sigma_{g}$ Linear]
{\psfig{figure=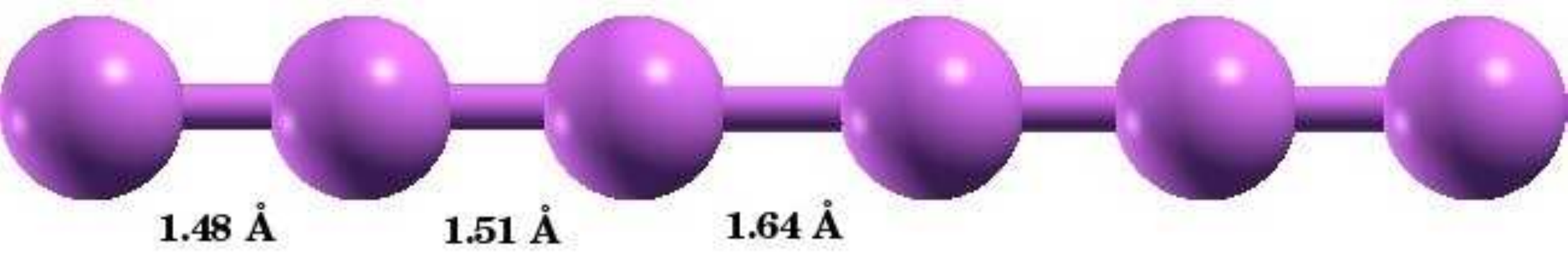,scale=0.26}}  
\end{center}
\caption{\label{fig:geometries-neutral}(Color online) Geometry optimized ground state structures
of different isomers of neutral B$_{6}$ clusters, along with the point group symmetries
obtained at the CCSD level. }
\end{figure*}

We have found a total of 11 isomers of neutral B$_{6}$ cluster with stable geometries 
as shown in the Fig. \ref{fig:geometries-neutral} \cite{gabedit}. The relative standings in energy 
are presented in the Table \ref{tab:energies-neutral}, along with point group symmetries and
electronic states.

\begin{table*}
\centering
\caption{Point group, electronic state and total energies of different isomers of B$_{6}$ cluster.}
\label{tab:energies-neutral}       
\begin{tabular}{clllc}
\hline\noalign{\smallskip}
Sr.    	& Isomer		& Point 	& Elect. 	& Total  	\\
no. 	& 			& group 	& State      	& Energy (Ha) 	\\  
\noalign{\smallskip}\hline\noalign{\smallskip}
1	& Planar ring (triplet)	& C$_{2h}$	& ${}^3 A_{u}$	&	-147.795051	\\
2	& Incomplete wheel 	& C$_{2v}$	& ${}^3 B_{1}$	& 	-147.774166   \\
3	& Bulged wheel 		& C$_{5v}$	& ${}^1 A_{1}$	&	-147.764477	\\
4	& Planar ring (singlet)	& C$_{s}$	& ${}^1 A^{'}$	&	-147.720277	\\
5	& Octahedron	 	& O$_{h}$	& ${}^3 A_{1g}$	&	-147.678302	\\
6	& Threaded tetramer 	& C$_{1}$	& ${}^3 A$	&	-147.676776	\\
7	& Threaded trimer 	& C$_{2v}$	& ${}^3 B_{1}$	&	-147.667709	\\
8	& Twisted trimers 	& C$_{1}$	& ${}^1 A$	&	-147.645847	\\
9	& Planar trimers 	& D$_{2h}$	& ${}^1 A_{g}$	&	-147.645522	\\
10	& Convex bowl	 	& C$_{1}$	& ${}^1 A$	&	-147.612607	\\
11	& Linear 		& D$_{\infty h}$& ${}^1 \Sigma_{g}$&	-147.449013	\\
\noalign{\smallskip}\hline
\end{tabular}
\end{table*}


\FloatBarrier

\begin{figure*}[!h]
\begin{center}
\subfigure[Planar Ring (Triplet) Isomer] 
{\psfig{figure=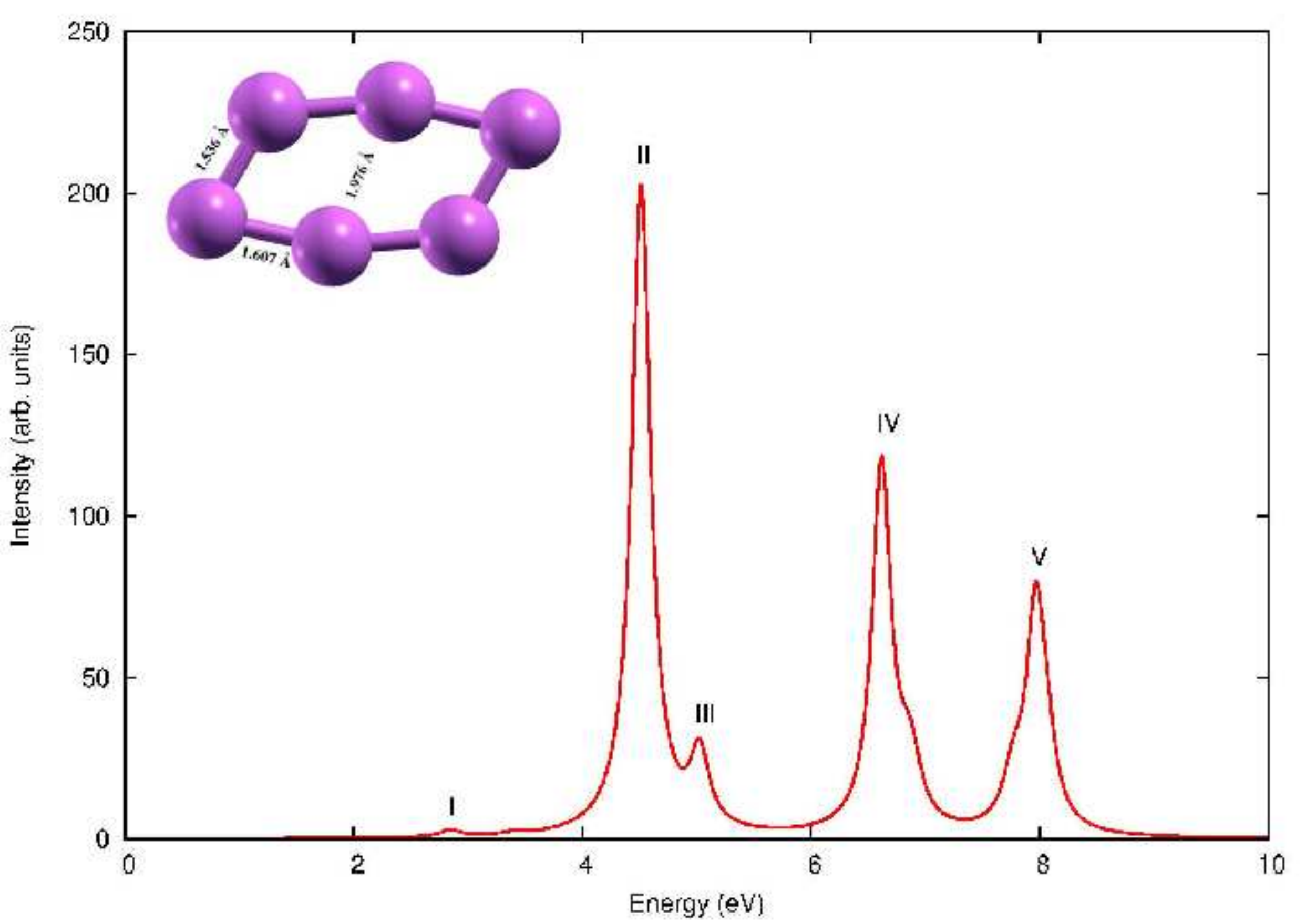,width=0.32\textwidth} \label{subfig:neutral-plot-planar-ring-triplet}} 
\subfigure[Incomplete Wheel]
 {\psfig{figure=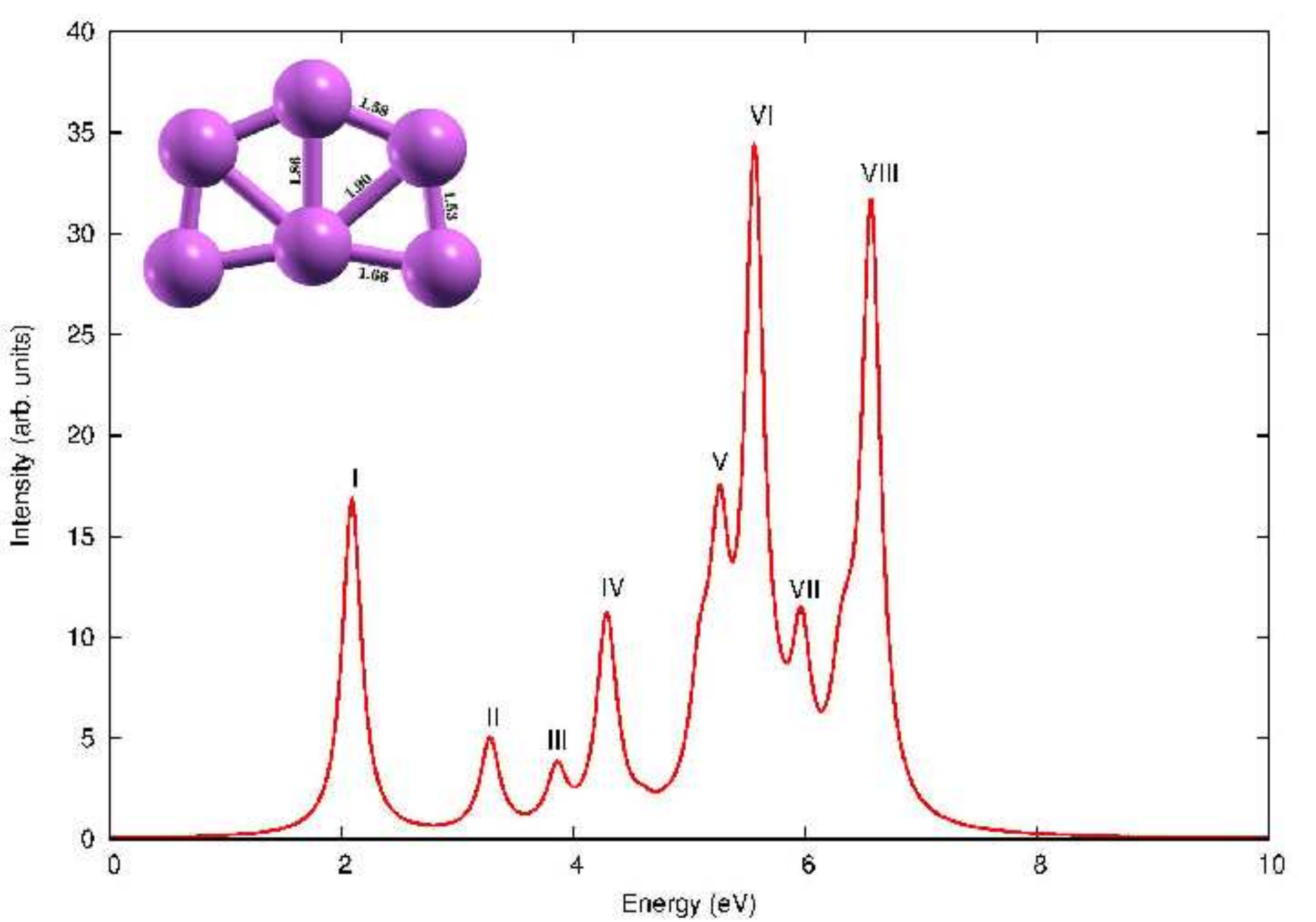,width=0.32\textwidth} \label{subfig:neutral-plot-incomplete-wheel}} 
\subfigure[Bulged wheel Isomer]
{\psfig{figure=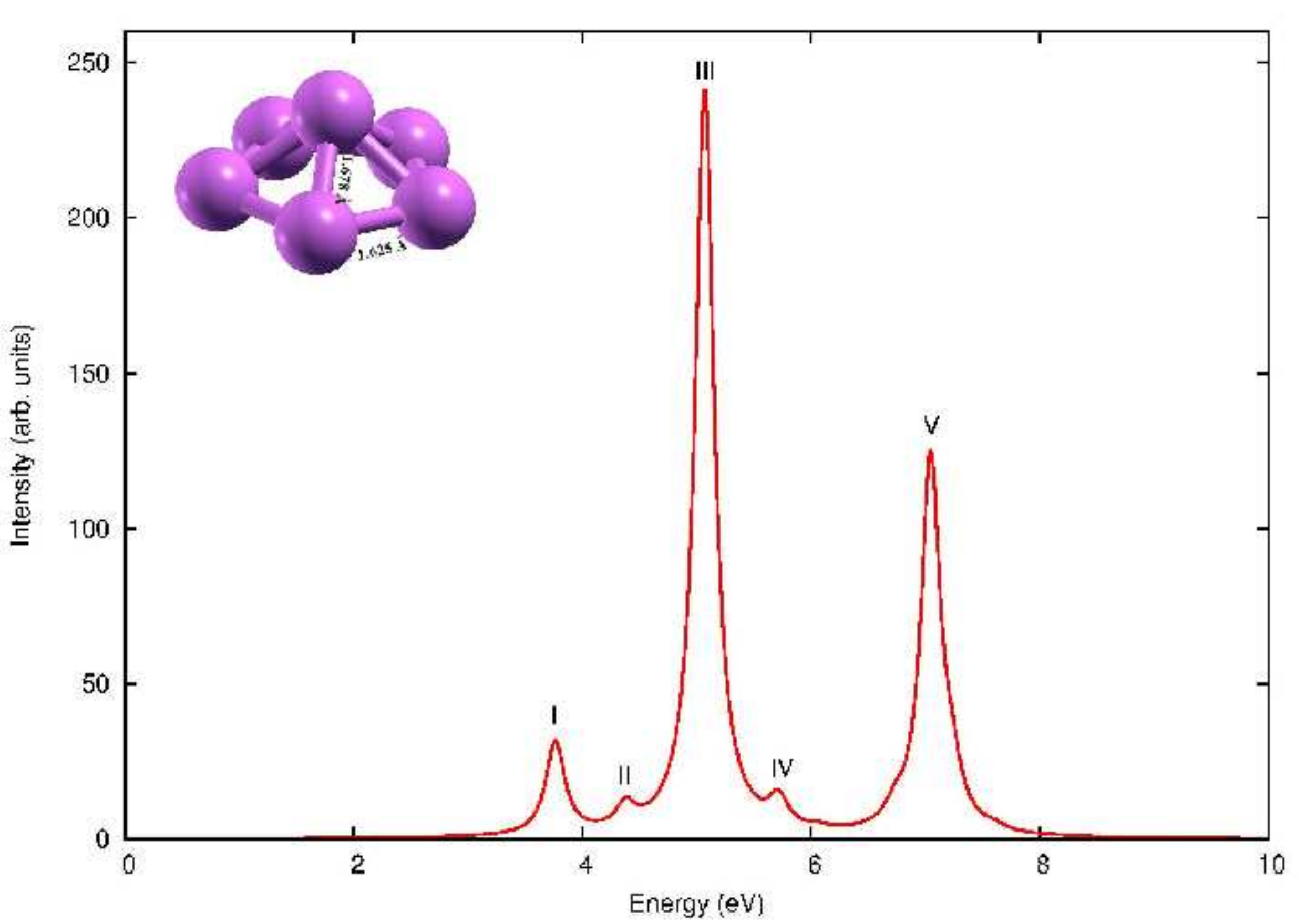,width=0.32\textwidth} \label{subfig:neutral-plot-bulged-wheel}} \\
\subfigure[Planar Ring (Singlet) Isomer] 
{\psfig{figure=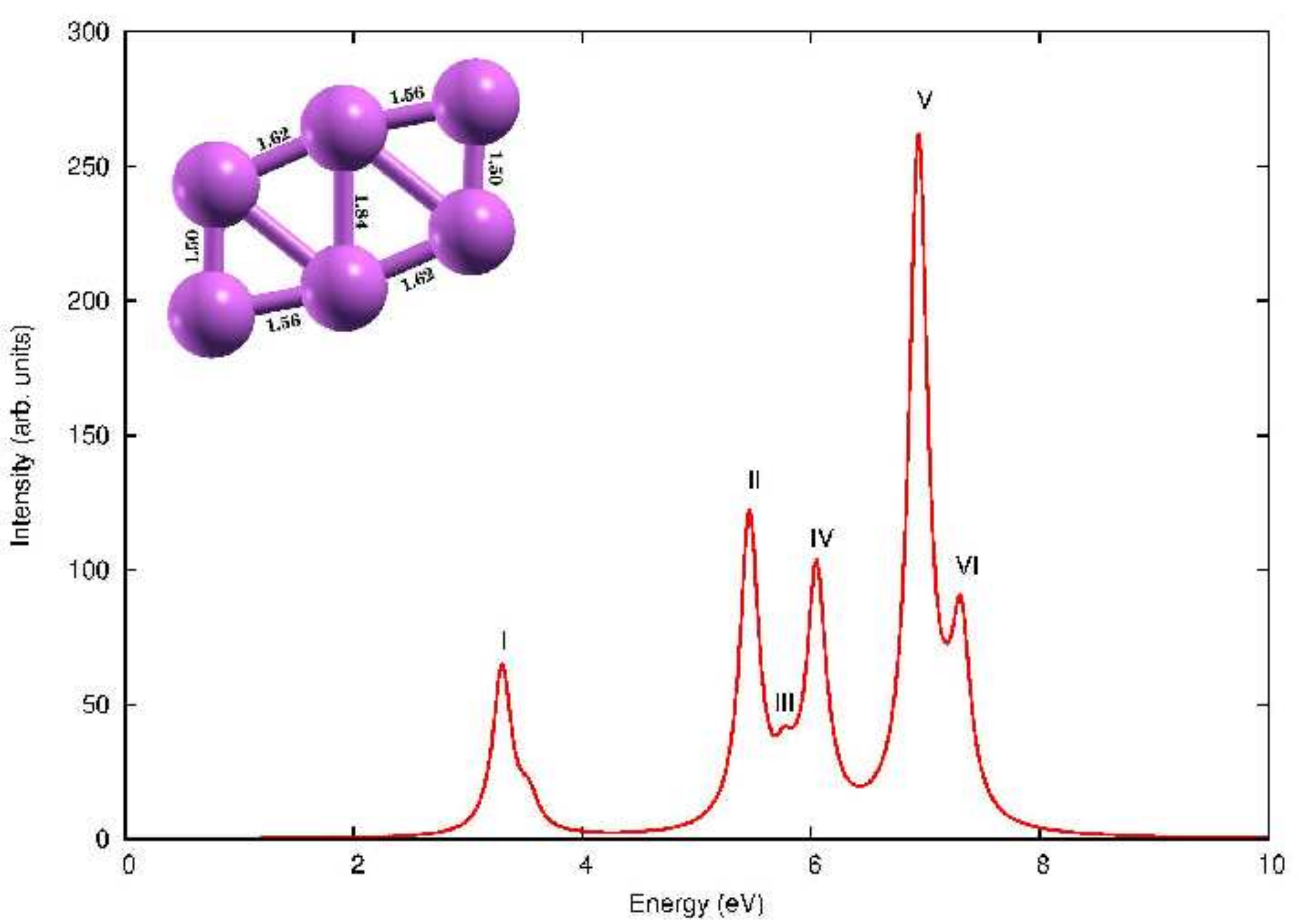,width=0.32\textwidth} \label{subfig:neutral-plot-planar-ring-singlet}} 
\subfigure[Octahedron]
{\psfig{figure=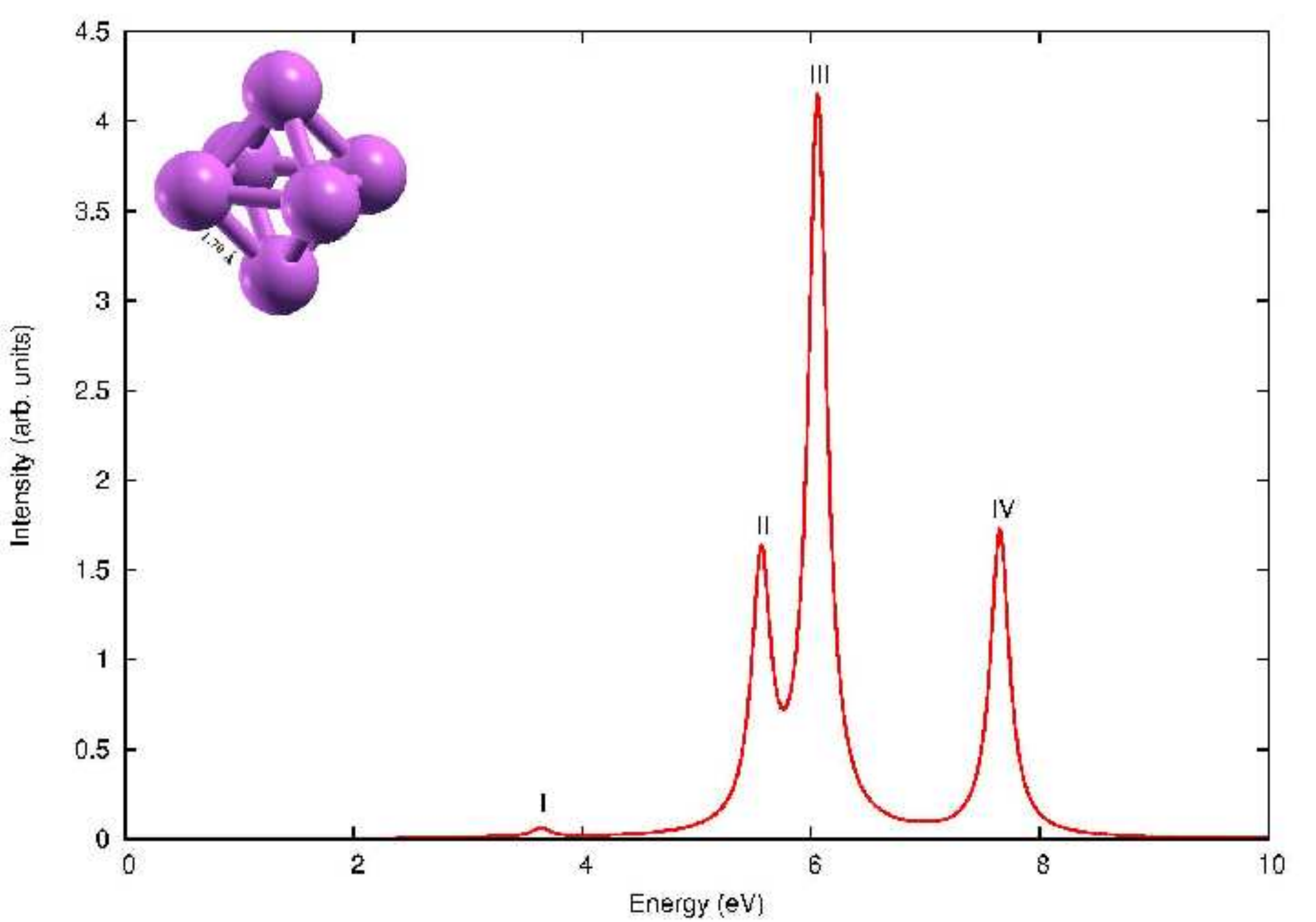,width=0.32\textwidth} \label{subfig:neutral-plot-octahedron} } 
\subfigure[Threaded tetramer]
{\psfig{figure=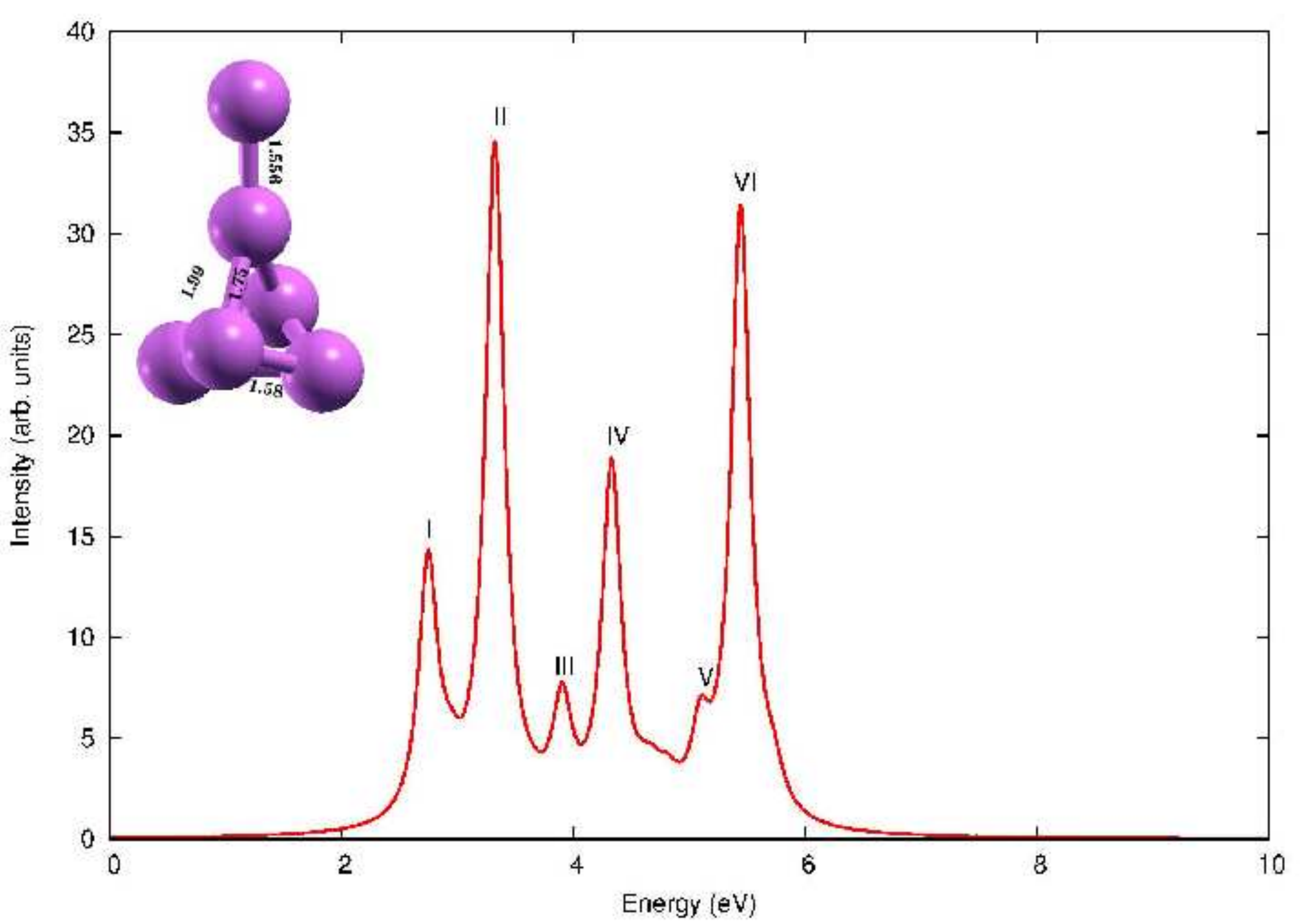,width=0.32\textwidth} \label{subfig:neutral-plot-threaded-tetramer}} \\
\subfigure[Threaded trimer]
{\psfig{figure=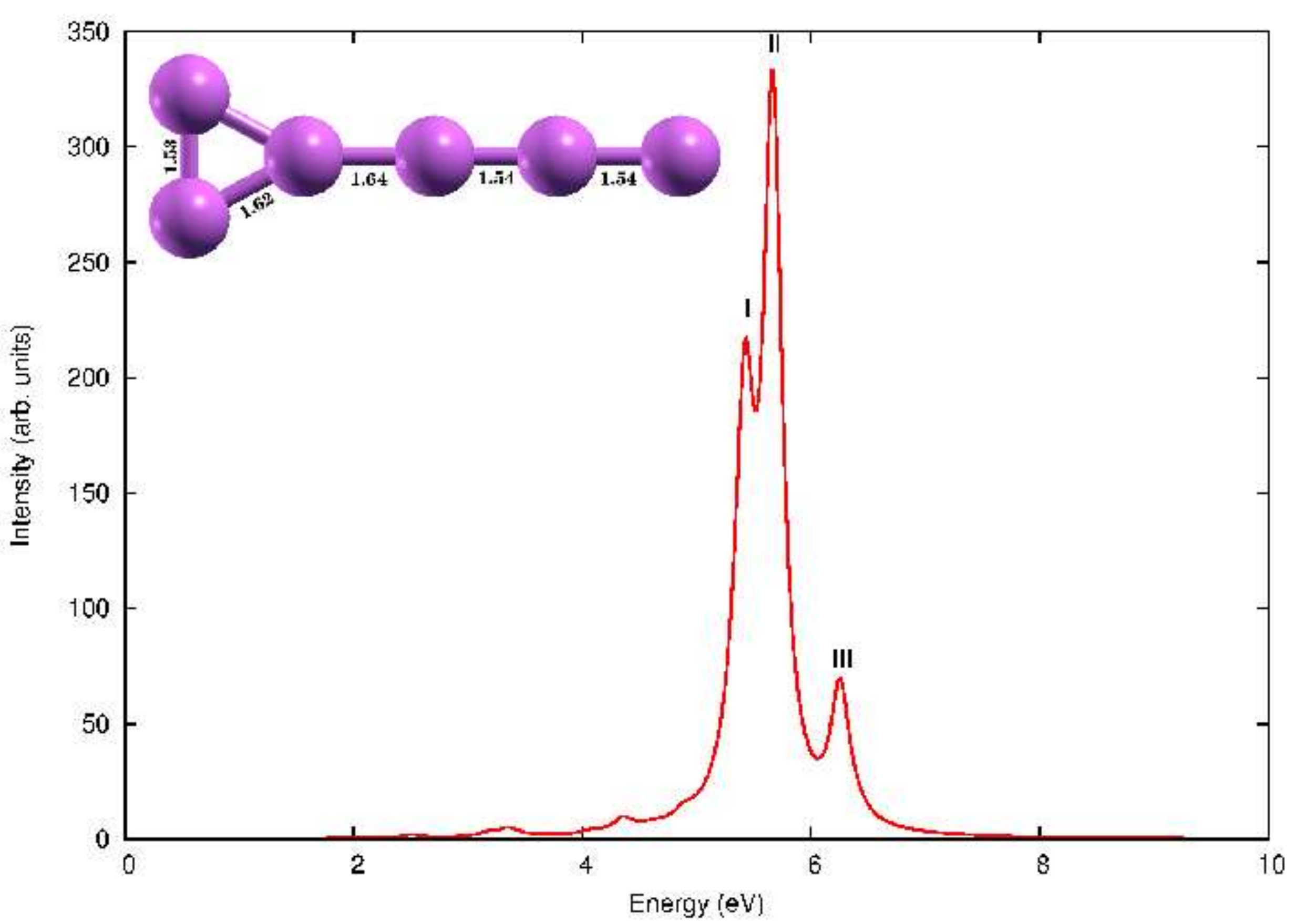,width=0.32\textwidth} \label{subfig:neutral-plot-threaded-trimer}} 
\subfigure[Twisted trimers]
{\psfig{figure=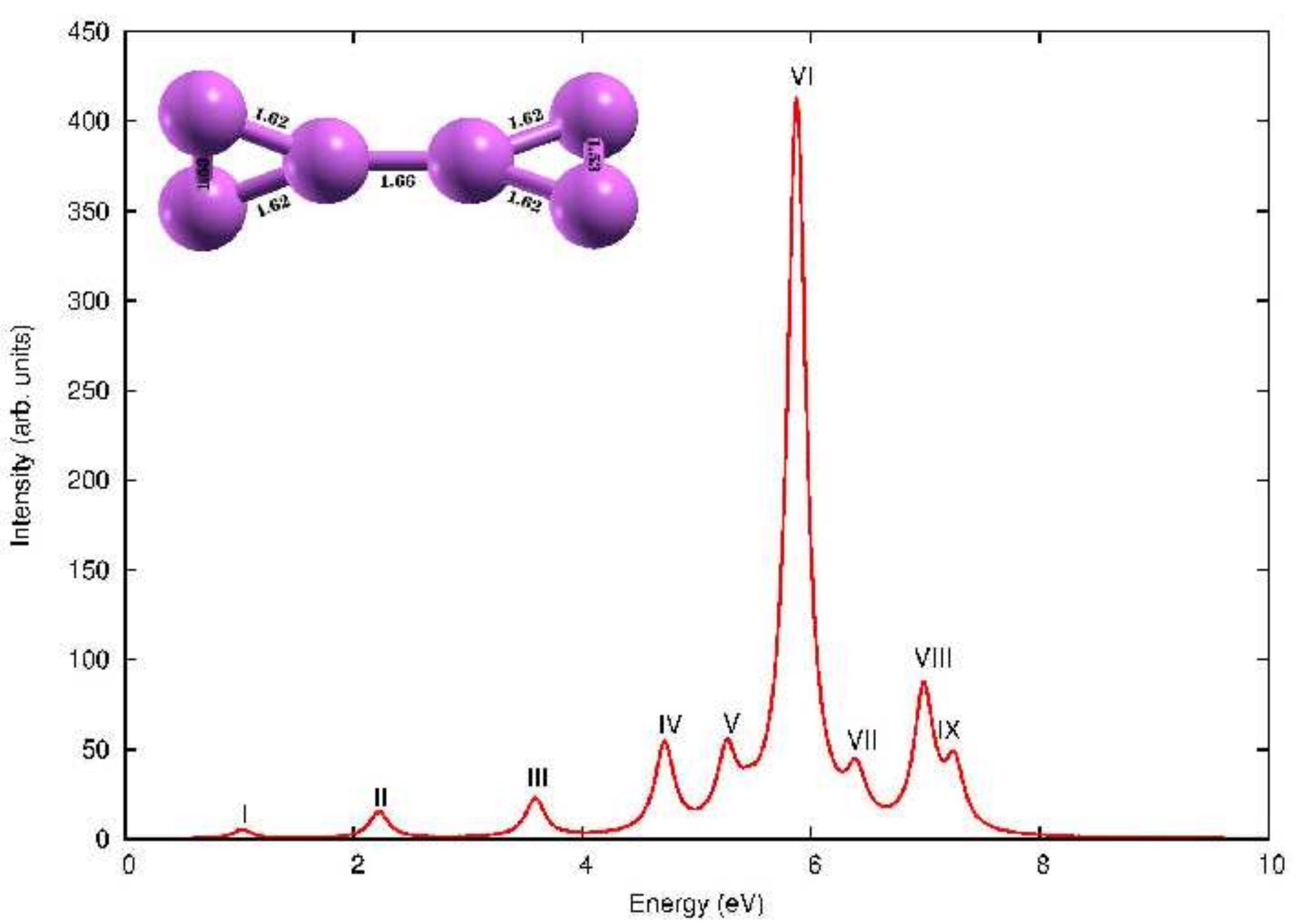,width=0.32\textwidth} \label{subfig:neutral-plot-twisted-trimers}}  
\subfigure[Planar trimers]
{\psfig{figure=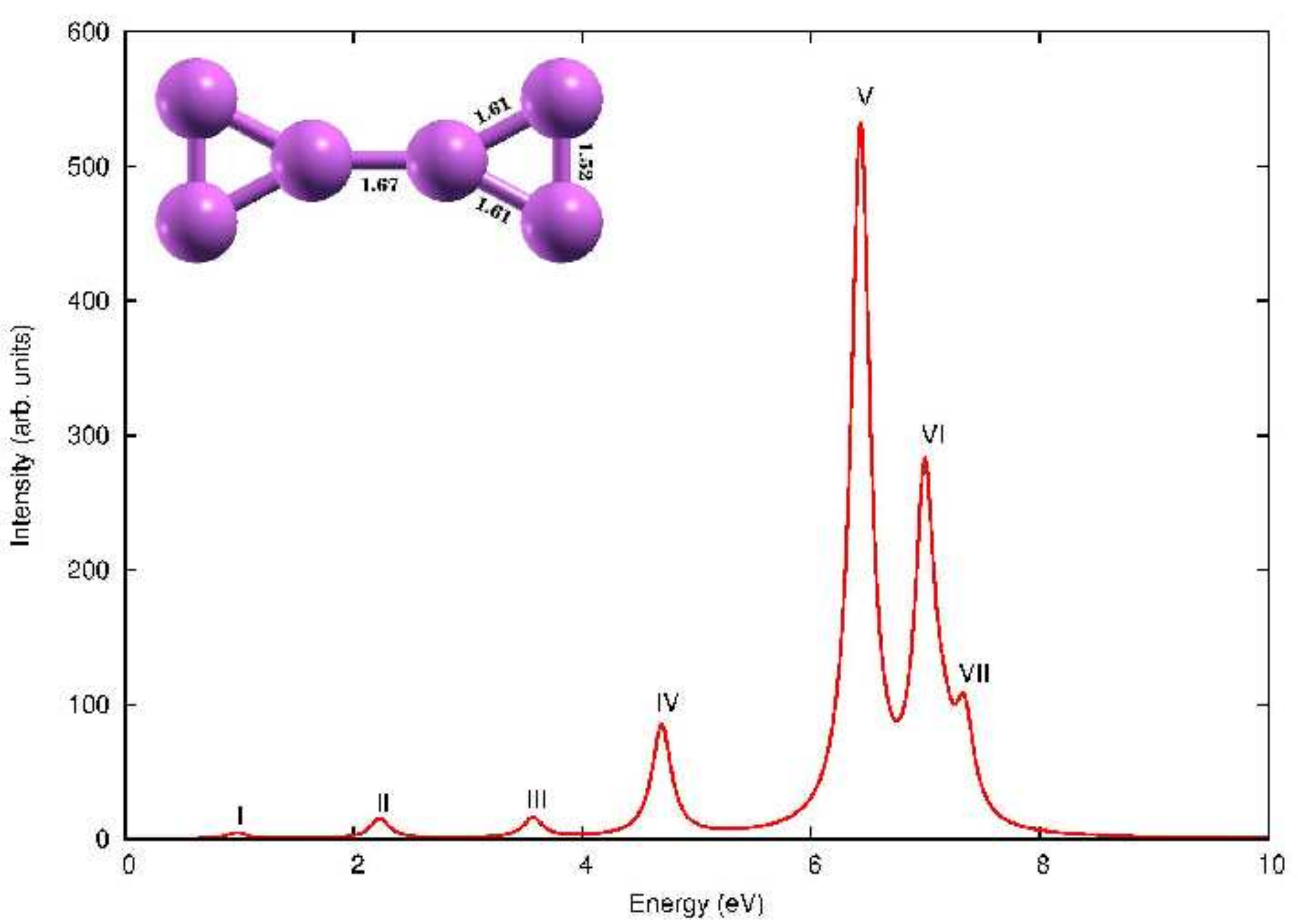,width=0.32\textwidth} \label{subfig:neutral-plot-planar-trimers}} \\
\subfigure[Convex bowl]
{\psfig{figure=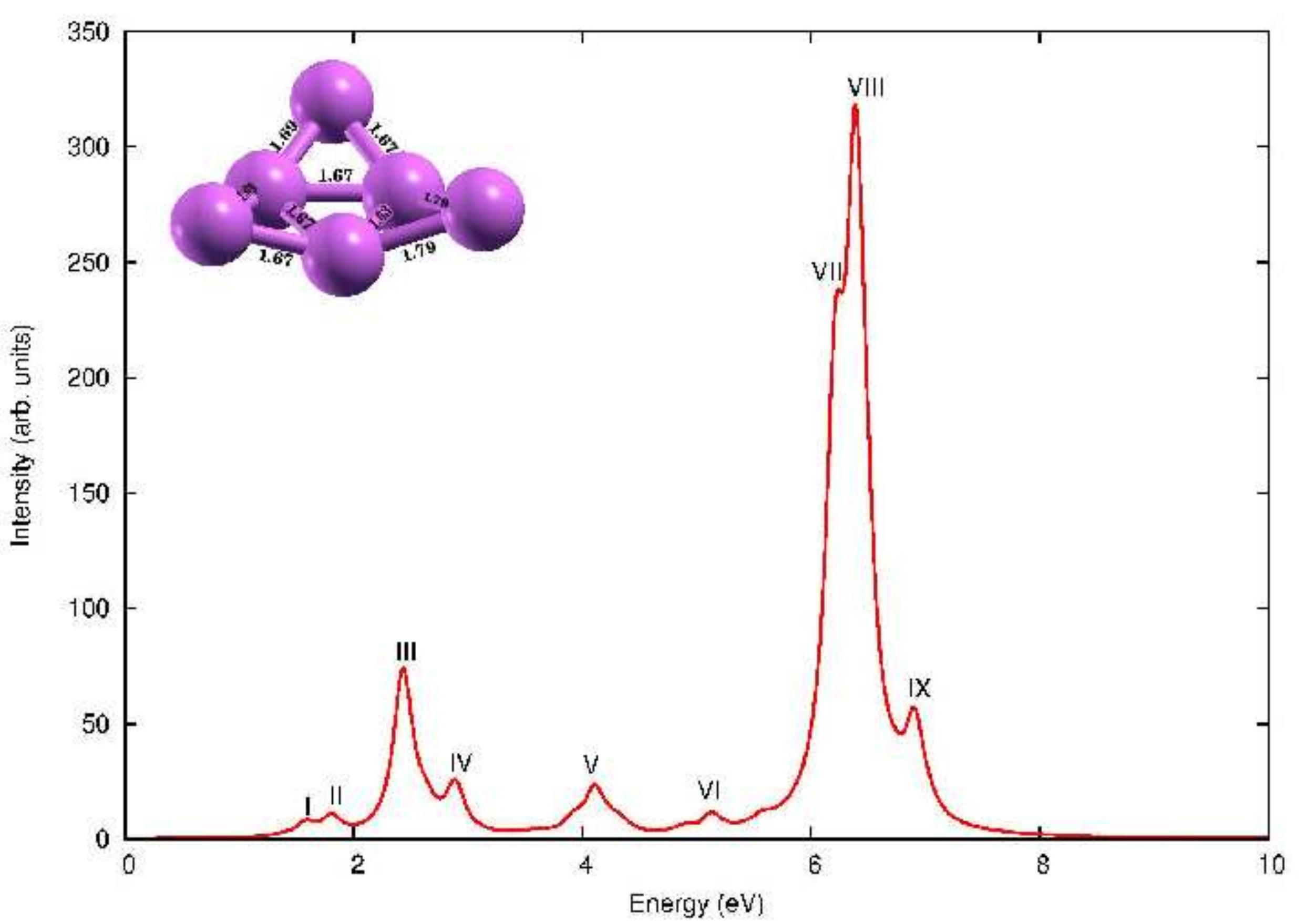,width=0.32\textwidth} \label{subfig:neutral-plot-convex-bowl}} 
\subfigure[Linear]
{\psfig{figure=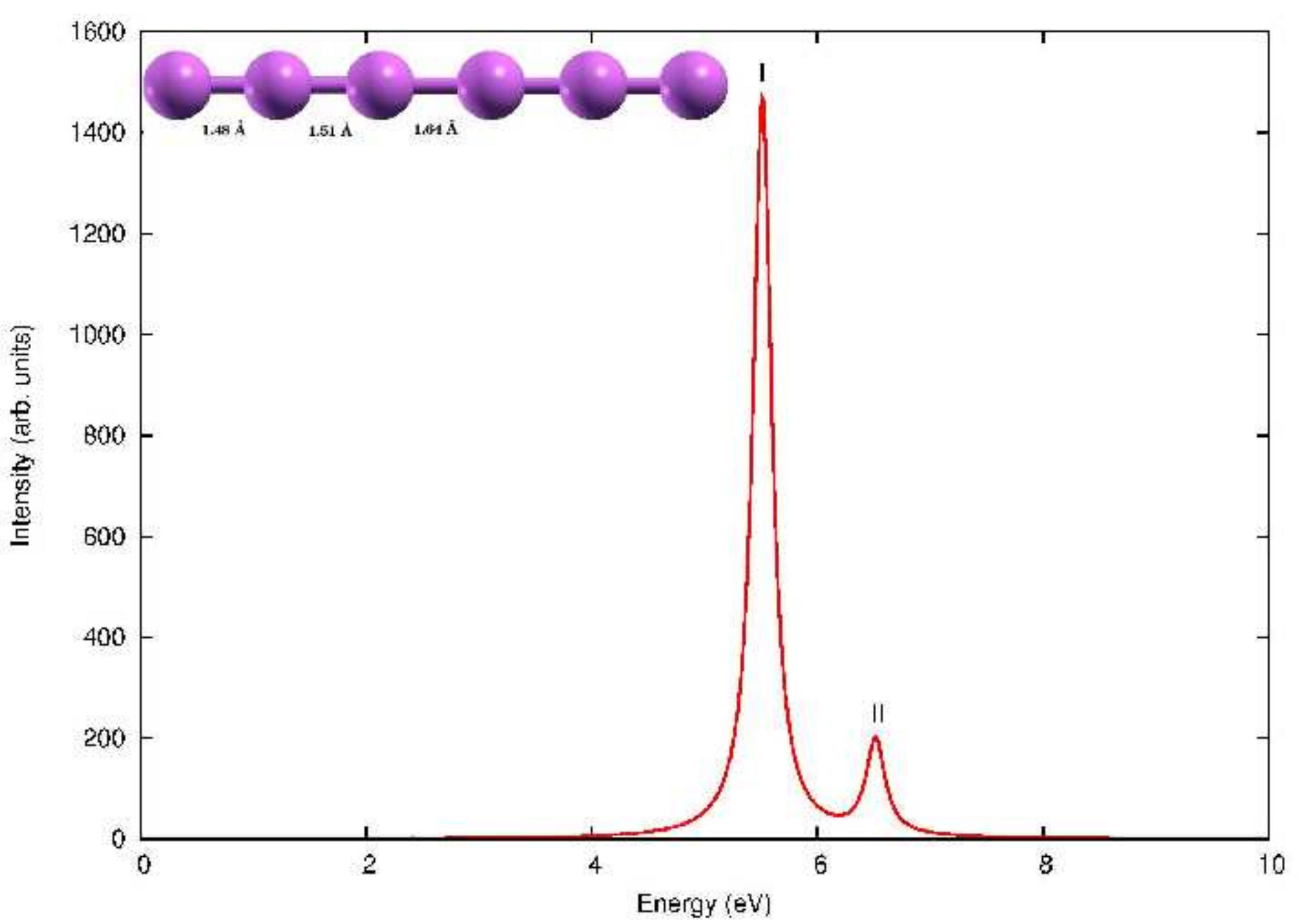,width=0.35\textwidth,height=4cm} \label{subfig:neutral-plot-linear} }
\end{center}
\caption{\label{fig:plots-neutral}(Color online) The linear optical absorption spectrum
of B$_{6}$, calculated using the SCI approach. The many-body wavefunctions of excited stated corresponding
to the peaks labeled are given in the Appendices 
\ref{Tab:table-neutral-planar-ring-triplet}--\ref{Tab:table-neutral-linear}. 
For plotting the spectrum, a uniform linewidth of 0.1 eV was assumed. }
\end{figure*}

\FloatBarrier

The most stable isomer of B$_{6}$ cluster has ring-like
planar structure, with C$_{2h}$ point group symmetry. Although
B$_{6}$ has an even number of electrons, the electronic ground
state of this isomer is a triplet -- an open shell system. The
equilibrium geometry obtained in our calculation is in good
agreement with the recently reported values \cite{structure-bonding-b6, 
turkish_boron, b6-isomerization,vlasta-chem-review}.
 The optical absorption spectrum calculated using the SCI approach
is as shown in the Fig. \ref{fig:plots-neutral}\subref{subfig:neutral-plot-planar-ring-triplet}.
It is mainly characterized by feeble absortion in the visible range, but much stronger absorption 
at higher energies. The many particle wave-functions of excited states contributing to various peaks 
are presented in Table \ref{Tab:table-neutral-planar-ring-triplet}. The first absorption peak at 2.84 eV
with very low intensity is characterized by $H - 3 \rightarrow H_{2}$ and $H - 1 \rightarrow H_{1}$ transitions.
Due to planar nature of the isomer, we can classify the absorption into two categories: (a) those 
with polarization along the direction of the plane and (b) polarization perpendicular to the plane.
In this case, it is seen that, in most of the cases, the absorption is due to polarizations along the plane of the isomer.
Also, instead of being dominated by single configurations, the wavefunctions of the excited states 
contributing to all the peaks exhibit strong configuration mixing. This is an indicator of 
plasmonic nature of the optical excitations \cite{plasmon}.

The second low lying isomer of B$_{6}$ is another planar structure resembing an incomplete wheel, \emph{i.e.}
one outer atom removed from B$_{7}$ wheel cluster.
This isomer is also a triplet system with C$_{2v}$ symmetry, lying 0.56 eV above the global minimum structure.
The optimized geometry is in good agreement with the other previously available reports \cite{b6-isomerization,b6-dft}. 
This is one of the isomers showing optical absorption at lower energies 
(\emph{cf.} Fig. \ref{fig:plots-neutral}\subref{subfig:neutral-plot-incomplete-wheel}). 
The many particle wave-functions of excited states contributing to various peaks 
are presented in Table \ref{Tab:table-neutral-incomplete-wheel}.
An intense peak at around 2.1 eV is seen in this isomer, characterized by $H_{1} \rightarrow L + 1$ and 
$H_{2} \rightarrow L $ transitions.

A wheel kind of structure, with its center slightly bulged out, is found to be the next stable isomer
of B$_{6}$. A singlet system with C$_{5v}$ point group symmetry, lies just 0.83 eV above in energy 
as compared to the most stable isomer. The pentagonal base has bond length of 1.625 \AA{} and the vertex 
atom is 1.678 \AA{} apart from the corners of the pentagon. Other reported values for those bond lengths are
1.61 \AA{}, 1.66 \AA{} \cite{turkish_boron, b6-dft} and 1.61 \AA{}, 1.659 \AA{} \cite{structure-bonding-b6,b6-isomerization} 
respectively. The optical absorption spectrum is presented in Fig. \ref{fig:plots-neutral}\subref{subfig:neutral-plot-bulged-wheel}.
The many-particle wave functions of excited states contributing to various peaks are presented in Table
\ref{Tab:table-neutral-bulged-wheel}. The onset of the spectrum occurs near 3.76 eV, with polarization in the plane
of the pentagonal base, characterized by excitations $H -1 \rightarrow L + 6$ and  $H \rightarrow L + 6$ with equal 
contribution.

A planar ring like structure, resembling the global minimum one; however, with singlet state and C$_{s}$ symmetry, 
is the next low lying isomer of B$_{6}$ cluster. The optimized geometry is in good agreement with Ref. \cite{b6-dft}.
The absorption spectrum is presented in Fig. \ref{fig:plots-neutral}\subref{subfig:neutral-plot-planar-ring-singlet} and
corresponding many-particle wavefunctions of various excited states are presented in Table 
\ref{Tab:table-neutral-planar-ring-singlet}. The absorption starts at the end of visible range, characterized by 
$H - 1 \rightarrow L $ and $H \rightarrow L + 23$.

An octahedron structure with O$_{h}$ point group symmetry is the next stable isomer of neutral B$_{6}$. Each side of the
octahedron is found to be 1.7 \AA{} as compared to the 1.68 \AA{} reported by \cite{turkish_boron} and \cite{b6-dft}.
The many-particle wave functions of the excited states correponding to various peaks 
(\emph{cf.} Fig. \ref{fig:plots-neutral}\subref{subfig:neutral-plot-octahedron}) are presented in Table
\ref{Tab:table-neutral-octahedron}. A very feeble absorption at 3.6 eV opens the spectrum, mainly 
characterized by excitations $H - 1 \rightarrow L$ and  $H - 1 \rightarrow L + 1$ with equal contribution.

Next isomer is previously unreported, with structure of a saddle threaded with dimer from top. It lies just 0.04 eV above 
the previous octahedron isomer. However the optical absorption spectrum 
(\emph{cf.} Fig. \ref{fig:plots-neutral}\subref{subfig:neutral-plot-threaded-tetramer}) is completely different. 
A narrow energy range hosts all the peaks. The onset of spectrum occurs near 2.7 eV, with major contribution from
$H - 1 \rightarrow H_{2}$ and $H_{1} \rightarrow L + 9$ (\emph{cf.} Table \ref{Tab:table-neutral-threaded-tetramer}).

An isosceles triangle connected to a linear chain of boron atoms forms the next isomer. This structure with C$_{2v}$
symmetry and a triplet electronic state have been reported in \cite{b6-dft}, which is in close agreement with our results.
The optical absorption spectrum (\emph{cf.} Fig. \ref{fig:plots-neutral}\subref{subfig:neutral-plot-threaded-trimer})
has distinctive closly lying peaks at 5.42 eV and 5.66 eV. The many-particle wavefunctions of excited states corresponding
to various peaks are presented in Table \ref{Tab:table-neutral-threaded-trimer}. Configurations $H - 3 \rightarrow L$
and $H - 1 \rightarrow L + 1$ contribute predominently to those closly lying peaks respectively.

A structure with two out of plane isosceles triangles joined together is found to be one of the isomers. 
The geometry has isosceles triangle with lengths 1.62 \AA{}, 1.62 \AA{} and 1.53 \AA{}, while two such triangles are
joined by a bond of length 1.66 \AA{}. The respective numbers reported by Ref. \cite{b6-dft} are 1.60 \AA{}, 1.60 \AA{},
1.50 \AA{} and 1.647 \AA{} respectively. The optical absorption spectrum contains many low intensity peaks except for 
strongest one at 5.87 eV, as presented in Fig. \ref{fig:plots-neutral}\subref{subfig:neutral-plot-twisted-trimers}. 
The onset of the spectrum occurs at 1 eV -- a peak equally dominated by $H\rightarrow L$ and $H - 1 \rightarrow L$
(\emph{cf.} Table \ref{Tab:table-neutral-twisted-trimers}). 

An almost degenerate structure forms the next isomer, lying just 0.009 eV above the previous isomer. Contrary to
the previous one, this geometry is completely planar and is a triplet system, with C$_{2v}$ point group symmetry. 
Probably because of such a strong near-degeneracy, this isomer has not been reported in the literature before.
The many-particle wave functions of the excited states correponding to various peaks 
(\emph{cf.} Fig. \ref{fig:plots-neutral}\subref{subfig:neutral-plot-planar-trimers}) are presented in Table
\ref{Tab:table-neutral-planar-trimers}. The spectrum opens with a very feeble peak $H \rightarrow L$ and 
$H - 1 \rightarrow L + 1 $ as dominant contributions to the excited state wavefunction.

Convex bowl shaped isomer and a perfect linear chain are found very high in energy, ruling out their existence at room
temperature. The optical spectra are presented in Figs. \ref{fig:plots-neutral}\subref{subfig:neutral-plot-convex-bowl}
and \ref{fig:plots-neutral}\subref{subfig:neutral-plot-linear} respectively. The corresponding many-particle wavefunctions 
of excited states of various peaks are presented in Table \ref{Tab:table-neutral-convex-bowl} and 
\ref{Tab:table-neutral-linear}. The onset of absorption spectrum of convex bowl isomer occurs at 1.58 eV with 
$H - 1 \rightarrow L  + 1 $ as the most dominant configuration. The bulk of the oscillator strength of the spectrum of linear 
isomer is carried by $ H - 1 \rightarrow L + 3$ and $H \rightarrow L + 2$ having equal contributions.


\subsection{B$_{6}^{+}$}

\begin{figure*}
\begin{center}
\subfigure[C$_{s}$, $^{2}A^{''}$ \newline Planar ring (I)]
{\psfig{figure=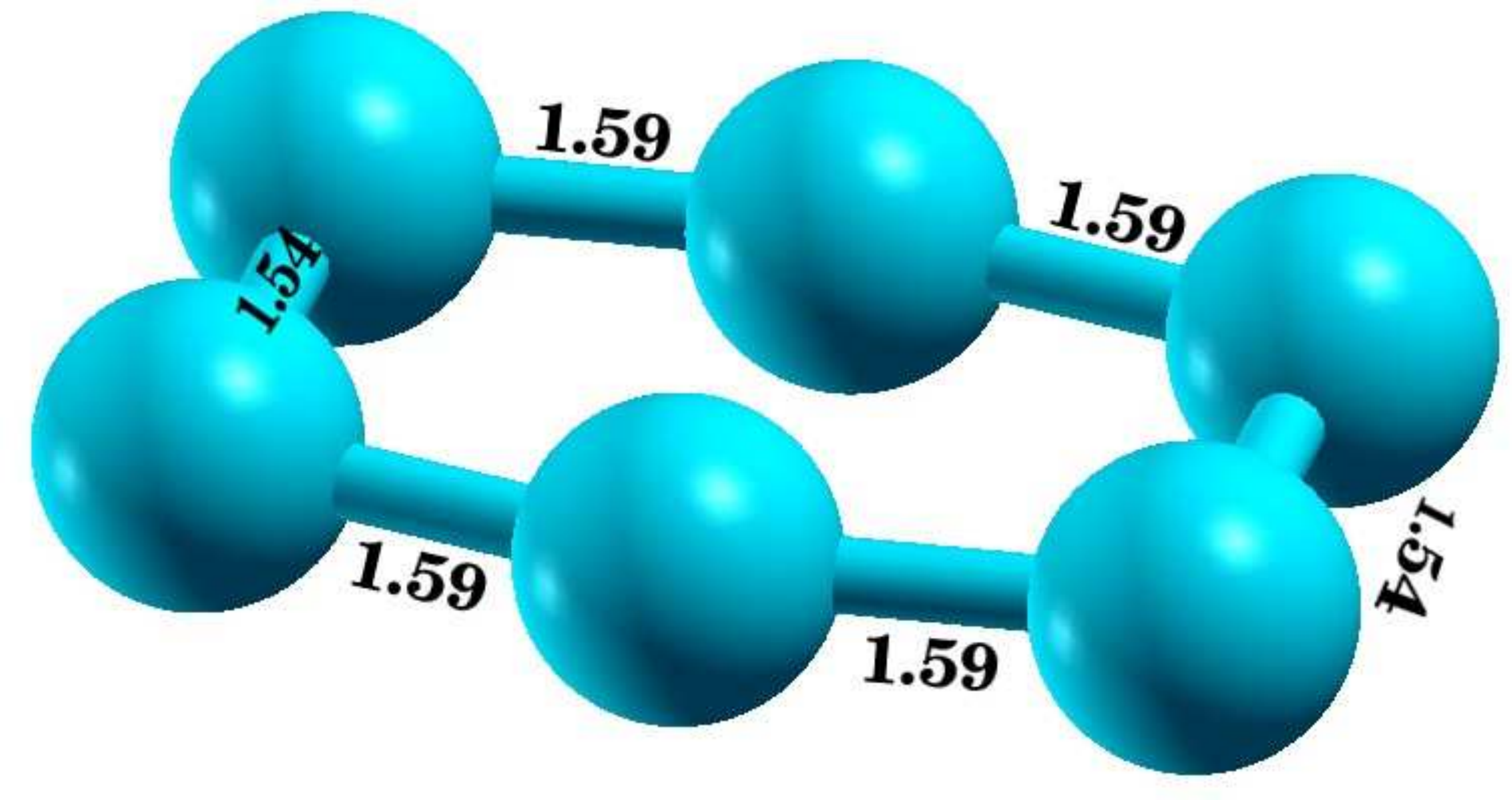,scale=0.12}} \hspace{0.5cm}
\subfigure[C$_{1}$, $^{2}A$ \newline Bulged wheel]
{\psfig{figure=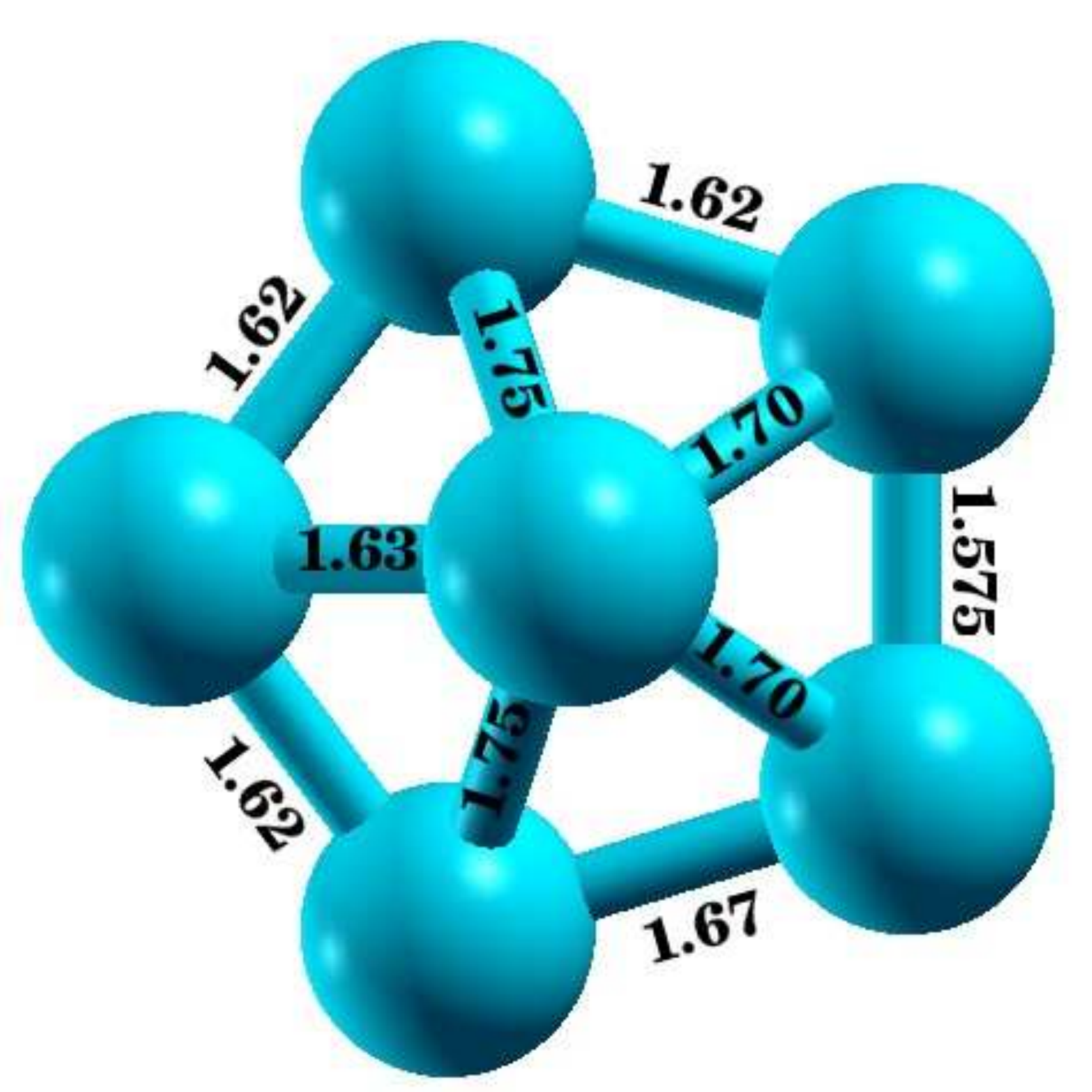,scale=0.15}} \hspace{0.5cm}
\subfigure[D$_{2h}$, $^{2}A_{g}$ \newline Planar ring (II)]
{\psfig{figure=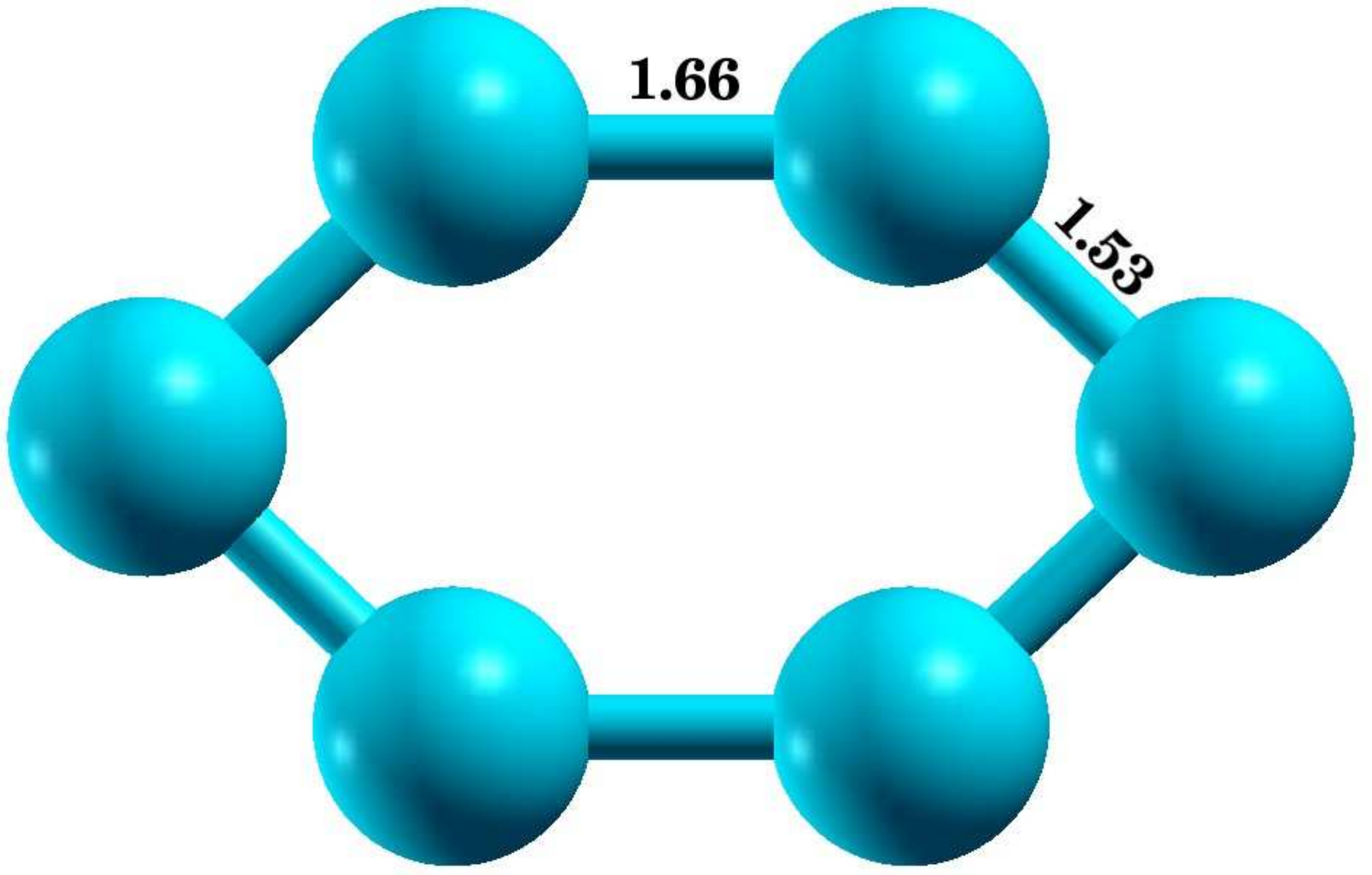,scale=0.1}} \hspace{0.5cm}
\subfigure[C$_{2v}$, $^{4}A_{2}$ \newline Incomplete wheel]
{\psfig{figure=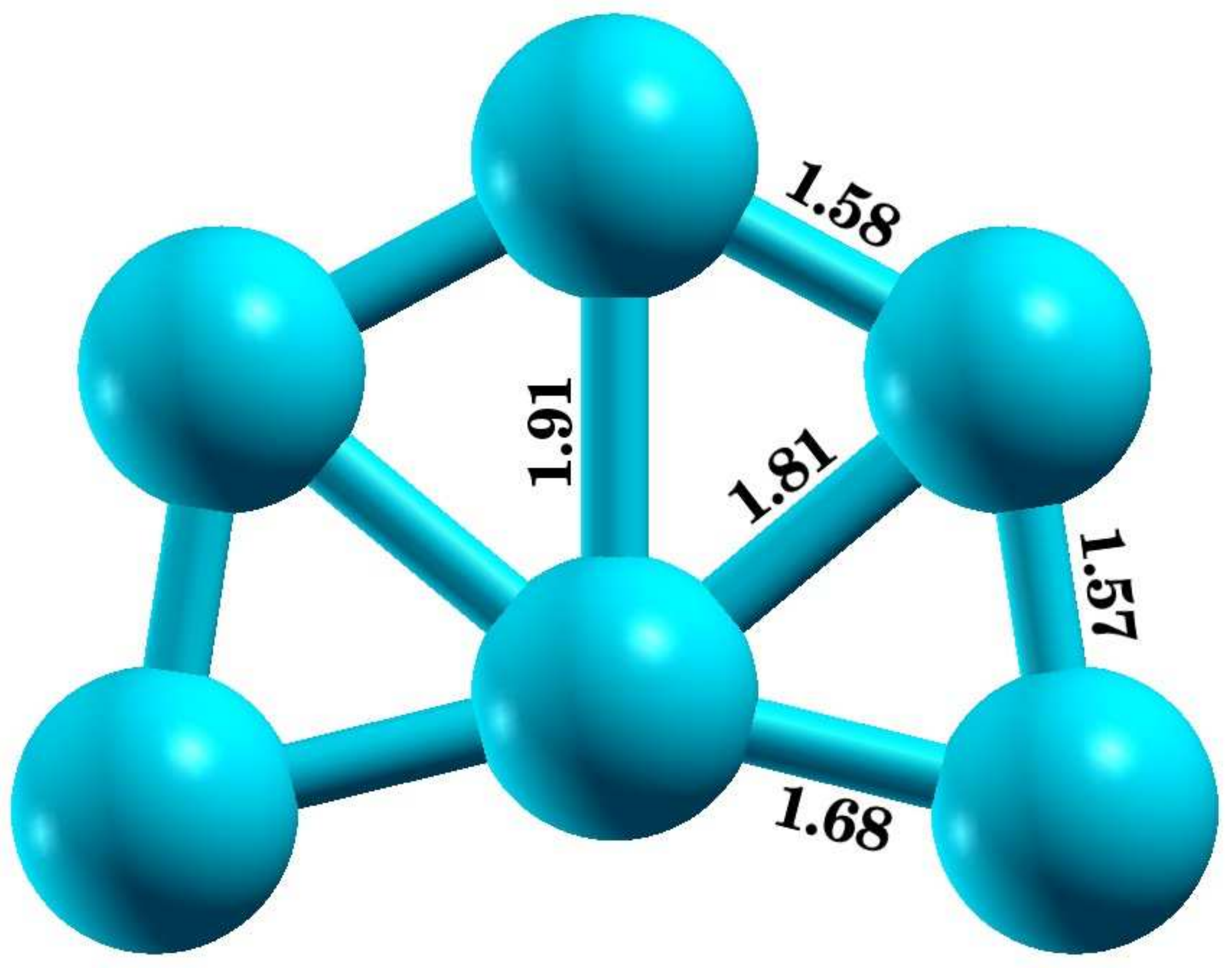,scale=0.1}} \\
\subfigure[C$_{2v}$, $^{4}A_{2}$ Threaded trimer]
{\psfig{figure=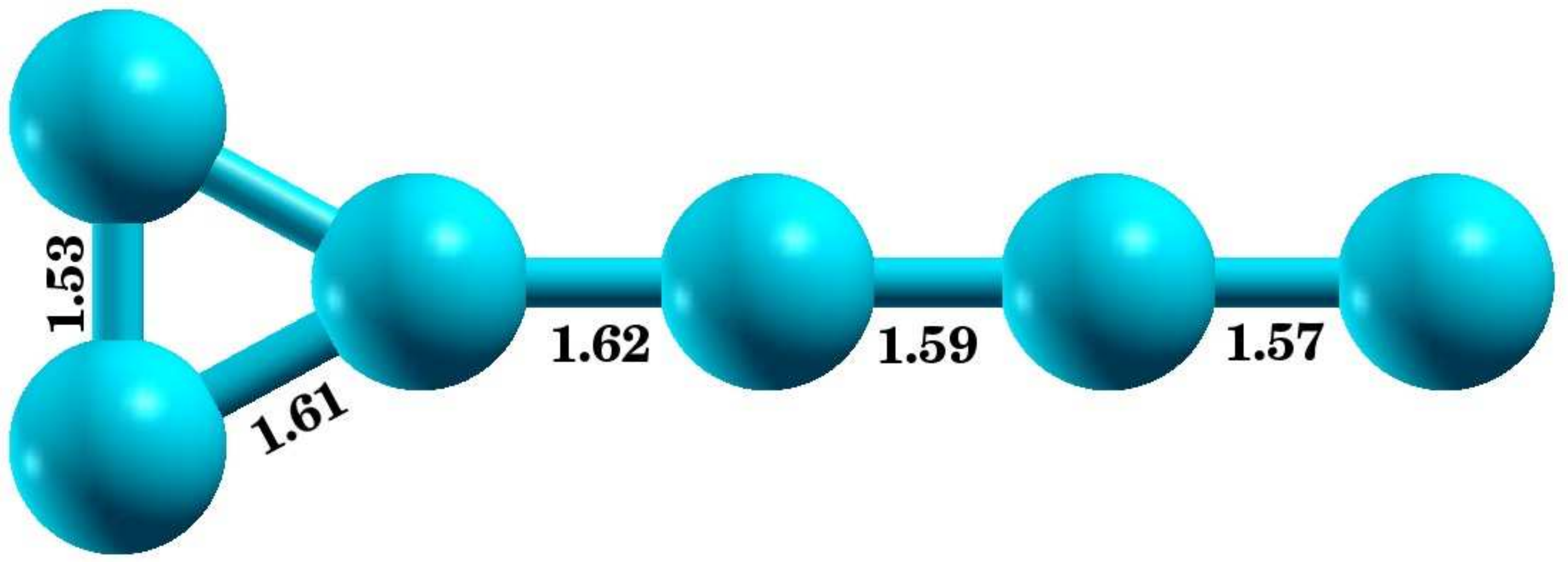,scale=0.13}} \hspace{0.5cm}
\subfigure[D$_{4h}$, $^{2}B_{1g}$ Bipyramid]
{\psfig{figure=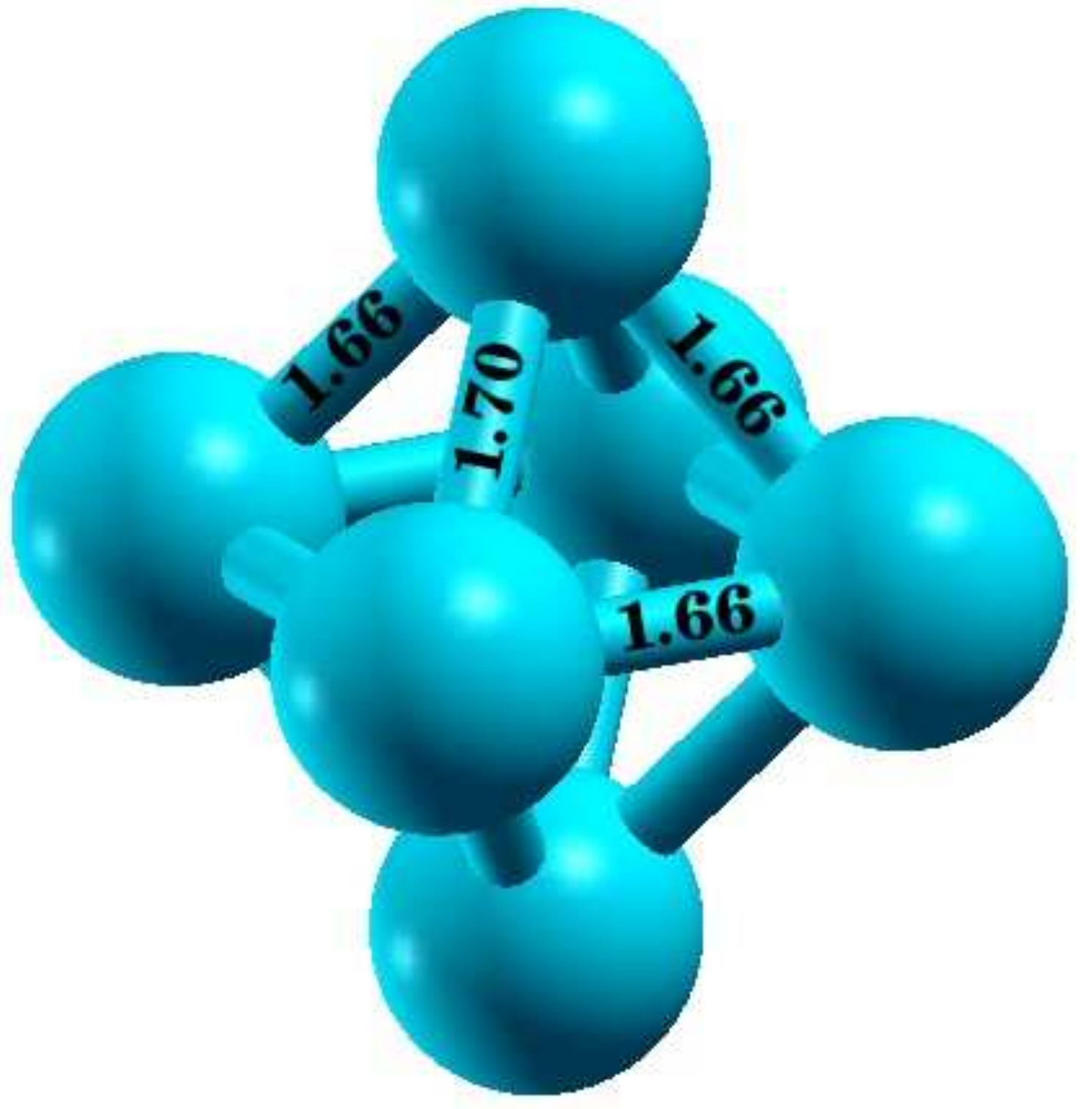,scale=0.15}} \hspace{0.5cm}
\subfigure[D$_{\infty h}$, $^{4}\Sigma_{u}$ Linear]
{\psfig{figure=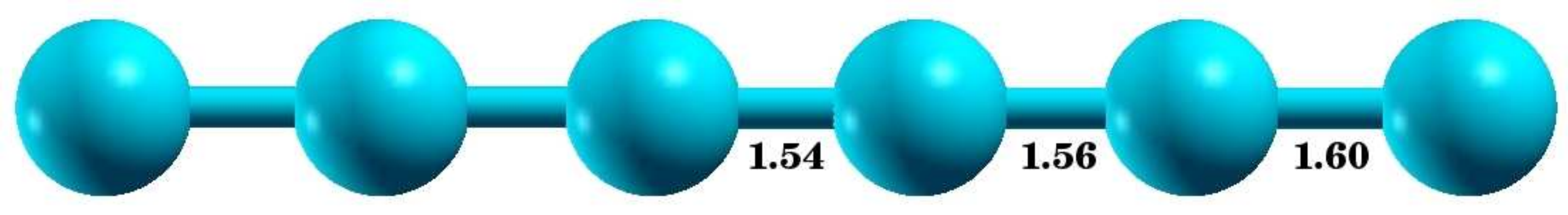,width=6.4cm}}   \hspace{0.5cm}
\subfigure[D$_{2h}$, $^{2}B_{2g}$ Planar trimers]
{\psfig{figure=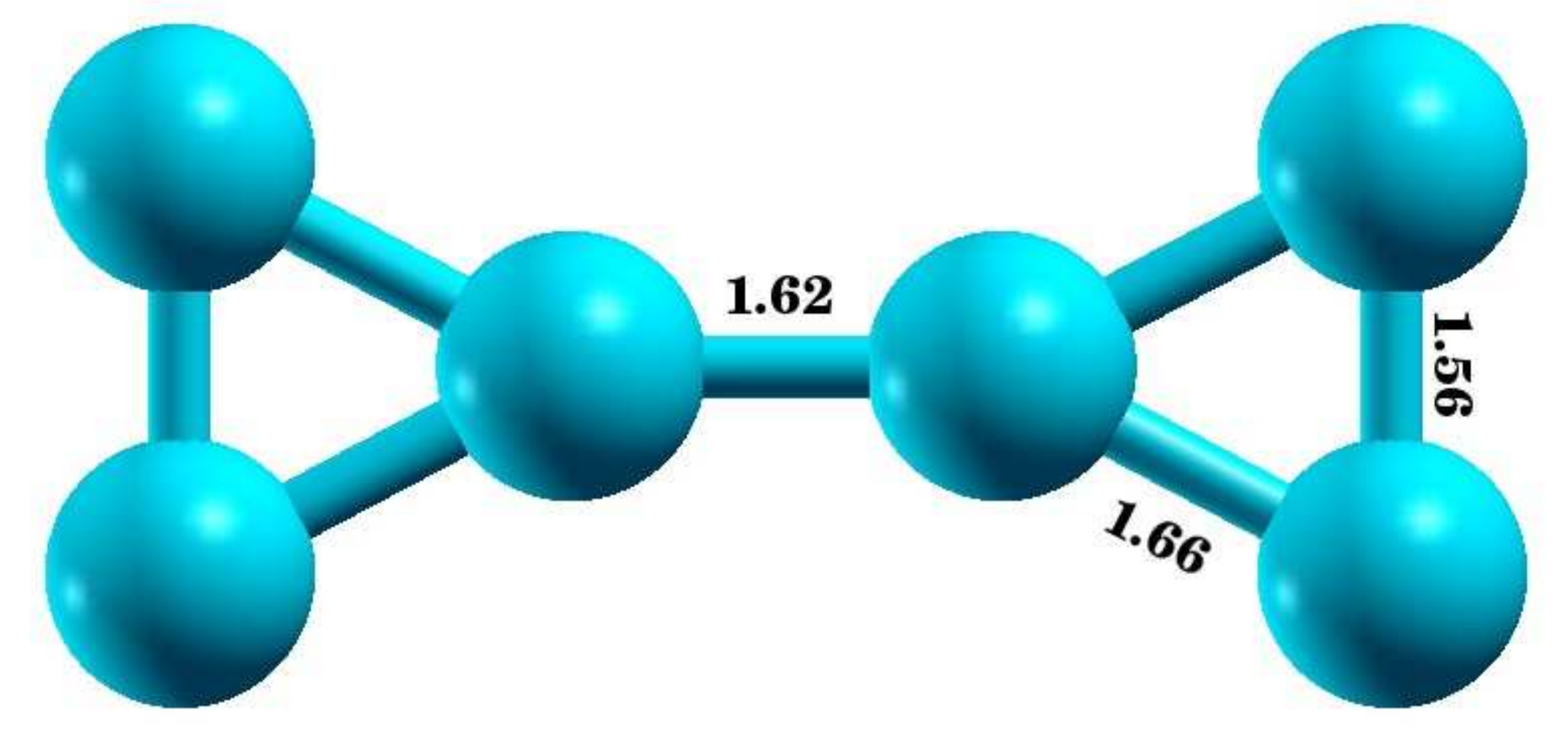,scale=0.15}} \\
\end{center}
\caption{\label{fig:geometries-cationic} (Color online) Geometry optimized ground state structures
of different isomers of cationic B$_{6}^{+}$ clusters, along with the point group symmetries
obtained at the CCSD level. }
\end{figure*}

We have found a total of 8 isomers of cationic (B$_{6}^{+}$) cluster with stable geometries 
as shown in the Fig. \ref{fig:geometries-cationic}. The relative standings in energy 
are presented in the Table \ref{tab:energies-cationic}, along with point group symmetries and
electronic states. Cationic clusters show activity in the visible range, contrary to their neutral counterpart.
Also, in most of the cases the geometry of the neutral isomer is retained, reflected in the fact that some peaks
show up in the optical absorption spectra at the same energies as those in the neutral cluster.

\begin{table}
\caption{Point group, electronic state and total energies of different isomers of B$_{6}^{+}$ cluster.}
\label{tab:energies-cationic}       
\begin{tabular}{clllc}
\hline\noalign{\smallskip}
Sr.    	& Isomer		& Point 	& Elect. 	& Total  	\\
no. 	& 			& group 	& State      	& Energy (Ha) 	\\  
\noalign{\smallskip}\hline\noalign{\smallskip}
1	& Planar ring (I)	& C$_{s}$	& ${}^2 A^{''}$	& -147.492831	\\
2	& Bulged wheel 		& C$_{1}$	& ${}^2 A$	& -147.491994	\\
3	& Planar ring (II)	& D$_{2h}$	& ${}^2 A_{g}$	& -147.480796	\\
4	& Incomplete wheel 	& C$_{2v}$	& ${}^4 A_{2}$	& -147.454234	\\
5	& Threaded trimer 	& C$_{2v}$	& ${}^4 A_{2}$	& -147.429627	\\
6	& Tetra. bipyramid 	& D$_{4h}$	& ${}^2 B_{1g}$	& -147.413145	\\
7	& Linear 		& D$_{\infty h}$& ${}^4\Sigma_{u}$&-147.392263	\\
8	& Planar trimers	& D$_{2h}$	& ${}^2 B_{2g}$	& -147.358494	\\
\noalign{\smallskip}\hline
\end{tabular}
\end{table}


\begin{figure*}[!t]
\begin{center}
\subfigure[Planar Ring (I) Isomer] 
{\psfig{figure=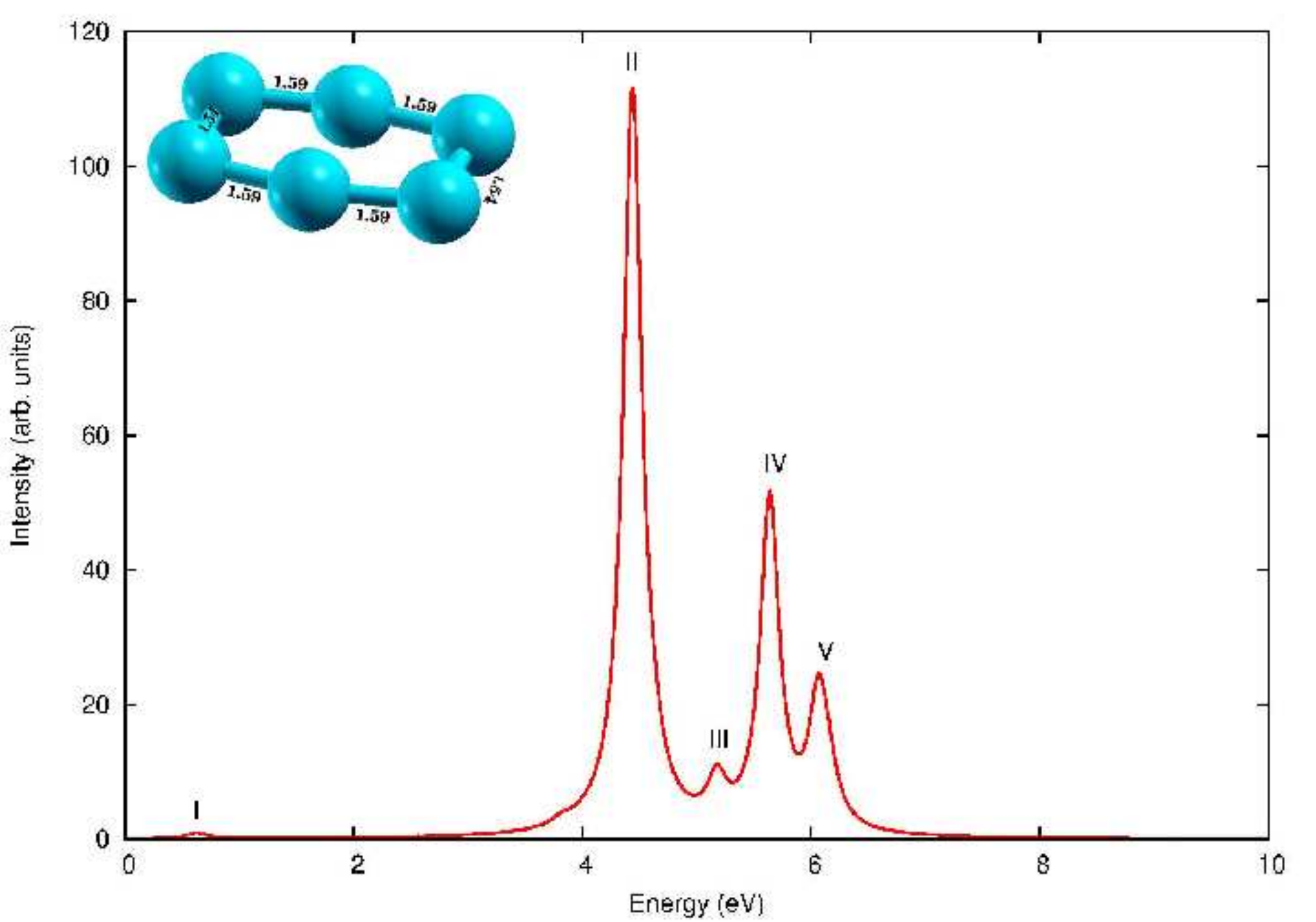,width=0.32\textwidth} \label{subfig:cationic-plot-planar-ring}} 
 \subfigure[Bulged wheel Isomer]
 {\psfig{figure=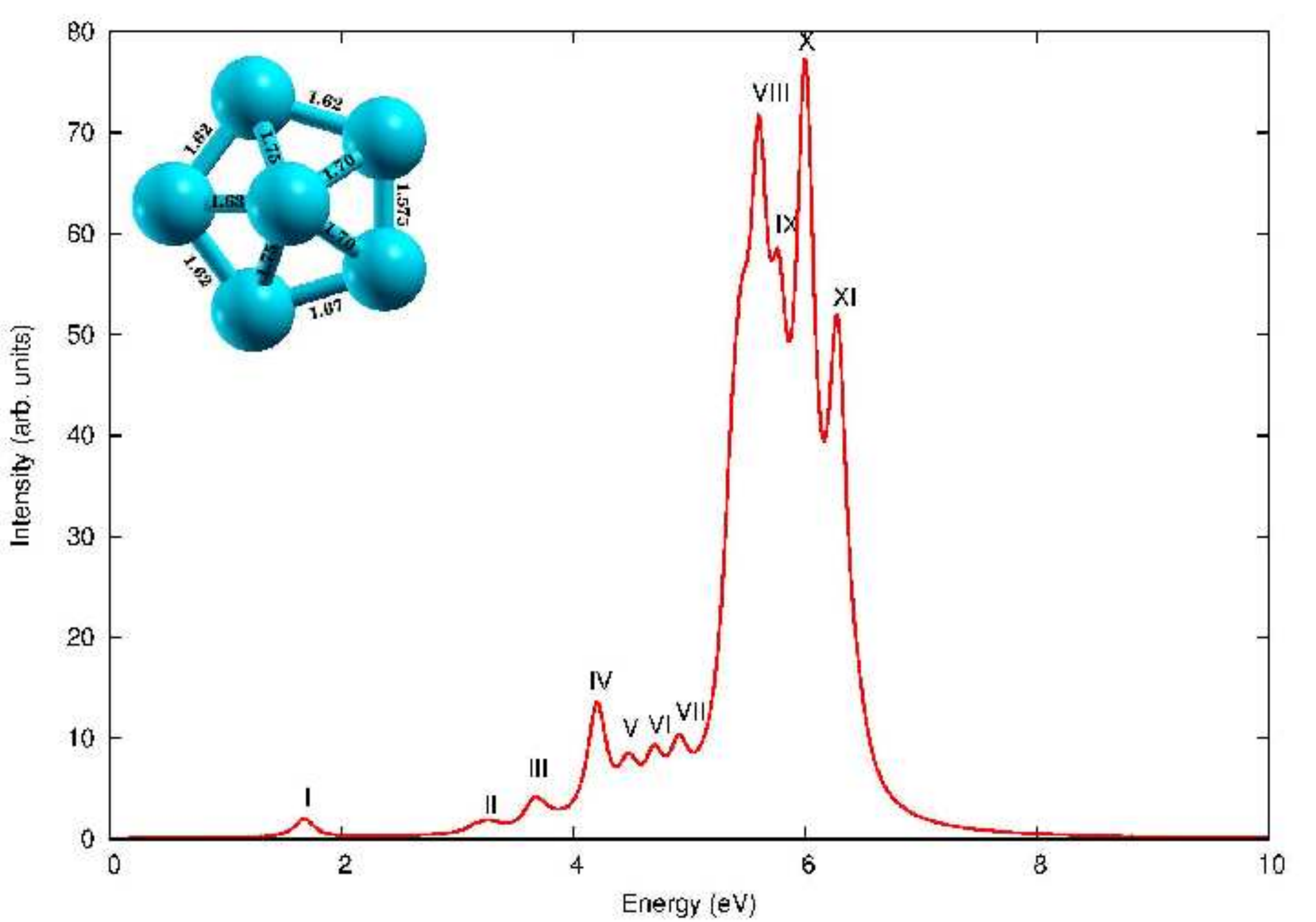,width=0.32\textwidth} \label{subfig:cationic-plot-bulged-wheel}}
\subfigure[Planar Ring (II) Isomer] 
{\psfig{figure=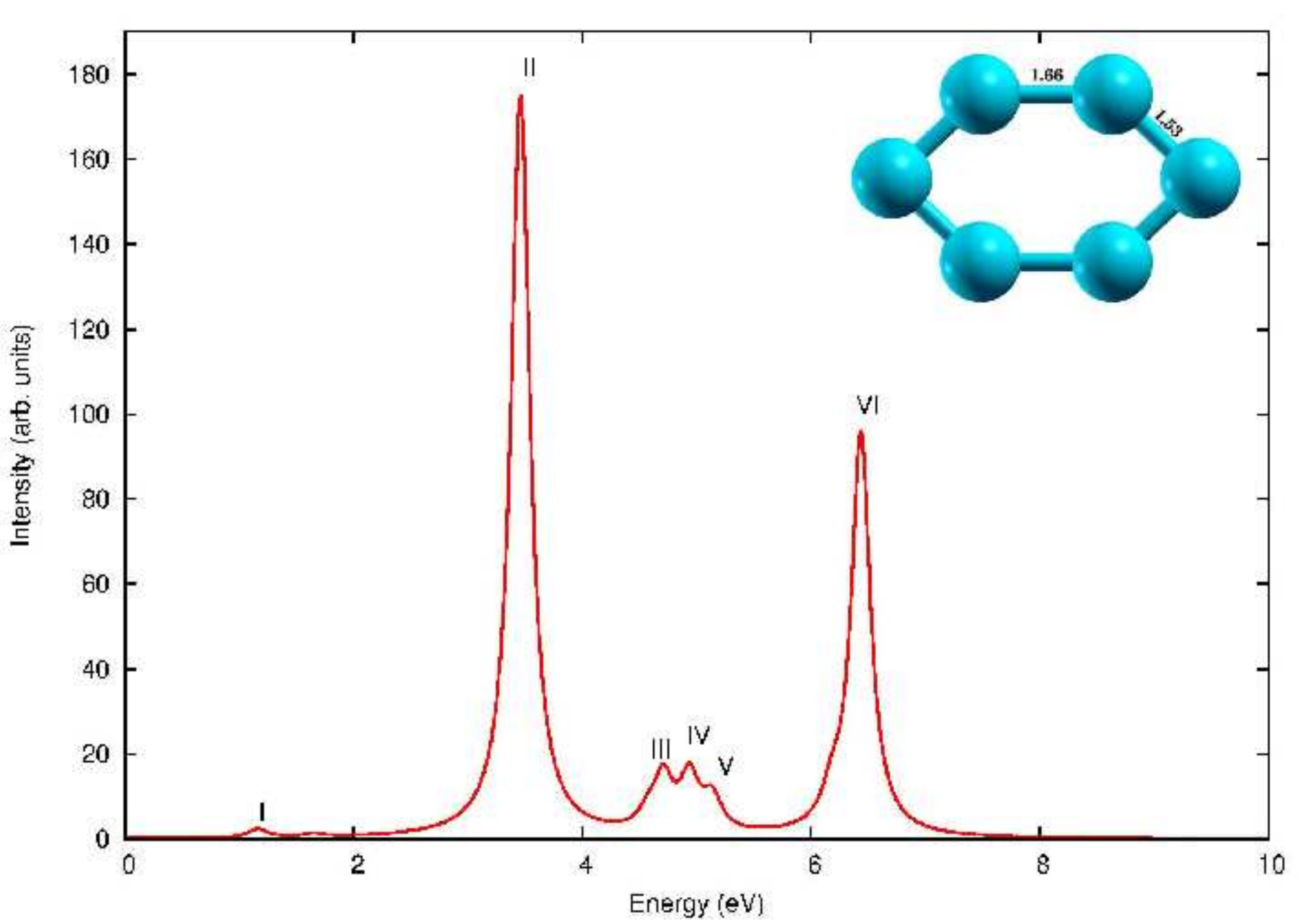,width=0.32\textwidth} \label{subfig:cationic-plot-planar-d2h-ring}} \\
\subfigure[Incomplete Wheel]
{\psfig{figure=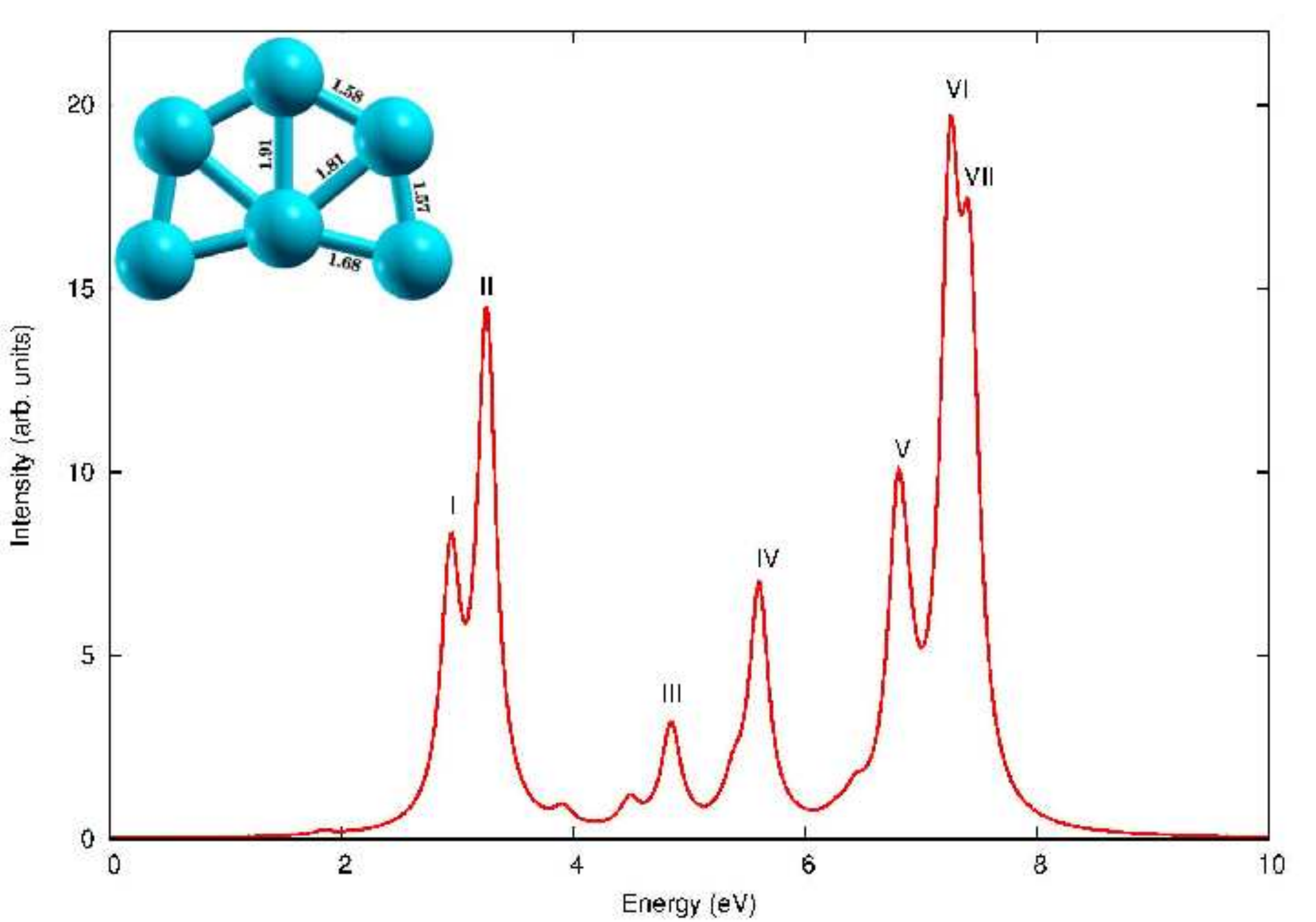,width=0.32\textwidth} \label{subfig:cationic-plot-incomplete-wheel}} 
\subfigure[Threaded trimer]
{\psfig{figure=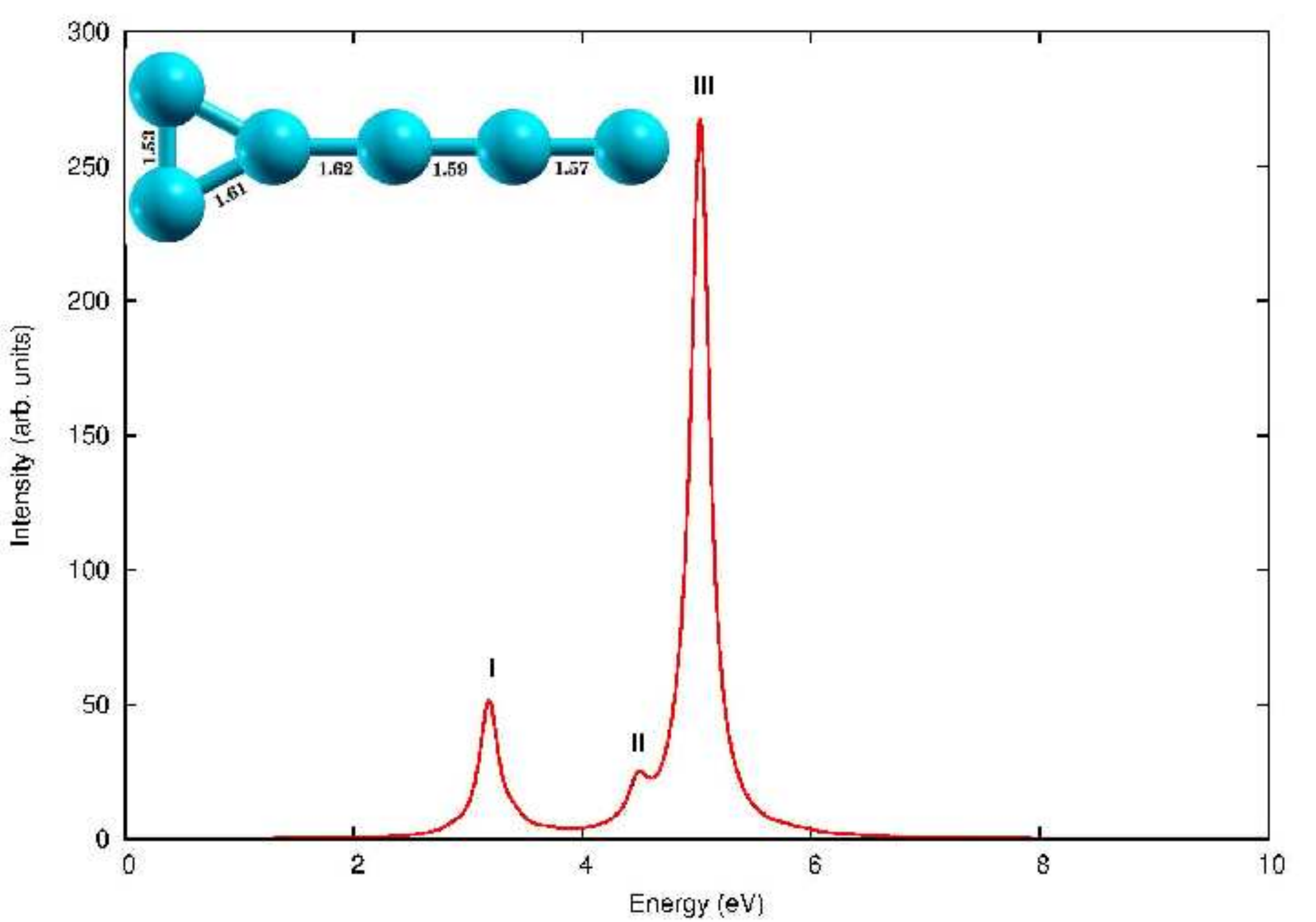,width=0.32\textwidth} \label{subfig:cationic-plot-threaded-trimer}} 
 \subfigure[Tetragonal bipyramid]
 {\psfig{figure=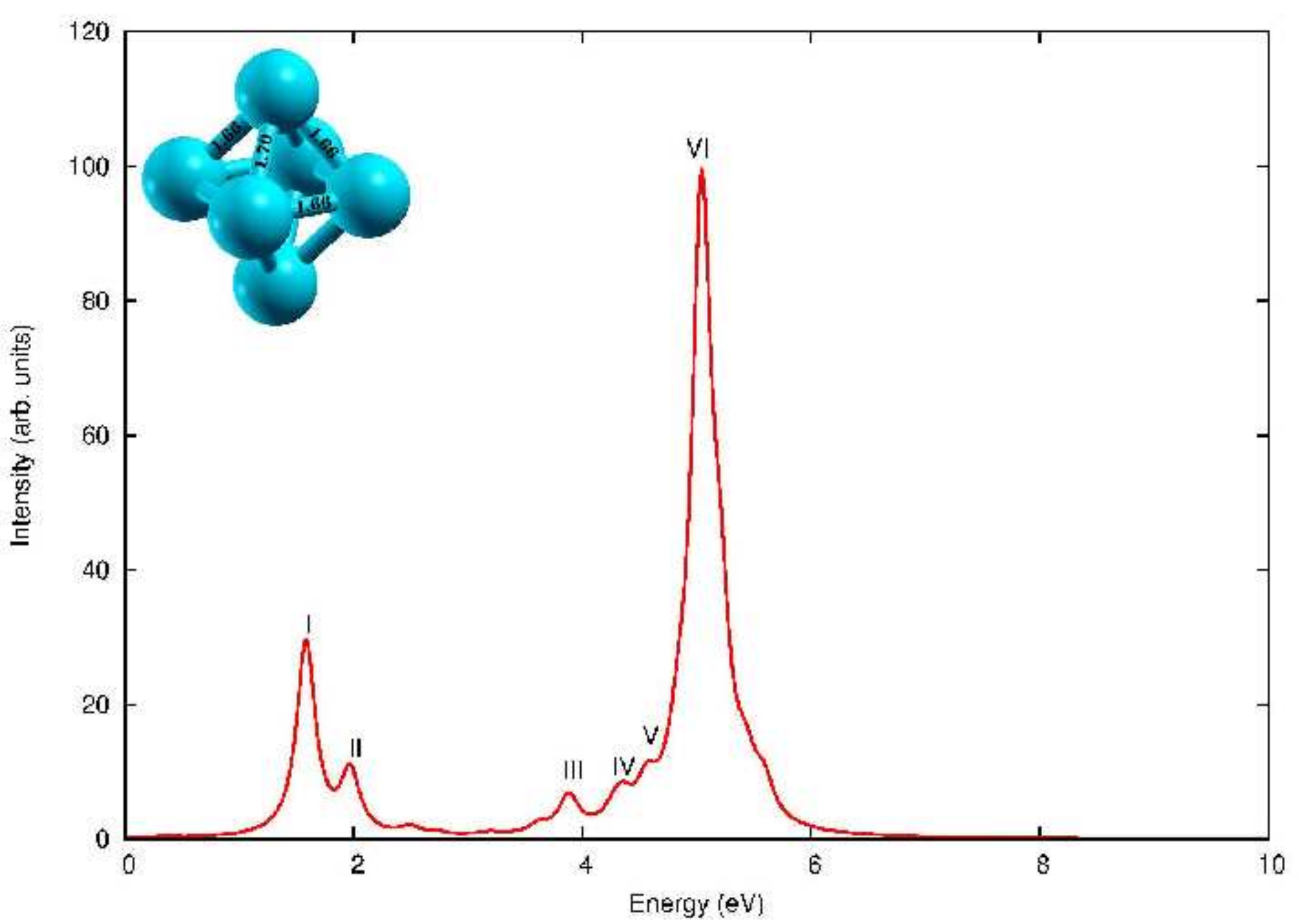,width=0.32\textwidth} \label{subfig:cationic-plot-bipyramid}} \\
 \subfigure[Linear]
 {\psfig{figure=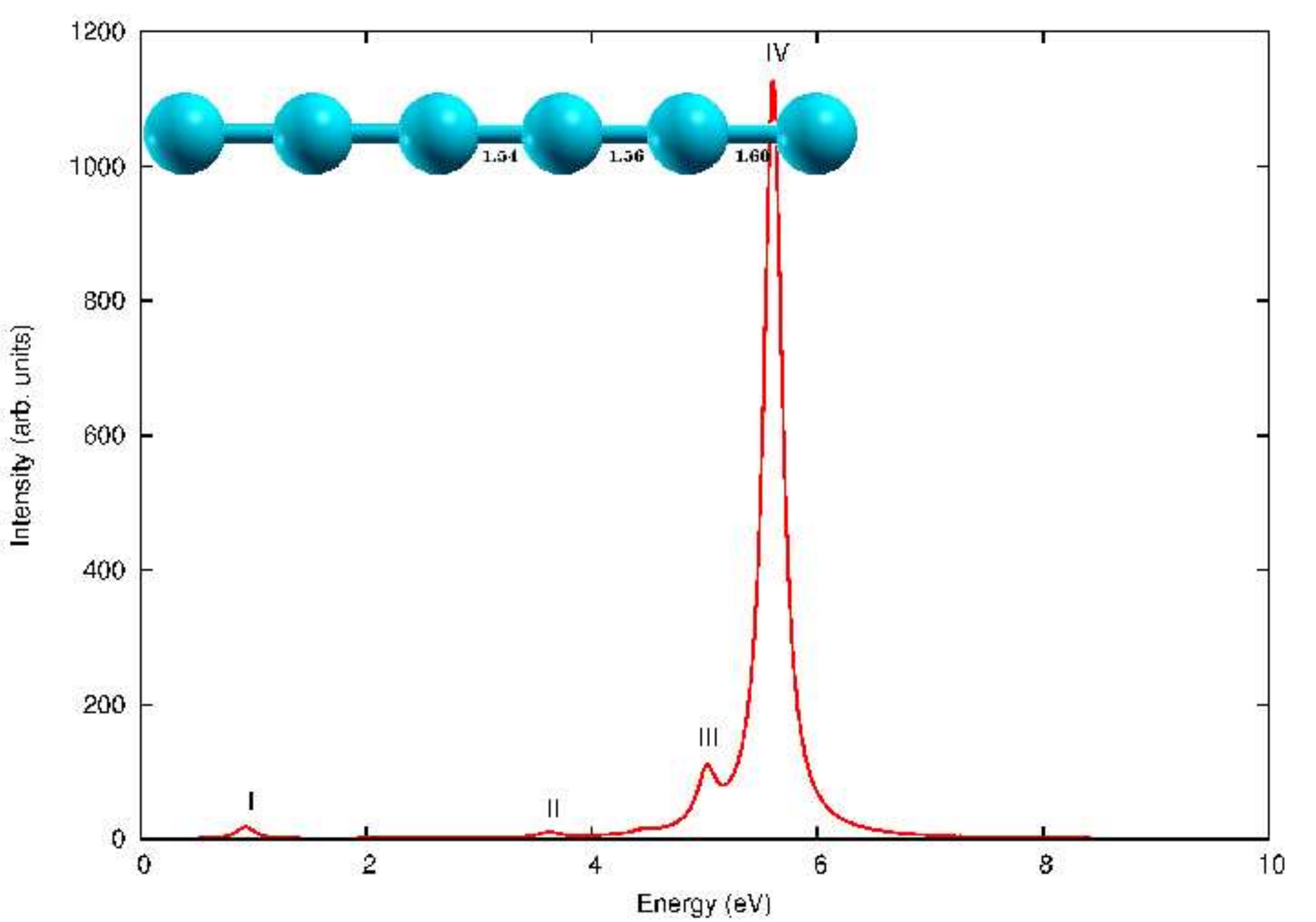,width=0.32\textwidth} \label{subfig:cationic-plot-linear} }
\subfigure[planar bingo]
 {\psfig{figure=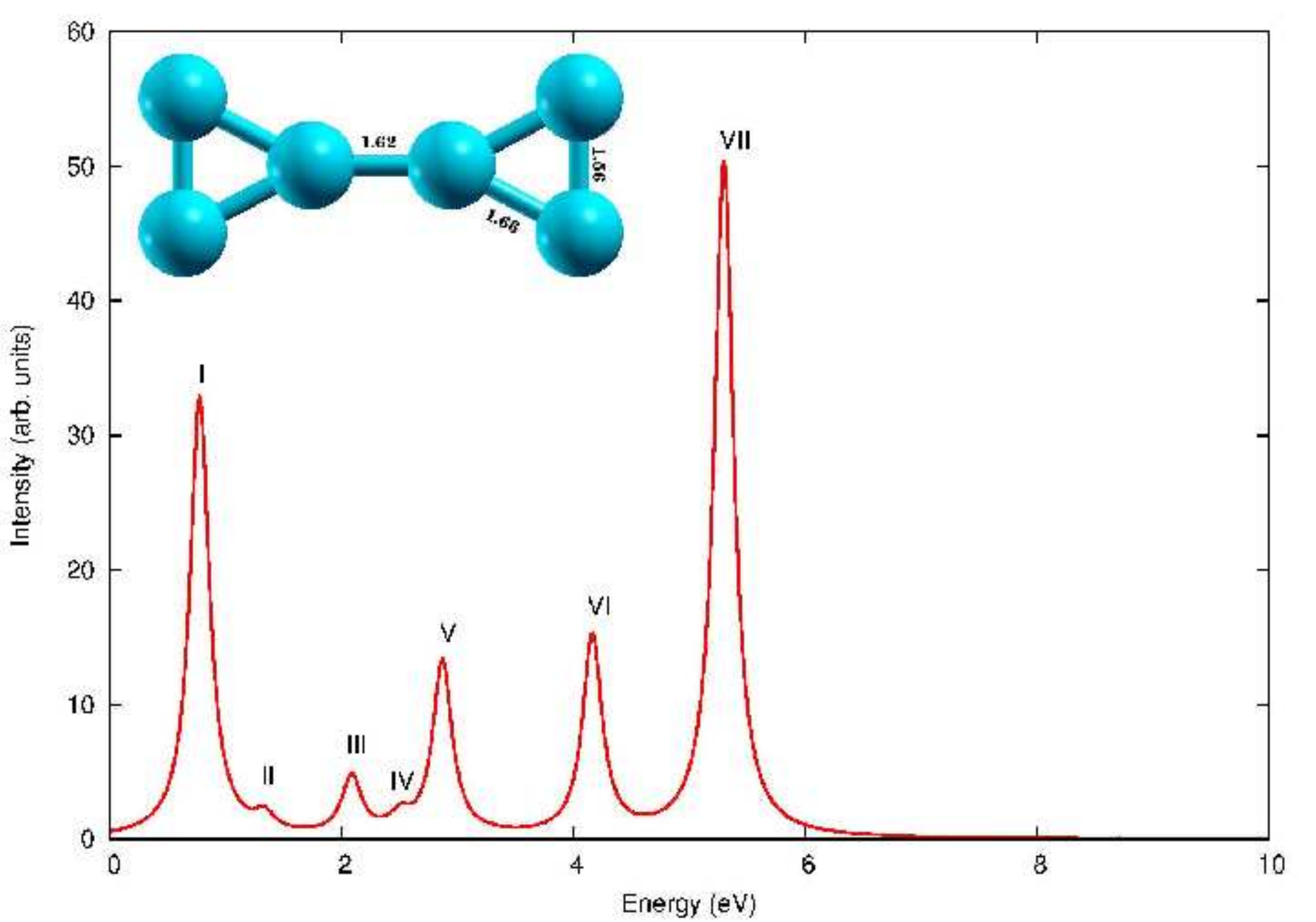,width=0.32\textwidth} \label{subfig:cationic-plot-planar-trimers}} 
\end{center}
\caption{\label{fig:plots-cationic}(Color online) The linear optical absorption spectrum
of B$_{6}^{+}$, calculated using the SCI approach. The many-body wavefunctions of excited stated corresponding
to the peaks labeled are given in the Appendices \ref{Tab:table-cationic-planar-ring}--
\ref{Tab:table-cationic-planar-trimers}. For plotting the spectrum, a uniform linewidth of 0.1 eV was assumed. }
\end{figure*}


The most stable isomer of B$_{6}^{+}$ cluster is a planar ring-type of structure, with C$_{s}$ point group symmetry.
This is in contrast to the other reported geometries which have D$_{2h}$ symmetry \cite{b6-isomerization,vlasta-chem-review,b6-dft}.
A slight difference in the orientation makes it less symmetric.
However, the bonds lengths obtained are in good agreement with those with D$_{2h}$ symmetric geometry cited above.
 The optical absorption spectrum calculated using SCI approach is as shown in the 
Fig. \ref{fig:plots-cationic}\subref{subfig:cationic-plot-planar-ring} and corresponding many particle 
wavefunctions of excited states contributing to the various peaks are presented in 
Table \ref{Tab:table-cationic-planar-ring}.
 Similar to the neutral counterpart, this isomer also has very feeble absorption in the 
visible range, with polarization perpendicular to the plane of the isomer. The dominant contribution to
the peak at 0.63 eV comes from $H_{1} \rightarrow L$ and $H - 2 \rightarrow L+ 16$ . The spectrum is slightly red-shifted
with respect to the neutral counterpart.

Bulged wheel structure is the next low lying isomer of B$_{6}^{+}$ with just 0.023 eV above the global minimum.
However, as compared to the neutral one, this geometry has C$_{1}$ symmetry due to the significant bond length 
reordering. Our results are consistent with the results of Refs. \cite{b6-isomerization, b6-dft}. 
The optical absorption spectrum is presented in Fig. \ref{fig:plots-cationic}\subref{subfig:cationic-plot-bulged-wheel}.
The many-particle wave functions of excited states contributing to various peaks are presented in Table
\ref{Tab:table-cationic-bulged-wheel}. The spectrum is distinctly different with bulk of the oscillator strength 
carried by peaks at 6 eV. The onset of spectrum occurs at 1.67 eV dominated by $H - 2 \rightarrow H_{1} $ 
and $H_{1} \rightarrow L + 3 $ configurations.

Another ring-like structure with D$_{2h}$ symmetry and doublet multiplicity lies next in the energy order. The six faced 
benzene type structure has 1.66 \AA{} and 1.53 \AA{} as unique bond lengths, which are somewhat larger than those 
reported in the literature \cite{b6-isomerization,b6-dft}.The optical absorption spectrum is presented in Fig. 
\ref{fig:plots-cationic}\subref{subfig:cationic-plot-planar-d2h-ring}. The many-particle wave functions 
of excited states contributing to various peaks are presented in Table \ref{Tab:table-cationic-planar-d2h-ring}.
All absorption peaks are due to the polarization along the plane of the isomer. The onset of spectrum occurs at 1.2 eV
with a very feeble peak with dominant contributions from $H - 1 \rightarrow L $ and $H - 2 \rightarrow H_{1} $.

Next low lying isomer is a planar incomplete wheel structure with C$_{2v}$ point group symmetry and quartet multiplicity.
This is consistent with results of Ref. \cite{b6-dft}. The optical absorption 
(\emph{cf.} Fig. \ref{fig:plots-cationic}\subref{subfig:cationic-plot-incomplete-wheel}) starts at 2.94 eV, with 
polarization transverse to the plane of the isomer. The many particle wave-functions of excited states contributing to various peaks 
are presented in Table \ref{Tab:table-cationic-incomplete-wheel}. The configurations contributing to the first peak are
$H_{1} \rightarrow L + 1 $ and $H_{2} \rightarrow L $, where $H_{1}$ and $H_{2}$ are singly occupied molecular orbitals of the quartet 
system.

Another isomer with the same point group symmetry and multiplicity as that of the previous one, but having a 
geometry of linear chain with an isosceles triangle at the end, is the next low lying isomer of cationic B$_{6}^{+}$.
Our results about geometry are in good agreement with the Ref. \cite{b6-dft}. 
The optical absorption spectrum (\emph{cf.} Fig. \ref{fig:plots-cationic}\subref{subfig:cationic-plot-threaded-trimer})
has very few peaks. The onset of spectrum occurs at 3.2 eV, with polarization along the plane of the isomer, and 
has dominant contribution from $H - 3 \rightarrow H_{3}$ and $H - 3 \rightarrow L$ (\emph{cf.} 
Table \ref{Tab:table-cationic-threaded-trimer}). 

Tetragonal bipyramid forms the next stable isomer of cationic B$_{6}^{+}$, with D$_{4h}$ point group symmetry 
and doublet multiplicity. This is in good agreement with the geometries reported in Refs. \cite{b6-isomerization}
and \cite{b6-dft}. The optical absorption spectrum 
(\emph{cf.} Fig. \ref{fig:plots-cationic}\subref{subfig:cationic-plot-bipyramid}) has 
peaks in the visible range at 1.58 eV and 1.97 eV, with dominant contribution from $H_{1} \rightarrow L + 2$ and
$H - 1 \rightarrow L + 2$ respectively (\emph{cf.} Table \ref{Tab:table-cationic-triangular-bipyramid}).

Two more structures were found stable \emph{i.e.} (a) a planar structure with two trimers joined together and,
(b) a linear one. These isomers are much above the global minimum energy, it rules out their room temperature existence.
In the linear isomer the absorption begins with a small peak at 0.93 eV, with dominant contribution from $H - 3 \rightarrow H_{3}$
and $H - 1 \rightarrow L $ configurations. In case of planar trimers structure, the spectrum seems to be red shifted as 
 compared to the neutral one. The first peak is found at 0.77 eV with $H - 2 \rightarrow H_{1} $ and $H_{1} \rightarrow L + 4 $
as dominant contribution to the wavefunction of the excited state.

\section{Conclusion and Outlook}
\label{sec:conclusion}
A large number of randomly selected initial structures of neutral B$_{6}$
and cationic B$_{6}^{+}$ clusters are taken into consideration for 
locating the global and local minimas on the potential energy curves. A careful 
geometry optimization is done for all those structures at a correlated level.
The optical absorption spectra of different low-lying isomers of both neutral
and cationic isomers are reported here. A singles configuration interaction approach was 
used to compute excited state energies and the absorption spectra of various clusters.
Spectra of cationic clusters appear slightly red-shifted with respect to the neutral one.
Different isomers exhibit distinct optical response, even though they are isoelectronic 
and many of them are almost degenerate. 
This signals a strong-structure property relationship, which can be exploited 
for experimental identification of these isomers; something which is not possible with the 
conventional mass spectrometry.
A strong mixture of configurations in the many-body wave functions of various excited states
are observed, indicating the plasmonic nature of the photoexcited states \cite{plasmon}.


Since aluminum also has the same number of valence electrons, it will be interesting to compute its
 optical absorption spectra, and compare them with those of boron cluster with the help of many-body 
wavefunctional analysis. The results of such calculations done by us will be 
communicated in near future \cite{aluminum-ravi}.

\acknowledgement
The author would like to acknowledge the Council of Scientific and Industrial Research (CSIR), India, 
for their financial support (SRF award No. 09/087/(0600)2010-EMR-I and travel grant no. TG/6939/12--HRD).

\appendix

\section{Excited State CI Wavefunctions, Energies and Oscillator Strengths}
\label{app:wavefunction}
In the following tables, we have given the
excitation energies (with respect to the ground state), and the many body
wavefunctions of the excited states, corresponding
to the peaks in the photoabsorption spectra of various isomers listed
in Fig. \ref{fig:geometries-neutral} and Fig. \ref{fig:geometries-cationic}
, along with the oscillator strength $f_{12}$ of the transitions,
\begin{equation}
f_{12}=\frac{2}{3}\frac{m_{e}}{\hbar^{2}}(E_{2}-E_{1})\sum_{i}|\langle m|d_{i}|G\rangle|^{2}
\end{equation}
where, $|m\rangle$ denotes the excited state in question, $|G\rangle$,
the ground state, and $d_{i}$ is the $i$-th Cartesian component
of the electric dipole operator. The single excitations are with respect to the reference state as given in
respective tables.

\FloatBarrier

\setcounter{table}{0}
\numberwithin{table}{section}
\begin{table*}
\caption{Excitation energies, $E$, and many-particle wave functions of excited
states corresponding to the peaks in the linear absorption spectrum
of B$_{6}$ -- planar ring (triplet) isomer (\emph{cf}. Fig. \ref{fig:plots-neutral}\subref{subfig:neutral-plot-planar-ring-triplet}).
The subscript $\Vert$ in the peak number denotes the absorption due to light polarized in the plane of isomer. 
In the wave function, the bracketed numbers are the CI coefficients of a given electronic configuration.
Symbols $H_{1}$,$H_{2}$ denote SOMOs discussed earlier, and $H$, and $L$, denote HOMO and LUMO orbitals respectively. Note
that, the reference state does not correspond to any peak, instead it represents the reference state from which singles excitations are occuring.}
\centering
\begin{tabular}{cccl}
\hline\noalign{\smallskip}
Peak	 	& $E$ (eV) & $f_{12}$ & Wave Function \\
\hline
\tabularnewline
Reference 	&	&	& $| H_{1}^{1}; H_{2}^1 \rangle $	\\
\tabularnewline
I$_{\Vert}$ 	& 2.84  &0.0187 & $| H - 3 \rightarrow H_{2} \rangle$(0.7353) \\
		&  	&	& $| H - 1 \rightarrow H_{1} \rangle$(0.5873) \\
\tabularnewline
II$_{\Vert}$ 	& 4.51 	&1.7647	& $| H - 1 \rightarrow H_{1} \rangle$(0.7214) \\
		&  	&	& $| H - 3 \rightarrow H_{2} \rangle$(0.4417) \\
\tabularnewline
III$_{\Vert}$ 	& 5.03 	&0.2023	& $| H - 3 \rightarrow  L \rangle $(0.5919) \\
		&  	&	& $| H - 1 \rightarrow L + 8 \rangle$(0.5137) \\
\tabularnewline
IV$_{\Vert}$ 	& 6.62 	&0.9942	& $| H - 1 \rightarrow L + 8 \rangle$(0.5440) \\
		&  	&	& $| H - 1 \rightarrow L + 3 \rangle$(0.4900) \\
\tabularnewline
V$_{\Vert}$ 	& 7.96 	&0.5799	& $| H - 1 \rightarrow L + 3 \rangle$(0.6800) \\
		&  	&	& $| H - 1 \rightarrow L + 8 \rangle$(0.5298) \\
\noalign{\smallskip}\hline
\end{tabular}\label{Tab:table-neutral-planar-ring-triplet}
\end{table*}

\begin{table*}
\caption{Excitation energies, $E$, and many-particle wave functions of excited
states corresponding to the peaks in the linear absorption spectrum
of B$_{6}$ -- incomplete wheel isomer (\emph{cf}. Fig. \ref{fig:plots-neutral}\subref{subfig:neutral-plot-incomplete-wheel}).
The subscripts $\Vert$ and $\bot$, in the peak number denote
 the absorption due to light polarized in, and perpendicular to the plane of wheel base, respectively. 
The rest of the information is the same as given in the caption for Table \ref{Tab:table-neutral-planar-ring-triplet}.}
\centering
\begin{tabular}{cccl}
\hline\noalign{\smallskip}
Peak	 	& $E$ (eV) & $f_{12}$ & Wave Function \\
\hline
\tabularnewline
Reference	&	&	& $| H_{1}^{1} ; H_{2}^{1} \rangle$ \\
\tabularnewline
I$_{\Vert}$ 	& 2.09  &0.1460	& $| H_{1} \rightarrow L + 1 \rangle$(0.8102) \\
		&  	&	& $| H_{2} \rightarrow L \rangle$(0.2871) \\
\tabularnewline
II$_{\Vert}$ 	& 3.28 	&0.0406	& $| H_{2} \rightarrow L \rangle$(0.9011) \\
		&  	&	& $| H - 1 \rightarrow  H_{1} \rangle$(0.2359) \\
\tabularnewline
III$_{\bot}$ 	& 3.86 	&0.0247	& $| H - 3 \rightarrow H_{2} \rangle $(0.9301) \\
		&  	&	& $| H - 3 \rightarrow L + 27  \rangle$(0.2154) \\
\tabularnewline
IV$_{\Vert}$ 	& 4.28 	&0.0849	& $| H - 2 \rightarrow H_{2} \rangle$(0.7926) \\
		&  	&	& $| H - 1 \rightarrow H_{1} \rangle$(0.3875) \\
\tabularnewline
V$_{\Vert}$ 	& 5.26 	&0.1008	& $| H - 5 \rightarrow H_{1} \rangle$(0.4241) \\
		&  	&	& $| H - 1 \rightarrow H_{1} \rangle$(0.4168) \\
\tabularnewline
VI$_{\Vert}$ 	& 5.56 	&0.2772	& $| H_{1} \rightarrow L + 16 \rangle $(0.5724) \\
		&  	&	& $| H_{1} \rightarrow L + 5 \rangle$(0.3863) \\
\tabularnewline
VII$_{\Vert}$ 	& 5.96 	&0.0677	& $| H_{1} \rightarrow L + 18 \rangle$(0.5153) \\
		&  	&	& $| H_{1} \rightarrow L + 8 \rangle$(0.3795) \\
\tabularnewline
VIII$_{\Vert}$ 	& 6.57 	&0.1286	& $| H - 1  \rightarrow L \rangle$(0.5208) \\
		&  	&	& $| H - 5 \rightarrow L \rangle$(0.5092) \\
\noalign{\smallskip}\hline
\end{tabular}\label{Tab:table-neutral-incomplete-wheel}
\end{table*}

\begin{table*}
\caption{Excitation energies, $E$, and many-particle wave functions of excited
states corresponding to the peaks in the linear absorption spectrum
of B$_{6}$ -- bulged wheel isomer (\emph{cf}. Fig. \ref{fig:plots-neutral}\subref{subfig:neutral-plot-bulged-wheel}).
The subscripts $\Vert$ and $\bot$, in the peak number denote
 the absorption due to light polarized in, and perpendicular to the plane of 
wheel base, respectively.
In the wave function, the bracketed numbers are the CI coefficients of a given electronic configuration.
Symbols $H$ and $L$ denote HOMO and LUMO orbitals respectively. Excitations are 
with respect to Hartree Fock reference state.}
\centering
\begin{tabular}{cccl}
\hline\noalign{\smallskip}
Peak	 	& $E$ (eV) & $f_{12}$ & Wave Function \\
\hline
\tabularnewline
I$_{\Vert}$ 	& 3.76  &0.2625	& $ | H - 1 \rightarrow L + 6 \rangle$(0.5622)  \\ 
		&	&	& $ | H \rightarrow L + 6 \rangle$(0.5622) \\
		&  	&	& $| H - 1 \rightarrow L + 2 \rangle$(0.4459)\\
		&	&	& $ | H \rightarrow L + 2 \rangle$(0.4459) \\
\tabularnewline
II$_{\Vert}$ 	& 4.38 	&0.0657	& $| H - 1 \rightarrow L + 3 \rangle $(0.4998)\\
		&	&	& $| H \rightarrow L + 3 \rangle$(0.4998) \\
		&  	&	& $| H - 1 \rightarrow L + 4 \rangle $(0.4981)\\
		&	&	& $| H \rightarrow L + 4 \rangle$(0.4981) \\
\tabularnewline
III$_{\Vert}$ 	& 5.07 	&2.1039	& $| H - 2 \rightarrow L + 1 \rangle $(0.7653)\\
		&	&	& $| H - 2 \rightarrow L \rangle$(0.7653) \\
		&  	&	& $| H - 2 \rightarrow L + 8 \rangle $(0.2857)\\
		&	&	& $| H - 2 \rightarrow L + 7 \rangle$(0.2857) \\
\tabularnewline
IV$_{\bot}$ 	& 5.71 	&0.0810	& $| H - 2 \rightarrow L + 6 \rangle$(0.6266) \\
		&  	&	& $| H - 2 \rightarrow L + 2 \rangle$(0.5564) \\
\tabularnewline
V$_{\Vert}$ 	& 7.05 	&1.0550	& $| H \rightarrow L + 5 \rangle $(0.7645)\\
		&	&	& $| H - 1 \rightarrow L + 5 \rangle$(0.7645) \\
		&  	&	& $| H \rightarrow L + 6 \rangle $(0.4763)\\
		&	&	& $| H - 1 \rightarrow L + 6 \rangle$(0.4763) \\
\noalign{\smallskip}\hline
\end{tabular}\label{Tab:table-neutral-bulged-wheel}
\end{table*}

\begin{table*}
\caption{Excitation energies, $E$, and many-particle wave functions of excited
states corresponding to the peaks in the linear absorption spectrum
of B$_{6}$ -- planar ring (singlet) isomer 
(\emph{cf}. Fig. \ref{fig:plots-neutral}\subref{subfig:neutral-plot-planar-ring-singlet}).
The subscripts $\Vert$ and $\bot$, in the peak number denote
 the absorption due to light polarized in, and perpendicular to the plane of 
the isomer, respectively.
The rest of the information is the same as given in the caption for Table \ref{Tab:table-neutral-bulged-wheel}.}
\centering
\begin{tabular}{cccl}
\hline\noalign{\smallskip}
Peak	 	& $E$ (eV) & $f_{12}$ & Wave Function \\
\hline
\tabularnewline
I$_{\Vert}$ 	& 3.29  &0.5397	& $| H - 1 \rightarrow L \rangle$(0.9174) \\
		&  	&	& $| H     \rightarrow L + 23 \rangle$(0.2504) \\
\tabularnewline
II$_{\Vert}$ 	& 5.46 	&1.0153	& $| H  \rightarrow L + 7\rangle$(0.4991) \\
		&  	&	& $| H  \rightarrow L + 23\rangle$(0.4093) \\
\tabularnewline
III$_{\Vert}$ 	& 5.77 	&0.1380	& $| H - 1 \rightarrow L + 2 \rangle$(0.5479) \\
		&  	&	& $| H - 1 \rightarrow L + 8 \rangle$(0.5181) \\
\tabularnewline
IV$_{\Vert,\bot}$& 6.04 &0.8199	& $| H \rightarrow L + 7 \rangle$(0.7472) \\
		&  	&	& $| H \rightarrow L + 12 \rangle$(0.3140) \\
\tabularnewline
V$_{\Vert}$ 	& 6.93 	&2.2212	& $| H \rightarrow L + 16 \rangle$(0.4623) \\
		&  	&	& $| H \rightarrow L + 12 \rangle$(0.4600) \\
\tabularnewline
VI$_{\Vert}$ 	& 7.31 	&0.6003	& $| H \rightarrow L + 16 \rangle$(0.6363) \\
		&  	&	& $| H \rightarrow L + 12 \rangle$(0.5687) \\
\noalign{\smallskip}\hline
\end{tabular}\label{Tab:table-neutral-planar-ring-singlet}
\end{table*}

\begin{table*}
\caption{Excitation energies, $E$, and many-particle wave functions of excited
states corresponding to the peaks in the linear absorption spectrum
of B$_{6}$ -- octahedron isomer 
(\emph{cf}. Fig. \ref{fig:plots-neutral}\subref{subfig:neutral-plot-octahedron}).
The subscripts $\Vert$ and $\bot$, in the peak number denote the 
absorption due to light polarized in, and perpendicular to the plane of 
pyramidal base, respectively.
The rest of the information is the same as given in the caption for Table \ref{Tab:table-neutral-planar-ring-triplet}.}
\centering
\begin{tabular}{cccl}
\hline\noalign{\smallskip}
Peak	 	& $E$ (eV) & $f_{12}$ & Wave Function \\
\hline
\tabularnewline
Reference		&	&	& $| H_{1}^{1} ; H_{2}^{1} \rangle $	\\
\tabularnewline
I$_{\Vert,\bot} $ 	& 3.63  &0.0047	& $| H - 1 \rightarrow L \rangle$(0.8716) \\
			&	&	& $| H - 1 \rightarrow L + 1 \rangle$(0.8716) \\
			&  	&	& $| H - 2 \rightarrow L + 1 \rangle$(0.2638)\\
			&	&	& $| H - 2 \rightarrow L \rangle$(0.2638) \\
\tabularnewline
II$_{\Vert,\bot}$ 	& 5.56 	&0.0129	& $| H - 1 \rightarrow L + 1 \rangle $ (0.5440)\\
			&	&	& $| H - 1 \rightarrow L \rangle$(0.5440) \\
			&  	&	& $| H - 1 \rightarrow L \rangle $(0.5277) \\
			&	&	& $| H - 1 \rightarrow L + 1 \rangle$(0.5277) \\
\tabularnewline
III$_{\Vert,\bot}$ 	& 6.05 	&0.0357	& $| H - 1 \rightarrow L \rangle $ (0.6603) \\
			&	&	& $| H - 1 \rightarrow L + 1 \rangle$(0.6603) \\
			&  	&	& $| H - 1 \rightarrow L + 1 \rangle $ (0.4383) \\
			&	&	& $| H - 1 \rightarrow L \rangle$(0.4383) \\
\tabularnewline
IV$_{\Vert,\bot}$ 	& 7.65 	&0.0149	& $| H - 1 \rightarrow L \rangle $(0.6741) \\
			&	&	& $| H - 1 \rightarrow L + 1 \rangle$(0.4383) \\
			&  	&	& $| H - 2 \rightarrow H_{2} \rangle $(0.5346) \\
			&	&	& $| H - 2 \rightarrow L + 1 \rangle$(0.5346) \\
\noalign{\smallskip}\hline
\end{tabular}\label{Tab:table-neutral-octahedron}
\end{table*}

\begin{table*}
\caption{Excitation energies, $E$, and many-particle wave functions of excited
states corresponding to the peaks in the linear absorption spectrum
of B$_{6}$ -- threaded tetramer isomer 
(\emph{cf}. Fig. \ref{fig:plots-neutral}\subref{subfig:neutral-plot-threaded-tetramer}).
The subscript $x$ in the peak number denote the 
absorption due to light polarized along the long axis, 
and, $y,z$ denotes polarization perpendicular to it.
The rest of the information is the same as given in the caption for Table \ref{Tab:table-neutral-planar-ring-triplet}.}
\centering
\begin{tabular}{cccl}
\hline\noalign{\smallskip}
Peak	 	& $E$ (eV) & $f_{12}$ & Wave Function \\
\hline
\tabularnewline
Reference	&	&	& $|H_{1}^{1}; H_{2}^{1} \rangle $ \\
\tabularnewline
I$_{x} $ 	& 2.74  &0.1116	& $| H -1 \rightarrow H_{2}\rangle$(0.5535) \\
		&  	&	& $| H_{1} \rightarrow L + 9\rangle$(0.3854) \\
\tabularnewline
II$_{x}$ 	& 3.33 	&0.2917	& $| H - 1 \rightarrow H_{2} \rangle$(0.7564) \\
		&  	&	& $| H_{1} \rightarrow L + 9\rangle$(0.2492) \\
		&  	&	& $| H_{1} \rightarrow L + 3\rangle$(0.2465) \\
\tabularnewline
III$_{z}$ 	& 3.90 	&0.0480	& $| H_{1} \rightarrow L \rangle$(0.8447) \\
		&  	&	& $| H - 1 \rightarrow L + 9 \rangle$(0.2022) \\
\tabularnewline
IV$_{z}$ 	& 4.33 	&0.1507	& $| H - 1 \rightarrow L + 9 \rangle$(0.4175) \\
		&  	&	& $| H - 1 \rightarrow L + 3 \rangle$(0.4043) \\
\tabularnewline
V$_{y}$ 	& 5.09 	&0.0338	& $| H - 3 \rightarrow L + 9 \rangle$(0.5013) \\
		&  	&	& $| H - 3 \rightarrow L + 3 \rangle$(0.4697) \\
\tabularnewline
VI$_{y}$ 	& 5.43 	&0.0852	& $| H - 1 \rightarrow L + 1 \rangle$(0.7137) \\
		&  	&	& $| H - 2 \rightarrow L \rangle$(0.2694) \\
\noalign{\smallskip}\hline
\end{tabular}\label{Tab:table-neutral-threaded-tetramer}
\end{table*}

\begin{table*}
\caption{Excitation energies, $E$, and many-particle wave functions of excited
states corresponding to the peaks in the linear absorption spectrum
of B$_{6}$ -- threaded trimer isomer 
(\emph{cf}. Fig. \ref{fig:plots-neutral}\subref{subfig:neutral-plot-threaded-trimer}).
The subscript $\Vert$ , in the peak number denotes the 
absorption due to light polarized along the long 
axis of the isomer.
The rest of the information is the same as given in the caption for Table \ref{Tab:table-neutral-planar-ring-triplet}.}
\centering
\begin{tabular}{cccl}
\hline\noalign{\smallskip}
Peak	 	& $E$ (eV) & $f_{12}$ & Wave Function \\
\hline
\tabularnewline
Reference	&	&	& $| H_{1}^{1} ; H_{2}^{1} \rangle $ \\
\tabularnewline
I$_{\Vert} $ 	& 5.42  &1.4376	& $| H - 3 \rightarrow L \rangle$(0.8262) \\
		&  	&	& $| H - 1 \rightarrow L + 1\rangle$(0.3212) \\
\tabularnewline
II$_{\Vert}$ 	& 5.66 	&2.6846	& $| H - 1 \rightarrow L + 1 \rangle$(0.6710) \\
		&  	&	& $| H - 3 \rightarrow L \rangle$(0.3680) \\
\tabularnewline
III$_{\Vert}$ 	& 6.25 	&0.5077	& $| H - 4 \rightarrow L + 1 \rangle$(0.6263) \\
		&  	&	& $| H - 1 \rightarrow L + 11 \rangle$(0.2702) \\
		&  	&	& $| H - 1 \rightarrow L + 1 \rangle$(0.2701) \\
\noalign{\smallskip}\hline
\end{tabular}\label{Tab:table-neutral-threaded-trimer}
\end{table*}

\begin{table*}
\caption{Excitation energies, $E$, and many-particle wave functions of excited
states corresponding to the peaks in the linear absorption spectrum
of B$_{6}$ -- twisted trimers isomer 
(\emph{cf}. Fig. \ref{fig:plots-neutral}\subref{subfig:neutral-plot-twisted-trimers}).
The subscripts $\Vert$ and $\bot$, in the peak number denote the 
absorption due to light polarized along, and perpendicular to 
the long axis of the isomer, respectively.
The rest of the information is the same as given in the caption for Table \ref{Tab:table-neutral-bulged-wheel}.}
\label{Tab:table-neutral-twisted-trimers}
\centering
\begin{tabular}{cccl}
\hline\noalign{\smallskip}
Peak	 	& $E$ (eV) & $f_{12}$ & Wave Function \\
\hline
\tabularnewline
I$_{\Vert}$ 	& 1.02  &0.0422	& $| H 	  \rightarrow L      \rangle$(0.6550) \\
		&	&	& $| H -1 \rightarrow L      \rangle$(0.6550) \\
		&  	&	& $| H    \rightarrow L + 1  \rangle$(0.5612) \\
		&	&	& $| H -1 \rightarrow L + 1  \rangle$(0.5612) \\
\tabularnewline
II$_{\bot}$ 	& 2.22 	&0.1279	& $| H - 2 \rightarrow L \rangle$(0.7650) \\
		&  	&	& $| H - 3 \rightarrow L + 1\rangle$(0.5316) \\
\tabularnewline
III$_{\Vert}$ 	& 3.58 	&0.1870	& $| H - 2 \rightarrow L + 2 \rangle$(0.4374) \\
		&	&	& $| H -5 \rightarrow L      \rangle$(0.4374) \\
		&  	&	& $| H - 4 \rightarrow L     \rangle$(0.4371) \\
		&	&	& $| H -2 \rightarrow L + 3  \rangle$(0.4371) \\
\tabularnewline
IV$_{\Vert}$	& 4.72 	&0.4337	& $| H - 2 \rightarrow L + 2 \rangle$(0.4874) \\
		&	&	& $| H - 2 \rightarrow L + 3 \rangle$(0.4874) \\
		&  	&	& $| H     \rightarrow L + 26\rangle$(0.3268) \\
		&	&	& $| H - 1 \rightarrow L + 26  \rangle$(0.3268)\\
\tabularnewline
V$_{\bot}$ 	& 5.26 	&0.3570	& $| H - 5 \rightarrow L + 3 \rangle$(0.5117) \\
		&  	&	& $| H - 4 \rightarrow L + 2 \rangle$(0.5010) \\
\tabularnewline
VI$_{\bot}$ 	& 5.87 	&3.4368	& $| H \rightarrow L + 3 \rangle$(0.5962) \\
		&  	&	& $| H - 1 \rightarrow L + 2 \rangle$(0.5962) \\
\tabularnewline
VII$_{\Vert}$ 	& 6.39  &0.2265 & $| H - 3 \rightarrow L + 2 \rangle$(0.4908) \\
		&	&	& $| H - 3 \rightarrow L + 3 \rangle$ (0.4908) \\
		&  	&	& $| H - 2 \rightarrow L + 3 \rangle$(0.4561) \\
		&	&	& $| H - 2 \rightarrow L + 2 \rangle$ (0.4561) \\
\tabularnewline
VIII$_{\bot}$ 	& 6.98 	&0.6645	& $| H - 3  \rightarrow L + 1\rangle$(0.6479) \\
		&  	&	& $| H - 2  \rightarrow L \rangle$(0.5508) \\
\tabularnewline
IX$_{\Vert}$ 	& 7.25 	&0.3140	& $| H \rightarrow L + 4 \rangle$ (0.5209) \\
		&	&	& $| H - 1 \rightarrow L + 4 \rangle$ (0.5209) \\
		&  	&	& $| H \rightarrow L + 7 \rangle$ (0.4072) \\
		&	&	& $| H - 1 \rightarrow L + 7 \rangle$ (0.4072) \\
\noalign{\smallskip}\hline
\end{tabular}
\end{table*}

\begin{table*}
\caption{Excitation energies, $E$, and many-particle wave functions of excited
states corresponding to the peaks in the linear absorption spectrum
of B$_{6}$ -- planar trimers isomer 
(\emph{cf}. Fig. \ref{fig:plots-neutral}\subref{subfig:neutral-plot-planar-trimers}).
The subscripts $\Vert$ and $\bot$, in the peak number denote the 
absorption due to light polarized in, and perpendicular to 
the plane of the isomer, respectively.
The rest of the information is the same as given in the caption for Table \ref{Tab:table-neutral-bulged-wheel}.}
\centering
\begin{tabular}{cccl}
\hline\noalign{\smallskip}
Peak	 	& $E$ (eV) & $f_{12}$ & Wave Function \\
\hline
\tabularnewline
I$_{\bot}$ 	& 0.97  &0.0393	& $| H \rightarrow L  \rangle$ (0.7804) \\
		&  	&	& $| H - 1 \rightarrow L + 1\rangle$(0.5529) \\
\tabularnewline
II$_{\Vert}$	& 2.22 	&0.1261	& $| H - 2 \rightarrow L \rangle$(0.7603) \\
		&  	&	& $| H - 4 \rightarrow L + 1\rangle$(0.5634) \\
\tabularnewline
III$_{\Vert}$ 	& 3.57 	&0.1293	& $| H - 2 \rightarrow L + 2 \rangle$ (0.6070) \\
		&  	&	& $| H - 3 \rightarrow L \rangle$(0.5440) \\
\tabularnewline
IV$_{\Vert}$	& 4.67 	&0.7257	& $| H - 3 \rightarrow L + 2 \rangle$ (0.7242) \\
		&  	&	& $| H \rightarrow L + 12 \rangle$ (0.3824) \\
		&  	&	& $| H - 2 \rightarrow L + 2 \rangle$ (0.5131) \\
		&  	&	& $| H - 3 \rightarrow L \rangle$ (0.4653) \\
\tabularnewline
V$_{\Vert}$ 	& 6.43 	&4.5721	& $| H \rightarrow L + 12 \rangle$(0.5256) \\
		&  	&	& $| H - 3 \rightarrow L + 2 \rangle$(0.4859) \\
\tabularnewline
VI$_{\Vert}$ 	& 6.99 	&1.8466	& $| H - 4 \rightarrow L + 1 \rangle$(0.5794) \\
		&  	&	& $| H \rightarrow L + 7 \rangle$(0.5156) \\
\tabularnewline
VII$_{\Vert}$ 	& 7.34  &0.6386	& $| H \rightarrow L + 12 \rangle$(0.6754) \\
		&  	&	& $| H \rightarrow L + 7 \rangle$(0.4922) \\
\noalign{\smallskip}\hline
\end{tabular}\label{Tab:table-neutral-planar-trimers}
\end{table*}

\begin{table*}
\caption{Excitation energies, $E$, and many-particle wave functions of excited
states corresponding to the peaks in the linear absorption spectrum
of B$_{6}$ -- convex bowl isomer 
(\emph{cf}. Fig. \ref{fig:plots-neutral}\subref{subfig:neutral-plot-convex-bowl}).
The subscripts $\Vert$ and $\bot$, in the peak number denote the 
absorption due to light polarized in, and perpendicular to 
the plane of the isomer, respectively.
The rest of the information is the same as given in the caption for Table \ref{Tab:table-neutral-bulged-wheel}.}
\centering
\begin{tabular}{cccl}
\hline\noalign{\smallskip}
Peak	 	& $E$ (eV) & $f_{12}$ & Wave Function \\
\hline
\tabularnewline
I$_{\Vert}$ 	& 1.58  &0.0486	& $| H - 1 \rightarrow L + 1     \rangle$ (0.9774) \\
\tabularnewline
II$_{\Vert,\bot}$& 1.80 &0.0679	& $| H \rightarrow L + 2 \rangle$(0.9079) \\
		&  	&	& $| H - 1 \rightarrow L + 3\rangle$(0.2748) \\
\tabularnewline
III$_{\Vert}$ 	& 2.43 	&0.6023	& $| H \rightarrow L \rangle$ (0.8644) \\
		&  	&	& $| H - 4 \rightarrow L + 1 \rangle$(0.4364) \\
\tabularnewline
IV$_{\Vert}$	& 2.89 	&0.1811	& $| H - 1 \rightarrow L  \rangle$ (0.8058) \\
		&  	&	& $| H - 3 \rightarrow L + 1\rangle$ (0.5421) \\
\tabularnewline
V$_{\Vert}$ 	& 4.09 	&0.1602	& $| H - 4 \rightarrow L \rangle$(0.5608) \\
		&  	&	& $| H - 3 \rightarrow L + 1 \rangle$(0.5347) \\
\tabularnewline
VI$_{\Vert,\bot}$ & 5.13 &0.0649& $| H - 4 \rightarrow L + 3 \rangle$(0.9070) \\
		&  	&	& $| H - 3 \rightarrow L + 2 \rangle$(0.1753) \\
\tabularnewline
VII$_{\Vert}$ 	& 6.21  &1.4363&  $| H \rightarrow L + 7 \rangle$(0.4118) \\
		&  	&	& $| H \rightarrow L + 4 \rangle$(0.3622) \\
\tabularnewline
VIII$_{\bot}$ 	& 6.39 	&2.2019	& $| H \rightarrow L + 6 \rangle$(0.4335) \\
		&  	&	& $| H - 2  \rightarrow L + 3 \rangle$(0.4081) \\
\tabularnewline
IX$_{\Vert}$ 	& 6.90 	&0.1476	& $| H \rightarrow L + 8 \rangle$(0.4435) \\
		&  	&	& $| H \rightarrow L + 13 \rangle$(0.3534) \\
\noalign{\smallskip}\hline
\end{tabular}\label{Tab:table-neutral-convex-bowl}
\end{table*}

\begin{table*}
\caption{Excitation energies, $E$, and many-particle wave functions of excited
states corresponding to the peaks in the linear absorption spectrum
of B$_{6}$ -- linear isomer 
 (\emph{cf}. Fig. \ref{fig:plots-neutral}\subref{subfig:neutral-plot-linear}).
The subscripts $\Vert$ and $\bot$, in the peak number denote the 
absorption due to light polarized along, and perpendicular to the axis of 
the isomer, respectively.
The rest of the information is the same as given in the caption for Table \ref{Tab:table-neutral-bulged-wheel}.}
\centering
\begin{tabular}{cccl}
\hline\noalign{\smallskip}
Peak	 	& $E$ (eV) & $f_{12}$ & Wave Function \\
\hline
\tabularnewline
I$_{\Vert} $ 	& 5.51	&12.8358& $| H - 1 \rightarrow L + 3 \rangle$(0.6510) \\
		&  	&	& $| H \rightarrow L + 2 \rangle$(0.6489) \\
\tabularnewline
II$_{\bot}$ 	& 6.51 	&1.6532 & $| H \rightarrow L + 4 \rangle $(0.7148)\\
		&	&	& $| H - 1 \rightarrow L + 4 \rangle$(0.7148) \\
		&  	&	& $| H - 3 \rightarrow L + 8 \rangle $(0.3847)\\
		&	&	& $| H - 2 \rightarrow L + 8 \rangle$(0.3847) \\
\noalign{\smallskip}\hline
\end{tabular}\label{Tab:table-neutral-linear}
\end{table*}

\FloatBarrier

\begin{table*}
\caption{Excitation energies, $E$, and many-particle wave functions of excited
states corresponding to the peaks in the linear absorption spectrum
of B$_{6}^{+}$ -- planar ring isomer (\emph{cf}. Fig. \ref{fig:plots-cationic}\subref{subfig:cationic-plot-planar-ring}).
The subscripts $\Vert$ and $\bot$, in the peak number denote the 
absorption due to light polarized in, and perpendicular to the plane of 
the isomer, respectively.
 In the wave function, the bracketed numbers are the CI coefficients of a given electronic configuration.
Symbols $H$, $H_{1}$ and $L$ denote HOMO, SOMO and LUMO orbitals respectively.}
\centering
\begin{tabular}{cccl}
\hline\noalign{\smallskip}
Peak	 	& $E$ (eV) & $f_{12}$ & Wave Function \\
\hline
\tabularnewline
Reference	&	&	& $|H_{1}^{1} \rangle $ \\
\tabularnewline
I$_{\bot}$ 	& 0.63  &0.0075	& $| H_{1} \rightarrow L \rangle$(0.9667) \\
		&  	&	& $| H - 2 \rightarrow L + 16 \rangle$(0.1229) \\
\tabularnewline
II$_{\Vert}$ 	& 4.43 	&0.9637	& $| H - 1 \rightarrow L \rangle$(0.7448) \\
		&  	&	& $| H_{1} \rightarrow L + 10 \rangle$(0.4227) \\
\tabularnewline
III$_{\Vert}$ 	& 5.18 	&0.0565	& $| H - 2 \rightarrow L + 1 \rangle $(0.6283) \\
		&  	&	& $| H - 1 \rightarrow L+1 \rangle$(0.5264) \\
\tabularnewline
IV$_{\Vert}$ 	& 5.64 	&0.4310	& $| H - 1 \rightarrow L+2 \rangle$(0.6350) \\
		&  	&	& $| H - 3 \rightarrow L+3 \rangle$(0.5173) \\
\tabularnewline
V$_{\Vert}$ 	& 6.06 	&0.1576	& $| H - 3 \rightarrow L+3 \rangle$(0.5624) \\
		&  	&	& $| H - 1 \rightarrow L+3 \rangle$(0.4278) \\
\noalign{\smallskip}\hline
\end{tabular}\label{Tab:table-cationic-planar-ring}
\end{table*}

\begin{table*}
\caption{Excitation energies, $E$, and many-particle wave functions of excited
states corresponding to the peaks in the linear absorption spectrum
of B$_{6}^{+}$ -- bulged wheel isomer (\emph{cf}. Fig. \ref{fig:plots-cationic}\subref{subfig:cationic-plot-bulged-wheel}).
The subscripts $\Vert$ and $\bot$, in the peak number denote the 
absorption due to light polarized in, and perpendicular to the plane 
of the wheel, respectively.
The rest of the information is the same as given in the caption for Table \ref{Tab:table-cationic-planar-ring}.}
\centering
\begin{tabular}{cccl}
\hline\noalign{\smallskip}
Peak	 	& $E$ (eV) & $f_{12}$ & Wave Function \\
\hline
\tabularnewline
Reference	&	&	& $|H_{1}^{1} \rangle $ \\
\tabularnewline
I$_{\Vert}$ 	& 1.67  &0.0159	& $| H - 2 \rightarrow H_{1} \rangle$(0.9568) \\
		&  	&	& $| H_{1} \rightarrow L + 3 \rangle$(0.1270) \\
\tabularnewline
II$_{\Vert}$ 	& 3.25 	&0.0061	& $| H_{1} \rightarrow L \rangle$(0.5877) \\
		&  	&	& $| H_{1} \rightarrow L + 3 \rangle$(0.3945) \\
\tabularnewline
III$_{\Vert}$ 	& 3.65 	&0.0021	& $| H_{1} \rightarrow L + 2 \rangle $(0.8106) \\
		&  	&	& $| H_{1} \rightarrow L + 1 \rangle $(0.2830) \\
\tabularnewline
IV$_{\Vert}$ 	& 4.20 	&0.0998	& $| H - 1 \rightarrow L \rangle$(0.5760) \\
		&  	&	& $| H - 2 \rightarrow L \rangle$(0.5743) \\
\tabularnewline
V$_{\Vert}$ 	& 4.46 	&0.0312	& $| H - 1 \rightarrow L+2 \rangle$(0.7945) \\
		&  	&	& $| H_{1} \rightarrow L \rangle$(0.2711) \\
\tabularnewline
VI$_{\Vert}$ 	& 4.69  &0.0417	& $| H - 1 \rightarrow L + 2 \rangle$(0.7945) \\
		&  	&	& $| H - 2 \rightarrow L \rangle$(0.2711) \\
\tabularnewline
VII$_{\Vert}$ 	& 4.91 	&0.0453	& $| H - 2 \rightarrow L + 1 \rangle$(0.6517) \\
		&  	&	& $| H - 1 \rightarrow L + 1 \rangle$(0.4364) \\
\tabularnewline
VIII$_{\Vert}$ 	& 5.60 	&0.4219	& $| H - 1 \rightarrow L + 3 \rangle $(0.6402) \\
		&  	&	& $| H - 2 \rightarrow L \rangle$(0.4002) \\
\tabularnewline
IX$_{\Vert,\bot}$& 5.77 &0.2700	& $| H - 2 \rightarrow L+2 \rangle$(0.8807) \\
		&  	&	& $| H - 1 \rightarrow L+4 \rangle$(0.6632) \\
		&  	&	& $| H - 2 \rightarrow L+1 \rangle$(0.3734) \\
		&  	&	& $| H - 2 \rightarrow L   \rangle$(0.2952) \\
\tabularnewline
X$_{\Vert}$ 	& 6.00 	&0.5544	& $| H - 3 \rightarrow L \rangle$(0.4788) \\
		&  	&	& $| H - 2 \rightarrow L+1 \rangle$(0.4263) \\
\tabularnewline
XI$_{\Vert}$ 	& 6.27 	&0.3573	& $| H - 2 \rightarrow L+3 \rangle$(0.6862) \\
		&  	&	& $| H - 4 \rightarrow L   \rangle$(0.4386) \\
\noalign{\smallskip}\hline
\end{tabular}\label{Tab:table-cationic-bulged-wheel}
\end{table*}


\begin{table*}
\caption{Excitation energies, $E$, and many-particle wave functions of excited
 states corresponding to the peaks in the linear absorption spectrum
 of B$_{6}^{+}$ -- planar ring (II) isomer (\emph{cf}. Fig. \ref{fig:plots-cationic}\subref{subfig:cationic-plot-planar-d2h-ring}).
The subscript $\Vert$ in the peak number denote the 
  absorption due to light polarized in the plane of 
  the isomer.
The rest of the information is the same as given in the caption for Table \ref{Tab:table-cationic-planar-ring}.}
  \centering
  \begin{tabular}{cccl}
 \hline\noalign{\smallskip}
 Peak	 	& $E$ (eV) & $f_{12}$ & Wave Function \\
 \hline
\tabularnewline
Reference	&	&	& $| H_{1}^{1} \rangle $ \\
 \tabularnewline
 I$_{\Vert}$ 	& 1.18  &0.0182	& $| H - 1 \rightarrow L \rangle$(0.6957) \\
 		&  	&	& $| H - 2 \rightarrow H_{1} \rangle$(0.6943) \\
 \tabularnewline
 II$_{\Vert}$ 	& 3.47 	&1.5345	& $| H - 1 \rightarrow L \rangle$(0.7424) \\
 		&  	&	& $| H - 2 \rightarrow H_{1} \rangle$(0.4272) \\
 \tabularnewline
 III$_{\Vert}$ 	& 4.70 	&0.1042	& $| H - 2 \rightarrow L + 1 \rangle $(0.7037) \\
 		&  	&	& $| H - 4 \rightarrow L \rangle$(0.4979) \\
 \tabularnewline
 IV$_{\Vert}$ 	& 4.93 	&0.1114	& $| H - 1 \rightarrow L+3 \rangle$(0.9077) \\
 		&  	&	& $| H - 5 \rightarrow L \rangle$(0.1606) \\
 \tabularnewline
 V$_{\Vert}$ 	& 5.12 	&0.0673	& $| H - 4 \rightarrow L \rangle$(0.7687) \\
 		&  	&	& $| H - 3 \rightarrow L+2 \rangle$(0.5208) \\
 \tabularnewline
 VI$_{\Vert}$ 	& 6.43 	&0.8286	& $| H_{1} \rightarrow L+6 \rangle$(0.6655) \\
		&  	&	& $| H - 1 \rightarrow L+3 \rangle$(0.4625) \\
\noalign{\smallskip}\hline
\end{tabular}\label{Tab:table-cationic-planar-d2h-ring}
\end{table*}

\begin{table*}
\caption{Excitation energies, $E$, and many-particle wave functions of excited
states corresponding to the peaks in the linear absorption spectrum
of B$_{6}^{+}$ -- incomplete wheel (quartet) isomer 
(\emph{cf}. Fig. \ref{fig:plots-cationic}\subref{subfig:cationic-plot-incomplete-wheel}).
The subscripts $\Vert$ and $\bot$, in the peak number denote the 
absorption due to light polarized in, and perpendicular 
to the plane of the isomer, respectively.
In the wave function, the bracketed numbers are the CI coefficients of a given electronic configuration.
Symbols $H$, $H_1$, $H_2$, $H_3$ and $L$ denote HOMO, SOMOs and LUMO orbitals respectively.}
\centering
\begin{tabular}{cccl}
\hline\noalign{\smallskip}
Peak	 	& $E$ (eV) & $f_{12}$ & Wave Function \\
\hline
\tabularnewline
Reference	&	&	& $|H_{1}^{1};H_{2}^{1};H_{3}^{1} \rangle $ \\
\tabularnewline
I$_{\bot}$ 	& 2.94  &0.0608	& $| H_{1} \rightarrow L + 1 \rangle$(0.7513) \\
		&  	&	& $| H_{2} \rightarrow L   \rangle$(0.6261) \\
\tabularnewline
II$_{\Vert}$	& 3.25 	&0.1207	& $| H - 5 \rightarrow H_{3} \rangle$(0.7994) \\
		&  	&	& $| H - 4 \rightarrow L  \rangle$(0.3377) \\
\tabularnewline
III$_{\bot}$ 	& 4.84 	&0.0248	&$| H_{3} \rightarrow L \rangle $(0.8592) \\
		&  	&	&$| H_{1} \rightarrow L + 1 \rangle $(0.2632) \\
\tabularnewline
IV$_{\Vert}$ 	& 5.60  &0.0552	& $| H - 3 \rightarrow H_{2} \rangle$(0.6397) \\
		&  	&	& $| H_{1} \rightarrow L + 3  \rangle$(0.4131) \\
\tabularnewline
V$_{\Vert}$	& 6.8	&0.0674	& $| H - 3 \rightarrow H_{1} \rangle$(0.6876) \\
		&  	&	& $| H - 4\rightarrow L  \rangle$(0.4045) \\
\tabularnewline
VI$_{\Vert}$ 	& 7.25 	&0.1363	&$| H - 5 \rightarrow H_{2} \rangle $(0.7232) \\
		&  	&	&$| H_{2} \rightarrow L \rangle$(0.2883) \\
\tabularnewline
VII$_{\bot}$ 	& 7.39  &0.0336	& $| H - 3 \rightarrow L  \rangle$(0.4522) \\
		&  	&	& $| H_{1} \rightarrow L + 5  \rangle$(0.3485) \\
\noalign{\smallskip}\hline
\end{tabular}\label{Tab:table-cationic-incomplete-wheel}
\end{table*}

\begin{table*}  
\caption{Excitation energies, $E$, and many-particle wave functions of excited
states corresponding to the peaks in the linear absorption spectrum
of B$_{6}^{+}$ -- threaded trimer (quartet) isomer 
(\emph{cf}. Fig. \ref{fig:plots-cationic}\subref{subfig:cationic-plot-threaded-trimer}).
The subscripts $\Vert$ and $\bot$, in the peak number denote the 
absorption due to light polarized along, and perpendicular 
to the long axis of the isomer, respectively.
The rest of the information is the same as given in the caption for Table \ref{Tab:table-cationic-incomplete-wheel}.}
\centering
\begin{tabular}{cccl}
\hline\noalign{\smallskip}
Peak	 	& $E$ (eV) & $f_{12}$ & Wave Function \\
\hline
\tabularnewline
Reference	&	&	& $|H_{1}^{1};H_{2}^{1};H_{3}^{1} \rangle $ \\
\tabularnewline
I$_{\Vert}$ 	& 3.18  &0.4369	& $| H - 3  \rightarrow H_{3} \rangle$(0.6993) \\
		&  	&	& $| H - 3  \rightarrow L   \rangle$(0.5802) \\
\tabularnewline
II$_{\Vert,\bot}$& 4.48 &0.1117	& $| H_{3} \rightarrow L + 2 \rangle$(0.8855) \\
		&  	&	& $| H - 3 \rightarrow L  \rangle$(0.7012) \\
		&  	&	& $| H - 4 \rightarrow L + 2 \rangle$(0.4816) \\
		&  	&	& $| H - 3 \rightarrow L + 3 \rangle$(0.3931) \\
\tabularnewline
III$_{\Vert}$ 	& 5.03 	&2.3151	&$| H_{3} \rightarrow L \rangle $(0.6337) \\
		&  	&	&$| H_{2} \rightarrow L + 1 \rangle$(0.5640) \\
\noalign{\smallskip}\hline
\end{tabular}\label{Tab:table-cationic-threaded-trimer}
\end{table*}

\begin{table*}
\caption{Excitation energies, $E$, and many-particle wave functions of excited
states corresponding to the peaks in the linear absorption spectrum
of B$_{6}^{+}$ -- tetragonal bipyramid isomer 
(\emph{cf}. Fig. \ref{fig:plots-cationic}\subref{subfig:cationic-plot-bipyramid}).
The subscripts $\Vert$ and $\bot$, in the peak number denote the 
absorption due to light polarized in, and perpendicular to the square plane of 
bipyramid, respectively.
The rest of the information is the same as given in the caption for Table \ref{Tab:table-cationic-planar-ring}.}
\centering
\begin{tabular}{cccl}
\hline\noalign{\smallskip}
Peak	 	& $E$ (eV) & $f_{12}$ & Wave Function \\
\hline
\tabularnewline
Reference	&	&	& $| H_{1}^{1} \rangle $ \\
\tabularnewline
I$_{\bot}$ 	& 1.58  &0.2525	& $| H_{1} \rightarrow L + 2 \rangle$(0.7189) \\
		&  	&	& $| H - 1 \rightarrow L + 2 \rangle$(0.6154) \\
\tabularnewline
II$_{\Vert}$ 	& 1.97 	&0.0803	& $| H_{1} \rightarrow L + 1 \rangle$(0.7492) \\
		&  	&	& $| H - 1 \rightarrow L + 1 \rangle$(0.5593) \\
\tabularnewline
III$_{\Vert,\bot}$ & 3.87 &0.0272&$| H - 2 \rightarrow L \rangle $(0.8717) \\
		&  	&	& $| H - 1 \rightarrow L \rangle$(0.8684) \\
		&  	&	& $| H - 2 \rightarrow L + 1  \rangle$(0.4082) \\
		&  	&	& $| H_{1} \rightarrow L \rangle$(0.3497) \\
\tabularnewline
IV$_{\Vert}$ 	& 4.36 	&0.0287	& $| H - 2 \rightarrow L+2 \rangle$(0.8419) \\
		&  	&	& $| H - 1 \rightarrow L   \rangle$(0.3385) \\
\tabularnewline
V$_{\Vert}$ 	& 4.55 	&0.0440	& $| H - 2 \rightarrow L + 2 \rangle$(0.9001) \\
		&  	&	& $| H - 4 \rightarrow L + 2 \rangle$(0.8807) \\
		&  	&	& $| H - 1 \rightarrow L     \rangle$(0.2555) \\
		&  	&	& $| H - 5 \rightarrow L     \rangle$(0.1997) \\
\tabularnewline
VI$_{\Vert,\bot}$& 5.04	&0.8028	& $| H - 4 \rightarrow L+2 \rangle$(0.8991) \\
		&  	&	& $| H - 1 \rightarrow L+2 \rangle$(0.6152) \\
		&  	&	& $| H - 4 \rightarrow L+3 \rangle$(0.5964) \\
		&  	&	& $| H - 1 \rightarrow L \rangle$(0.5334) \\
		&  	&	& $| H - 2 \rightarrow L + 2 \rangle$(0.4299) \\
		&  	&	& $| H - 1 \rightarrow L + 3 \rangle$(0.3269) \\
\noalign{\smallskip}\hline
\end{tabular}\label{Tab:table-cationic-triangular-bipyramid}
\end{table*}

\begin{table*}
\caption{Excitation energies, $E$, and many-particle wave functions of excited
states corresponding to the peaks in the linear absorption spectrum
of B$_{6}^{+}$ -- linear (quartet) isomer 
(\emph{cf}. Fig. \ref{fig:plots-cationic}\subref{subfig:cationic-plot-linear}).
The subscripts $\Vert$ and $\bot$, in the peak number denote the 
absorption due to light polarized along, and perpendicular 
to the long axis of the isomer, respectively.
The rest of the information is the same as given in the caption for Table \ref{Tab:table-cationic-incomplete-wheel}.}
\centering
\begin{tabular}{cccl}
\hline\noalign{\smallskip}
Peak	 	& $E$ (eV) & $f_{12}$ & Wave Function \\
\hline
\tabularnewline
Reference	&	&	& $| H_{1}^{1};H_{2}^{1};H_{3}^{1} \rangle $ \\
\tabularnewline
I$_{\Vert}$ 	& 0.93  &0.1533	& $| H - 3 \rightarrow H_{3} \rangle$(0.8624) \\
		&  	&	& $| H_{2} \rightarrow L   \rangle$(0.3187) \\
\tabularnewline
II$_{\bot}$	& 3.62 	&0.0677	& $| H_{2} \rightarrow L \rangle$(0.7478) \\
		&  	&	& $| H_{1} \rightarrow L \rangle$(0.6474) \\
\tabularnewline
III$_{\Vert}$ 	& 5.02 	&0.6861	&$| H - 3 \rightarrow L + 4  \rangle $(0.9263) \\
		&  	&	&$| H - 3 \rightarrow L + 12 \rangle$(0.2264) \\
\tabularnewline
IV$_{\Vert}$ 	& 5.60 	&9.8432 &$| H - 4 \rightarrow L \rangle $(0.6486) \\
		&  	&	&$| H - 3 \rightarrow L + 4 \rangle$(0.3521) \\
\noalign{\smallskip}\hline
\end{tabular}\label{Tab:table-cationic-linear}
\end{table*}

\begin{table*}
\caption{Excitation energies, $E$, and many-particle wave functions of excited
states corresponding to the peaks in the linear absorption spectrum
of B$_{6}^{+}$ -- planar trimers isomer 
(\emph{cf}. Fig. \ref{fig:plots-cationic}\subref{subfig:cationic-plot-planar-trimers}).
The subscripts $\Vert$ and $\bot$, in the peak number denote the 
absorption due to light polarized in, and perpendicular to the plane of 
the isomer, respectively.
The rest of the information is the same as given in the caption for Table \ref{Tab:table-cationic-planar-ring}.}
\centering
\begin{tabular}{cccl}
\hline\noalign{\smallskip}
Peak	 	& $E$ (eV) & $f_{12}$ & Wave Function \\
\hline
\tabularnewline
Reference	&	&	& $| H_{1}^{1} \rangle $ \\
\tabularnewline
I$_{\Vert}$ 	& 0.77  &0.2915	& $| H - 2 \rightarrow H_{1} \rangle$(0.9647) \\
		&  	&	& $| H_{1} \rightarrow L + 4 \rangle$(0.1605) \\
\tabularnewline
II$_{\bot}$ 	& 1.33 	&0.0108	& $| H_{1} \rightarrow L  \rangle$(0.8340) \\
		&  	&	& $| H - 2 \rightarrow L + 1 \rangle$(0.5093) \\
\tabularnewline
III$_{\bot}$ 	& 2.08 	&0.0378	& $| H - 2 \rightarrow L + 1 \rangle $(0.9744) \\
		&  	&	& $| H_{1} \rightarrow L \rangle$(0.1888) \\
\tabularnewline
IV$_{\bot}$ 	& 2.51 	&0.0103	& $| H - 3 \rightarrow H_{1} \rangle$(0.9414) \\
		&  	&	& $| H - 1 \rightarrow L + 4  \rangle$(0.2905) \\
\tabularnewline
V$_{\Vert}$ 	& 2.86 	&0.1131	& $| H - 1 \rightarrow L  \rangle$(0.7701) \\
		&  	&	& $| H - 3 \rightarrow L + 1 \rangle$(0.5892) \\
\tabularnewline
VI$_{\Vert}$	& 4.16	&0.1267	& $| H - 1 \rightarrow L+2 \rangle$(0.6627) \\
		&  	&	& $| H - 4 \rightarrow L \rangle$(0.5105) \\
\tabularnewline
VII$_{\Vert}$ 	& 5.30 	&0.4383	& $| H_{1} \rightarrow L + 9 \rangle$(0.5023) \\
		&  	&	& $| H - 1 \rightarrow L + 2 \rangle$(0.4505) \\
\noalign{\smallskip}\hline
\end{tabular}\label{Tab:table-cationic-planar-trimers}
\end{table*}

\FloatBarrier
 
 \bibliographystyle{epj}
\providecommand{\noopsort}[1]{}\providecommand{\singleletter}[1]{#1}%


\end{document}